\documentclass[a4paper,fleqn,usenatbib]{mnras}

\usepackage{newtxtext,newtxmath}

\usepackage{amssymb}
\usepackage{graphicx}

\title[SMBH spin in galactic nuclei simulations]{Galactic nuclei evolution with spinning black holes: method and implementation}

\author[Fiacconi, Sijacki, \& Pringle]{Davide Fiacconi$^{1,2}$\thanks{E-mail: fiacconi@ast.cam.ac.uk}, Debora Sijacki$^{1,2}$, and J. E. Pringle$^{1}$\\
$^{1}$Institute of Astronomy, University of Cambridge, Madingley Road, Cambridge CB3 0HA, UK\\
$^{2}$Kavli Institute for Cosmology, University of Cambridge, Madingley Road, Cambridge CB3 0HA, UK\\
}

\begin{document}

\date{\today}

\pagerange{\pageref{firstpage}--\pageref{lastpage}} \pubyear{2018}

\maketitle

\label{firstpage}


\begin{abstract}
Supermassive black holes at the centre of galactic nuclei mostly grow in mass through gas accretion over cosmic time.
This process also modifies the angular momentum (or spin) of black holes, both in magnitude and in orientation.
Despite being often neglected in galaxy formation simulations, spin plays a crucial role in modulating accretion power, driving jet feedback, and determining recoil velocity of coalescing black hole binaries.
We present a new accretion model for the moving-mesh code {\sc arepo} that incorporates (i) mass accretion through a thin $\alpha$-disc, and (ii) spin evolution through the Bardeen-Petterson effect.
We use a diverse suite of idealised simulations to explore the physical connection between spin evolution and larger scale environment.
We find that black holes with mass $\lesssim 10^{7}$~M$_{\sun}$ experience quick alignment with the accretion disc.
This favours prolonged phases of spin-up, and the spin direction evolves according to the gas inflow on timescales as short as $\lesssim 100$~Myr, which might explain the observed jet direction distribution in Seyfert galaxies.
Heavier black holes ($\gtrsim 10^{8}$~M$_{\sun}$) are instead more sensitive to the local gas kinematic.
Here we find a wider distribution in spin magnitudes: spin-ups are favoured if gas inflow maintains a preferential direction, and spin-downs occur for nearly isotropic infall, while the spin direction does not change much over short timescales $\sim 100$~Myr.
We therefore conclude that supermassive black holes with masses $\gtrsim 5 \times 10^{8}$~M$_{\sun}$ may be the ideal testbed to determine the main mode of black hole fuelling over cosmic time.
\end{abstract}



\section{Introduction}

Firm observational evidence attests that supermassive black holes with masses $10^{6} \lesssim M_{\bullet}/{\rm M}_{\sun} \lesssim 10^{10}$ ubiquitously reside in the nuclei of massive galaxies.
Their presence may be indirectly inferred via the gravitational imprint that they leave on the motion of nearby stars (e.g. \citealt{eckart+97,ferrarese+05,ghez+05,vandenbosch+10}) or gas (e.g \citealt{defrancesco+08,greene+10,kuo+11,vandenbosch+16}), as well as by the widely accepted idea that supermassive black holes power the emission of active galactic nuclei (AGN) through mass accretion (e.g. \citealt{lyndenbell+69,urry+95}).
Such emission not only carries information about the central engine, but probably has a direct impact on the properties of the host galaxy, i.e. what is customarily called ``AGN feedback'' (e.g. \citealt{springel+05b,sijacki+07,dubois+12,fabian+12}).
This is suggested by the discovery of scaling relations between the supermassive black hole mass and the bulge mass and velocity dispersion of the host (e.g. \citealt{ferrarese+00,tremaine+02,gultekin+09,kormendy+13,mcconnell+13}), indicating a possible mutual connection between supermassive black holes and their hosts (e.g. \citealt{silk+98,king+03,king+15,sijacki+15}).

Gas accretion appears to be the main mechanism to grow supermassive black holes over cosmic time (e.g. \citealt{soltan+82}), as black hole - black hole mergers have a subdominant contribution.
It is believed that several distinct physical processes contribute to gas transport from kpc scales all the way to the event horizon of supermassive black holes.
For example, galaxy mergers (e.g. \citealt{barnes+91,springel+05b,hopkins+06}) or galactic bars (e.g \citealt{laine+02,laurikainen+04,hopkins+10,fanali+15}) may efficiently funnel gas toward the galactic nucleus through gravitational torques over a few dynamical times. 
When the gas finally reaches the proximity of the supermassive black hole, it settles into an accretion disc and mass transport proceeds at the lower pace imposed by the effective viscosity of the accretion disc \citep{shakura+73,king+07}.

The assembly of supermassive black holes has been extensively studied within the theoretical framework of galaxy formation by means of cosmological simulations \citep{sijacki+07, dimatteo+12, sijacki+15, rosasguevara+16,volonteri+16, weinberger+17}.
However, the details of gas accretion are necessarily encapsulated in simplified sub-grid recipes based on radial accretion solutions that directly connect the mass accretion rate to the large scale properties of the gas \citep{hoyle+41,bondi+52}.
These recipes usually neglect gas angular momentum (e.g. \citealt{booth+09,biernacki+17}), except for some recent attempts to include it \citep{angles-alcazar+13,angles-alcazar+15,rosasguevara+15,curtis+16b}.

While often uniquely considered, the mass is not the only fundamental quantity of astrophysical black holes that may influence the details of accretion.
The second quantity is the black hole angular momentum, often dubbed ``spin''.
Over the last ten years, there have been several attempts to measure the spin of supermassive black holes in nearby galaxies through X-ray spectroscopy by modelling the shape of the reflected iron K$\alpha$ line at 6.4 keV \citep{fabian+00}.
However, the observational inference of the spin is possibly even more challenging than estimating the mass of a supermassive black hole and the results are still widely debated (e.g. \citealt{brenneman+06,schmoll+09,delacalleperez+10,patrick+11,brenneman+13,reynolds+14}).

Constraining the distribution and understanding the evolution of supermassive black hole spins is of fundamental relevance to understand the assembly of supermassive black holes over cosmic time and their connection with the parent galaxies.
Indeed, spin significantly modifies the radiative efficiency to convert mass accretion energy in radiation, going from $\sim 5\%$ for non-rotating black holes to about 40\% for maximally spinning black holes.
This effectively modulates not only the black hole mass growth but the energy at disposal for AGN feedback, and therefore the potential impact on the host galaxy (e.g. \citealt{king+06, sijacki+09}).
The rotational energy of a spinning black hole is also thought to be the reservoir of energy to launch relativistic jets that likely contribute to the evolution of the intracluster
medium in massive galaxy clusters (e.g. \citealt{blandford+77,tchekhovskoy+11,fabian+12}).
Moreover, the gravitational recoil kick after the emission of gravitational waves from a coalescing supermassive black hole binary is strongly dependent on the amplitude and relative alignment of the two black hole angular momenta and it can range from less than $100$~km~s$^{-1}$ to a few thousands km~s$^{-1}$ (e.g. \citealt{schnittman+07,baker+08,lousto+12}), potentially affecting the occupation fraction of supermassive black holes in their host galaxies \citep{schnittman+07b,sijacki+09,volonteri+10,gerosa+15}. 

Most of the theoretical work on supermassive black hole spin either focused on analytical (e.g. \citealt{king+05,martin+07,perego+09,dotti+13}) and numerical (e.g. \citealt{fragile+07,tchekhovskoy+11,nixon+12b}) calculations of the interaction between the spin and the accretion disc, or on semi-analytic models attempting to explore the role of spin in the broader context of the assembly of galaxies (e.g. \citealt{berti+08, fanidakis+11,volonteri+13}).
However, less has been done in trying to include spin evolution in detailed hydrodynamical simulations \citep{sijacki+09,maio+13,dubois+14a, dubois+14b}.
Building up on previous theoretical work, in this paper we incorporate the results of the analytical theory into a new model suitable for galaxy formation simulations to self-consistently describe (i) mass accretion and angular momentum transfer from large scales to the accretion disc, and (ii) the interplay between accretion disc and the black hole spin.
While here we describe the main properties of the model and we show applications to idealised simulations to study the physical mechanisms responsible for linking the spin evolution with the local environment, our aim is to apply this model to cosmological simulations of galaxy formation. 

The paper is organised as follows.
Section~\ref{sec_model_description} provides a detailed theoretical description of the accretion model, the physical processes that connect the accretion disc and the black hole spin, and the numerical implementation of the model in the moving-mesh code {\sc arepo}. 
Section~\ref{sec_results} presents our results based on an extensive suite of simulations to test the capabilities of the model and to explore how a variety of physical conditions of the gas distribution in the nucleus of galaxies may impact the spin evolution.
We discuss our results in Section~\ref{sec_discussion}, also highlighting possible shortcomings of the model and future directions for improvements, before we summarise our findings in Section~\ref{sec_conclusions}.


\begin{figure}
\begin{center}
\includegraphics[width=\columnwidth]{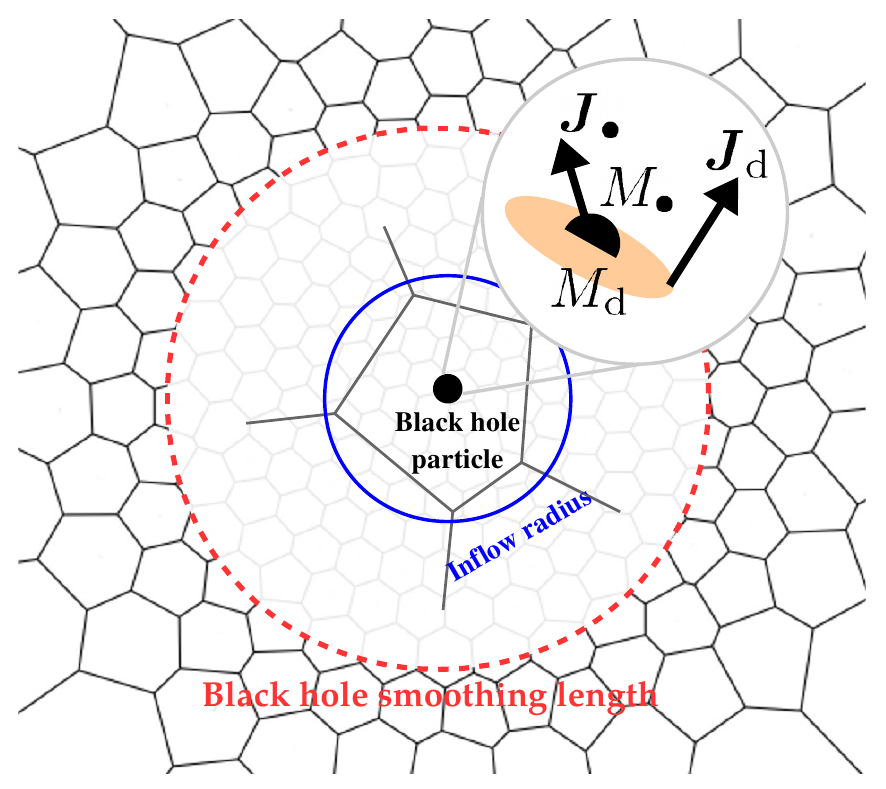}
\caption{
Schematic view of the black hole accretion model. 
The Voronoi tessellation shows the discretisation of the computational domain within the {\sc arepo} code surrounding a given black hole particle shown in the centre.
The red dashed circle marks the smoothing length associated with the black hole, i.e. the radius that contains $N_{\rm neighbours}$ mesh-generating points.
The central part of the figure shows a zoom-in on the black hole particle enclosed by the parent hydro cell. 
The blue solid circle denotes the inflow radius $\langle r_{\rm inflow} \rangle$ described in Section~\ref{subsec_inflow}.
The inset within the grey circle shows a schematic view of the quantities stored in the black hole particle, namely the black hole mass $M_{\bullet}$, the black hole angular momentum $\bmath{J}_{\bullet}$, the accretion disc mass $M_{\rm d}$, and the accretion disc angular momentum $\bmath{J}_{\rm d}$.
}
\label{fig_model_scheme}
\end{center}
\end{figure}

\section{Modelling black hole spin evolution through accretion and disc coupling} \label{sec_model_description}


\subsection{Model overview and simulation code}

We proceed to describe our new sub-grid black hole accretion model that tracks the evolution of the mass and angular momentum of a central black hole surrounded by a thin accretion disc experiencing inflow from the outer environment.
Throughout the paper, we adopt the following notation.
We refer to any quantity associated with the central black hole by means of the subscript $_{\bullet}$, whereas we identify the respective quantities for the accretion disc with the subscript $_{\rm d}$.
The quantities related to inflow have the subscript $_{\rm inflow}$.
The angular momentum and the specific angular momentum are $\bmath{J}$ and $\bmath{L}$, while the associated versors are $\bmath{j}$ and $\bmath{l}$, respectively. 

Figure~\ref{fig_model_scheme} shows a schematic view of our model further described below.
In essence, we consider a black hole surrounded by an accretion disc on a sub-grid scale which we model analytically.
We evolve the mass and angular momentum vector of both the black hole and the accretion disc subject to mutual interaction and external inflow.
We have implemented the model in the moving-mesh code {\sc arepo} \citep{springel+10, pakmor+16}, which adopts a second-order, finite-volume solver on a Voronoi tessellation for the equations of inviscid hydrodynamics, and a hierarchical octtree algorithm for gravity \citep{barnes+86,springel+05} supplemented with the PM method.
The unstructured mesh evolves as the mesh-generating points follow the fluid motion, providing nearly Lagrangian adaptivity and the capability to locally refine and derefine the mesh according to different criteria.


\subsection{Mass accretion through a thin accretion disc}\label{subsec_accretion}

We start from a modified version of the accretion disc particle method proposed by \citet{power+11}.
We assume that a black hole particle represents a central black hole surrounded by a sub-grid, unresolved accretion disc.
The black hole is described by the mass $M_{\bullet}$ and the angular momentum
\begin{equation}
\bmath{J}_{\bullet} = \frac{G M_{\bullet}^2 a_\bullet}{c} \bmath{j}_{\bullet},
\end{equation}
where $0 \leq a_{\bullet} \lesssim 1$ is the dimensionless spin, or spin parameter, while $G$ and $c$ retain the usual meaning of gravitational constant and speed of light, respectively.
\citet{thorne+74} has shown that the spin up of the black hole is limited to $a_{\bullet} \approx 0.998$ because of torques induced by photons produced in the inner accretion disc and swallowed by the black hole. Therefore, we cap the spin parameter at 0.998.
The accretion disc is globally parametrised by its total mass $M_{\rm d}$ and its total angular momentum $\bmath{J}_{\rm d}$.
The structure of the disc follows the geometrically thin $\alpha$-disc solution by \citet{shakura+73} under the following assumptions: (i) steady state; (ii) the gas pressure is much higher than the radiation pressure, and (iii) free-free absorption dominates the disc opacity\footnote{Our assumptions correspond to part c) of the solution presented by  \citet{shakura+73}. While the disc structure may change in the inner region, for the sake of simplicity we assume that the same solution and the implied scalings are valid everywhere in the disc.}.
Under these assumptions, we calculate the mass accretion rate $\dot{M}$ from the accretion disc on to the black hole as the unique value (once the free parameter $\alpha$ is specified) that is consistent with a thin $\alpha$-disc solution that has a total mass $M_{\rm d}$ and total angular momentum $J_{\rm d}$, and we express $\dot{M}$ in units of the Eddington accretion rate $\dot{M}_{\rm Edd}$ as
\begin{equation} \label{eq_f_edd}
f_{\rm Edd} \approx 0.76 \left( \frac{\eta}{0.1} \right) \left( \frac{M_{\rm d}}{10^{4}~{\rm M}_{\sun}} \right)^5 \left(\frac{M_{\bullet}}{10^6~{\rm M}_{\sun}} \right)^{-47/7} \left( \frac{a_{\bullet} J_{\rm d} / J_{\bullet}}{3} \right)^{-25/7}.
\end{equation}
The latter equation is derived explicitly in Appendix~\ref{appendix_formulae}, where we also discuss the underlying assumptions.
Here $\eta$ is the $a_{\bullet}$-dependent radiative efficiency of a thin disc (\citealt{novikov+73}; see Appendix~\ref{appendix_afunc} for the explicit definition) and we have implicitly assumed $\alpha = 0.1$, in agreement with the available observational constraints on thin, fully-ionised accretion discs \citep{king+07}.
Nonetheless, we note that the exact value of $\alpha$ enters the normalisation of the thin disc quantities only weakly (see e.g. \citealt{frank+02}).
We then calculate $\dot{M}$ as
\begin{equation}
\dot{M} = \dot{M}_{\rm Edd} \min(f_{\rm Edd}, 1) = \frac{M_{\bullet}}{\eta \tau_{\rm Sal}} \min(f_{\rm Edd}, 1),
\end{equation}
where $\tau_{\rm Sal} = \kappa_{\rm es} c / (4 \pi G) \approx 4.5 \times 10^{8}$~yr is the Salpeter time calculated with the electron scattering opacity $\kappa_{\rm es} \approx 0.4$~cm$^{2}$~g$^{-1}$, and the term $\min(f_{\rm Edd}, 1)$ limits the accretion rate to $\dot{M}_{\rm Edd}$, consistently with the assumptions behind the thin disc model.
In addition to that, we also impose a lower limit $f_{\rm Edd}^{\rm (min)} = 10^{-4}$ to $f_{\rm Edd}$; we discuss the limitations of this choice in Section~\ref{sec_discussion}, but we do not observe any significant effect of this choice as long as $f_{\rm Edd}^{\rm (min)} \lesssim 10^{-3}$.

Once we have specified the mass accretion rate $\dot{M}$ from the accretion disc to the black hole, the evolution of $M_{\bullet}$ and $M_{\rm d}$ follows from the conservation of mass between the two components with the additional contribution from the large scale inflow $\dot{M}_{\rm inflow}$, i.e.
\begin{equation}\label{eq_bh_mass_evol}
\frac{{\rm d} M_{\bullet}}{{\rm d}t} = (1 - \eta) \dot{M},
\end{equation}
and
\begin{equation}\label{eq_disc_mass_evol}
\frac{{\rm d} M_{\rm d}}{{\rm d}t} = -\dot{M} + \dot{M}_{\rm inflow}.
\end{equation}
We note that the black hole accretes only a fraction $(1 - \eta)$ of the available mass as the remaining is released in radiation and can be in principle coupled with a feedback model.
Equation~(\ref{eq_bh_mass_evol}) describes the transfer of mass from the sub-grid accretion disc to the black hole, whereas equation~(\ref{eq_disc_mass_evol}) dictates the evolution of the accretion disc mass.
The latter links the sub-grid model to the large-scale hydrodynamical simulation through the mass inflow rate $\dot{M}_{\rm inflow}$, i.e. the mass inflow that eventually circularises and joins the accretion disc before being accreted by the black hole, as further discussed in Section~\ref{subsec_angmom} and \ref{subsec_inflow}.

The balance between accretion on to the black hole and inflow from larger scales will induce the growth of $M_{\rm d}$ if $\dot{M}_{\rm inflow} > \dot{M}$.
As the disc mass increases, its self-gravity may become significant and the outer parts of the accretion disc, where the Toomre parameter $Q \lesssim 1$, undergo gravitational instabilities \citep{kolykhalov+80,pringle+81, lodato+07, perego+09}. 
Specifically, gravitational instabilities are likely to occur when the accretion disc mass becomes larger than the limiting mass $M_{\rm sg}$, i.e. the mass that an arbitrarily extended thin $\alpha$-disc would contain within the radius where $Q(R) = 1$, 
\begin{equation} \label{eq_M_sg}
M_{\rm sg} \approx 2 \times 10^{4} \left(\frac{f_{\rm Edd}}{ \eta_{0.1}} \right)^{4 / 45} \left( \frac{M_{\bullet}}{10^{6}~{\rm M}_{\sun}} \right)^{34 / 45}~{\rm M}_{\sun},
\end{equation}
where $\eta_{0.1} = \eta/0.1$. We derive explicitly equation~(\ref{eq_M_sg}) in Appendix~\ref{appendix_formulae}.
If $M_{\rm d} > M_{\rm sg}$, the outer disc fragments and eventually may form stars.
As a result, the mass and angular momentum transport through the disc may significantly depart from the viscous one assumed by the thin disc model; however, the exact details of this
process are not clear \citep{goodman+03}.
On one hand, the development of spiral arms and clumps in the outer regions of the disc could induce gravitational torques able to remove angular momentum from the inner regions, thus enhancing the accretion rate on to the black hole (e.g. \citealt{hopkins+11}).
Moreover, the migration of gaseous clumps caused by dynamical friction may contribute with bursts of accretion.
On the other hand, dense gaseous clumps may form stars and the most massive ones will eventually explode as supernovae.
Such explosions could clear out the gas in the accretion disc, therefore halting the accretion on to the central black hole.
For the sake of simplicity, we circumvent the uncertainties of this regime by preventing the accretion disc from reaching it.
Specifically, we cap at every timestep the inflow rate $\dot{M}_{\rm inflow}$ in order to satisfy the inequality $M_{\rm d} \leq M_{\rm sg}$ at all times\footnote{We have verified that none of the simulations presented here actually reached the $M_{\rm sg}$ threshold.}.
We provide additional details on the numerical implementation in Section~\ref{subsec_implem}. 


\subsection{Black hole spin evolution}\label{subsec_angmom}

The angular momentum $\bmath{J}_{\bullet}$ of the black hole evolves both in magnitude and orientation because of the interaction with the accretion disc.
The main mechanisms that set the evolution of $\bmath{J}_{\bullet}$ are mass accretion and gravito-magnetic coupling \citep{bardeen+75,scheuer+96}. 

The total angular momentum $\bmath{J}_{\rm d}$ is dominated by the outer, extended regions of the accretion disc that carry the largest amount of angular momentum.
The initial direction of $\bmath{J}_{\rm d}$ is set by the occurrence of a generic accretion event (e.g. the disruption of a gaseous cloud by tidal forces) that leads to the formation of an accretion disc which in principle is unrelated to the direction of $J_{\bullet}$.
The central spinning black hole induces Lense-Thirring precession around $\bmath{j}_{\bullet}$ on the accretion disc gas.
The rotating plane of the gas precesses at a frequency $\omega_{\rm LT} = (2G/c^2) J_{\bullet} / R^3$, where $R$ is the distance from the central black hole along the disc plane, i.e. faster if closer to the central black hole.

The precession motion is hindered by the vertical viscosity in the disc.
Within the framework of a thin $\alpha$-disc model, the vertical viscosity $\nu_2$ can be related to the radial viscosity $\nu_1 = \alpha c_{\rm s} H$, where $c_{\rm s}$ is the gas sound speed and $H$ is the disc vertical scale height, as $\nu_{2}/\nu_{1} = \xi / (2 \alpha^2)$, where $\xi \sim O(1)$ is a parameter that can be determined numerically \citep{papaloizou+83, lodato+07b}.
If the disc is sufficiently viscous, i.e. $H/R \lesssim \alpha$ (a condition that is met everywhere in $\alpha$-disc models with typical parameters that describe supermassive black hole accretion discs; \citealt{frank+02}), the interplay between Lense-Thirring precession and viscosity forces the inner region of the accretion disc to align with $\bmath{J}_{\bullet}$ (or anti-align if initially counter-rotating, i.e. if $\bmath{j}_{\bullet} \cdot \bmath{j}_{\rm d} < 0$).
The central region remains misaligned with respect to the outer disc, creating a warp in the disc that diffuses outward to about the warping radius $R_{\rm warp}$, i.e the disc
location where the Lense-Thirring precession period equals the vertical warp propagation timescale (see Appendix~\ref{appendix_formulae}; \citealt{pringle+92, lodato+06, lodato+07b, martin+07, perego+09,dotti+13}).
The warp propagates on timescale much shorter than the local radial viscous time that determines mass transport, therefore the disc can attain a steady warped state \citep{lodato+06, martin+07}.

The angular momentum direction of the gas inside the accretion disc changes as it flows through the warp around $R_{\rm warp}$, i.e. the gas effectively experiences a torque. 
If we focus for the moment only on the black hole+accretion disc system as if it were an isolated system\footnote{We stress that here we consider the black hole and accretion disc system as isolated only for the sake of explanation clarity.
In practice, the system is not isolated and the accretion disc angular momentum can also change because of external inflow, as indicated by equation~(\ref{eq_disc_am_evol}) and further detailed below.},  conservation of the total angular momentum $\bmath{J}_{\rm tot} = \bmath{J}_{\bullet} + \bmath{J}_{\rm d}$ requires an  opposite torque to act on the central black hole.
As a response, $J_{\rm d}$ and $J_{\bullet}$ precess and (counter)align with respect to $\bmath{J}_{\rm tot}$.
This process is known as Bardeen-Petterson effect \citep{bardeen+75}.
The torque felt by the black hole may be calculated after knowing the shape of the warped disc.
\citet{pringle+92} derived the partial differential equation to calculate the angular momentum density $\Sigma(R) \bmath{L}(R)$ across the accretion disc, where $\Sigma(R)$ is the disc surface mass density at the (spherical) radius $R$ from the central black hole.
Once the structure of the disc is known, the torque due to the Bardeen-Petterson effect can be expressed as
\begin{equation}\label{eq_torque_BP}
\left. \frac{{\rm d}\bmath{J}_{\bullet}}{{\rm d}t} \right|_{\rm BP} = \frac{4 \pi G}{c^2} \int_{\rm disc} \frac{\Sigma(R) \bmath{L}(R) \times \bmath{J}_{\bullet}}{R^2}~{\rm d}R.
\end{equation}
We note that the torque above does not modify the magnitude of $\bmath{J}_{\bullet}$ but only its orientation because it is proportional to a cross product with $\bmath{J}_{\bullet}$ itself.

The inner region of the disc also provides matter for accretion on to the black hole.
Matter is accreted when it reaches the inner edge of the accretion disc, corresponding to the innermost stable circular orbit $R_{\rm isco}$, whose extent depends on $a_{\bullet}$ as described in Appendix~\ref{appendix_afunc}.
Then, gas falls on to the black hole carrying along the specific angular momentum $\bmath{L}_{\rm isco}$ at $R_{\rm isco}$ (see Appendix~\ref{appendix_afunc} for the explicit definition of $L_{\rm isco}$).
This not only contributes to the growth of $M_{\bullet}$, but it also modifies $J_{\bullet}$ and the spin parameter $a_{\bullet}$.
Specifically, accreted material can spin up or down the black hole depending on the orientation of the accretion disc close to $R_{\rm isco}$ with respect to $J_{\bullet}$.
As discussed above, the central region of the accretion disc aligns or counter-aligns with $\bmath{J}_{\bullet}$ if initially $\bmath{j}_{\bullet} \cdot \bmath{j}_{\rm d} > 0$ or $\bmath{j}_{\bullet} \cdot \bmath{j}_{\rm d} < 0$, respectively.
Therefore, accretion modifies the magnitude of $\bmath{J}_{\bullet}$ as 
\begin{equation}\label{eq_torque_acc}
\left. \frac{{\rm d}\bmath{J}_{\bullet}}{{\rm d}t} \right|_{\rm acc} = \dot{M} L_{\rm isco}~{\rm sign}(\bmath{j}_{\bullet} \cdot \bmath{j}_{\rm d})~\bmath{j}_{\bullet}.
\end{equation}

The total evolution equation for $\bmath{J}_{\bullet}$ can be obtain by summing up the contributions from equation~(\ref{eq_torque_BP}) and (\ref{eq_torque_acc}).
However, evaluating equation~(\ref{eq_torque_BP}) would require the full solution for the accretion disc structure that is not easily achievable within a sub-grid model; instead, we adopted a different strategy.
\citet{king+05} provide a general expression for the torque on $\bmath{J}_{\bullet}$ resulting from the Bardeen-Petterson effect, 
\begin{equation}\label{eq_torque_BP_general}
\begin{aligned}
\left. \frac{{\rm d} \bmath{J}_{\bullet}}{{\rm d}t} \right|_{\rm BP} & = -K_{1} \left[\bmath{J}_{\bullet} \times \bmath{J}_{\rm d} \right] - K_{2} \left[ \bmath{J}_{\bullet} \times \left( \bmath{J}_{\bullet} \times \bmath{J}_{\rm d} \right) \right] \\
& = J_{\bullet} \left\{-\tilde{K}_{1} \left[\bmath{j}_{\bullet} \times \bmath{j}_{\rm d} \right] - \tilde{K}_{2} \left[ \bmath{j}_{\bullet} \times \left( \bmath{j}_{\bullet} \times \bmath{j}_{\rm d} \right) \right]\right\}.
\end{aligned}
\end{equation}
The first term on the right-hand side induces precession around $\bmath{J}_{\rm tot}$, while the second is related to the alignment or counter-alignment of $\bmath{J}_{\bullet}$ at the expense of $\bmath{J}_{\rm disc}$ to conserve the total angular momentum \citep{king+05}.
We have here redefined for convenience the unknown coefficients $K_{1}$ and $K_{2}$ as precession and alignment rates $\tilde{K}_{1}$ and $\tilde{K}_{2}$, respectively.
$\tilde{K}_{1}$ and $\tilde{K}_{2}$ can be in general arbitrarily complicated functions of the black hole and disc properties.
We constrain the values of $\tilde{K}_{1}$ and $\tilde{K}_{2}$ by expanding equation~(\ref{eq_torque_BP_general}) in the same limit of existing analytical solutions for the disc structure and for the torque on the black hole \citep{scheuer+96,martin+07,perego+09}.
Specifically, \citet{martin+07} have calculated the analytical expression of ${\rm d}\bmath{J}_{\bullet} / {\rm d}t|_{\rm BP}$ for an arbitrary viscosity law $\nu_{1} \propto R^{-\beta}$, under the following assumptions: (i) small initial misalignment between $\bmath{J}_{\bullet}$ and $\bmath{J}_{\rm d}$, and (ii) $J_{\rm d} \gg
J_{\bullet}$.
The latter assumption means that the disc is extended and the outer regions, which effectively dominate the direction of $J_{\rm d}$, define a fixed direction in space\footnote{We note that, despite this assumption allows a simple matching with the analytic theory, it might break down for $J_{\bullet} \gtrsim J_{\rm d}$.
However, as discussed further below in the current section, $J_{\rm d} < J_{\bullet}$ is typically achieved when the black hole mass becomes large and the dynamics changes.};
for instance we can simply consider $\bmath{j}_{\rm d}$ along the $z$ axis without loss of generality.
Assumptions (i) and (ii) correspond to $j_{\bullet, x} \sim j_{\bullet, y} \ll j_{\bullet, z} \sim 1$ and $\bmath{j}_{\rm d} = \bmath{e}_{z} = constant$, respectively.
At first order in $j_{\bullet, x}, j_{\bullet, y}$, equation~(\ref{eq_torque_BP_general}) can be expanded in  
\begin{equation}
\begin{aligned}
\frac{{\rm d}j_{\bullet, x}}{{\rm d}t} & \approx -\tilde{K}_{2} j_{\bullet, x} - \tilde{K}_{2} j_{\bullet, y}, \\
\frac{{\rm d}j_{\bullet, y}}{{\rm d}t} & \approx \tilde{K}_{1} j_{\bullet, x} - \tilde{K}_{2} j_{\bullet, y}, \\
\frac{{\rm d}j_{\bullet, z}}{{\rm d}t} & \approx \tilde{K}_{2} (j_{\bullet, x}^2 + j_{\bullet, y}^2) \approx 0.
\end{aligned}
\end{equation}
The torque in equation~(50) of \citet{martin+07} can be expressed in an equivalent form and, assuming $\beta = 3/4$ appropriate for the $\alpha$-disc model, we can match the coefficients $\tilde{K}_{1} = \sin(\pi/7) / \tau_{\rm align}$ and $\tilde{K}_{2} = \cos(\pi/7) / \tau_{\rm align}$, where the timescale $\tau_{\rm align}$ for the torque to modify the black hole angular momentum is \citep{martin+07, perego+09, dotti+13}
\begin{equation} \label{eq_t_align}
\tau_{\rm align} \approx 0.17~\left( \frac{M_{\bullet}}{10^6~{\rm M}_{\sun}} \right)^{-2/35} \left( \frac{f_{\rm Edd}}{\eta_{0.1}} \right)^{-32/35} a_{\bullet}^{5/7}~{\rm Myr}.
\end{equation}
We explicitly derive this expression in Appendix~\ref{appendix_formulae}.
This timescale relates both to the precession timescale $\tau_{\rm align}/\sin(\pi/7)$ and the alignment timescale $\tau_{\rm align}/\cos(\pi/7)$ and it is physically determined by the mass flow through the warped region, which explains the almost inversely linear dependence on $f_{\rm Edd}$.
We note that the viscosity law of a standard $\alpha$-disc model implies a slightly shorter timescale for alignment than precession by a factor $\tan(\pi/7) \approx 0.48$. 
We further discuss the limitations of this approach in Section~\ref{sec_discussion}. 

We can summarise the evolution equations for the black hole and the accretion disc angular momenta:
\begin{equation}\label{eq_bh_am_evol}
\begin{aligned}
\frac{{\rm d}\bmath{J}_{\bullet}}{{\rm d}t} = & \dot{M} L_{\rm isco}~{\rm sign}(\bmath{j}_{\bullet} \cdot \bmath{j}_{\rm d})~\bmath{j}_{\bullet} - \\
& J_{\bullet} \left\{\frac{\sin(\pi/7)}{\tau_{\rm align}} \left[\bmath{j}_{\bullet} \times \bmath{j}_{\rm d} \right] + \frac{\cos(\pi/7)}{\tau_{\rm align}} \left[ \bmath{j}_{\bullet} \times \left( \bmath{j}_{\bullet} \times \bmath{j}_{\rm d} \right) \right]\right\},
\end{aligned}
\end{equation}
and
\begin{equation}\label{eq_disc_am_evol}
\frac{{\rm d}\bmath{J}_{\rm d}}{{\rm d}t} = - \frac{{\rm d}\bmath{J}_{\bullet}}{{\rm d}t} + \dot{\bmath{J}}_{\rm inflow}.
\end{equation}
The evolution equation for $\bmath{J}_{\rm d}$ includes an external torquing term $\dot{\bmath{J}}_{\rm inflow}$.
This term captures the effect of inflowing material that joins the accretion disc not only adding mass, but also carrying angular momentum and therefore modifying $\bmath{J}_{\rm tot}$. 
It is related to the mass inflow $\dot{M}_{\rm inflow}$ as $\dot{\bmath{J}}_{\rm inflow} = \dot{M}_{\rm inflow} \bmath{L}_{\rm inflow}$, where $\bmath{L}_{\rm inflow}$ is the specific angular momentum of the inflowing gas joining the accretion disc and it can be calculated directly from the simulation, as described in Section~\ref{subsec_inflow}.

The set of equations~(\ref{eq_bh_mass_evol}), (\ref{eq_disc_mass_evol}), (\ref{eq_bh_am_evol}) and (\ref{eq_disc_am_evol}) completely specifies the evolution of the masses and angular momenta of a black hole and a surrounding thin accretion disc owing to the mutual coupling provided by accretion and the Bardeen-Petterson effect, as well as due to mass inflow from larger scales.
However, such a description may break down for a black hole of large mass.
This is because the maximum mass $M_{\rm sg}$ of an accretion disc grows sub-linearly with $M_{\bullet}$, implying that $M_{\rm d}/M_{\bullet}$ becomes progressively smaller for larger $M_{\bullet}$.
As a consequence, the disc also carries less angular momentum relatively to the black hole and it becomes more compact, whereas the larger mass and angular momentum of the black hole can induce Lense-Thirring precession at larger radii.
Therefore, the disc cannot reach the warped steady state when $R_{\rm warp}$ becomes larger than the disc radius; the latter condition translates into a critical black hole mass \citep{martin+07, dotti+13}, 
\begin{equation} \label{eq_M_warp}
M_{\bullet}^{\rm (warp)} \approx 10^{7}~\left( \frac{M_{\rm d}}{10^4~{\rm M}_{\sun}} \right)^{35 / 82} \left( \frac{f_{\rm Edd}}{\eta_{0.1}} \right)^{-17 / 82} a_{\bullet}^{-25 / 82}~{\rm M}_{\sun}.
\end{equation}
We explicitly derive this expression in Appendix~\ref{appendix_formulae}.
Beyond this mass, which depends on the black hole and disc properties, the description of equation~(\ref{eq_bh_am_evol}) is not valid anymore; instead, the disc aligns (or counter-aligns) with the black hole over a very short timescale set by the diffusive propagation of the warp that effectively interests the whole accretion disc.
A proper modelling of this phenomena is beyond the purpose of our sub-grid model, therefore we apply a simplified approach.
We assume that the alignment is instantaneous and we align or counter-align the black hole and the accretion disc along the direction of $\bmath{J}_{\rm tot}$ according to the criterion derived by \citet{king+05}: $\bmath{J}_{\bullet}$ and $\bmath{J}_{\rm d}$ end up being aligned if $\bmath{j}_{\bullet} \cdot \bmath{j}_{\rm d} > - J_{\bullet} / (2 J_{\rm d})$, otherwise they counter-align.
This general criterion, that can be derived from equation~(\ref{eq_torque_BP_general}), shows that alignment is the final configuration whenever $J_{\rm d} > 2 J_{\bullet}$ since $-1 \leq \bmath{j}_{\bullet} \cdot \bmath{j}_{\rm d} \leq 1$, while counter-alignment is possible and becomes equally likely when the black hole dominates the total angular momentum of the system.


\subsection{Connecting the sub-grid model to the simulation}\label{subsec_inflow}

Our sub-grid model for black hole accretion disc described by equations~(\ref{eq_bh_mass_evol}), (\ref{eq_disc_mass_evol}), (\ref{eq_bh_am_evol}) and (\ref{eq_disc_am_evol}) is connected to the hydrodynamical simulation through the boundary conditions provided by $\dot{M}_{\rm inflow}$ and $\dot{\bmath{J}}_{\rm inflow} = \dot{M}_{\rm inflow} \bmath{L}_{\rm inflow}$.
Therefore, we have devised robust estimators of these two quantities for the implementation of the model in {\sc arepo}.
The inflow rate may be measured from the mass flux $\rho \bmath{u}$ on to the black hole particle, namely 
\begin{equation}
\dot{M}_{\rm inflow} = - \oint_{A} \rho \bmath{u} \cdot {\rm d} \bmath{S} \approx -A~\langle \Phi_{M} \rangle(\bmath{x}_{\bullet}),
\end{equation}
where $\langle \Phi_{M} \rangle(\bmath{x}_{\bullet})$ is the numerical estimate of the mass flux at the position of the black hole $\bmath{x}_{\bullet}$ and $A$ is an effective area through which mass is accreted.
The mass flux is computed via a smoothed-particle hydrodynamic interpolation of the local flux on to the black hole within a smoothing length $h_{\bullet}$ that encompasses the $N_{\rm neighbour}$ closest mesh-generating points, 
\begin{equation} \label{eq_mass_flux_sph}
\langle \Phi_{M} \rangle(\bmath{x}_{\bullet}) = \frac{\sum_{j=1}^{N_{\rm neighbour}} \rho_{j} u_{r, j} W(d_j)}{\sum_{j=1}^{N_{\rm neighbour}} W(d_j)},
\end{equation}
where $d_j = |\bmath{x}_{j} - \bmath{x}_{\bullet}|/h_{\bullet}$ is the distance between the black hole and the centre of mass $\bmath{x}_j$ of the $j$-th gas cell divided by the black hole smoothing length, $u_{r,j} = (\bmath{x}_j - \bmath{x}_{\bullet}) \cdot (\bmath{u}_{j} - \bmath{u}_{\bullet}) / |\bmath{x}_j - \bmath{x}_{\bullet}|$, and $W(x)$ is a cubic spline kernel with compact support over $h_{\bullet}$.
According to our definition, $\langle \Phi_{M} \rangle < 0$ corresponds to inflow.

\begin{figure*}
\begin{center}
\includegraphics[width=2.1\columnwidth]{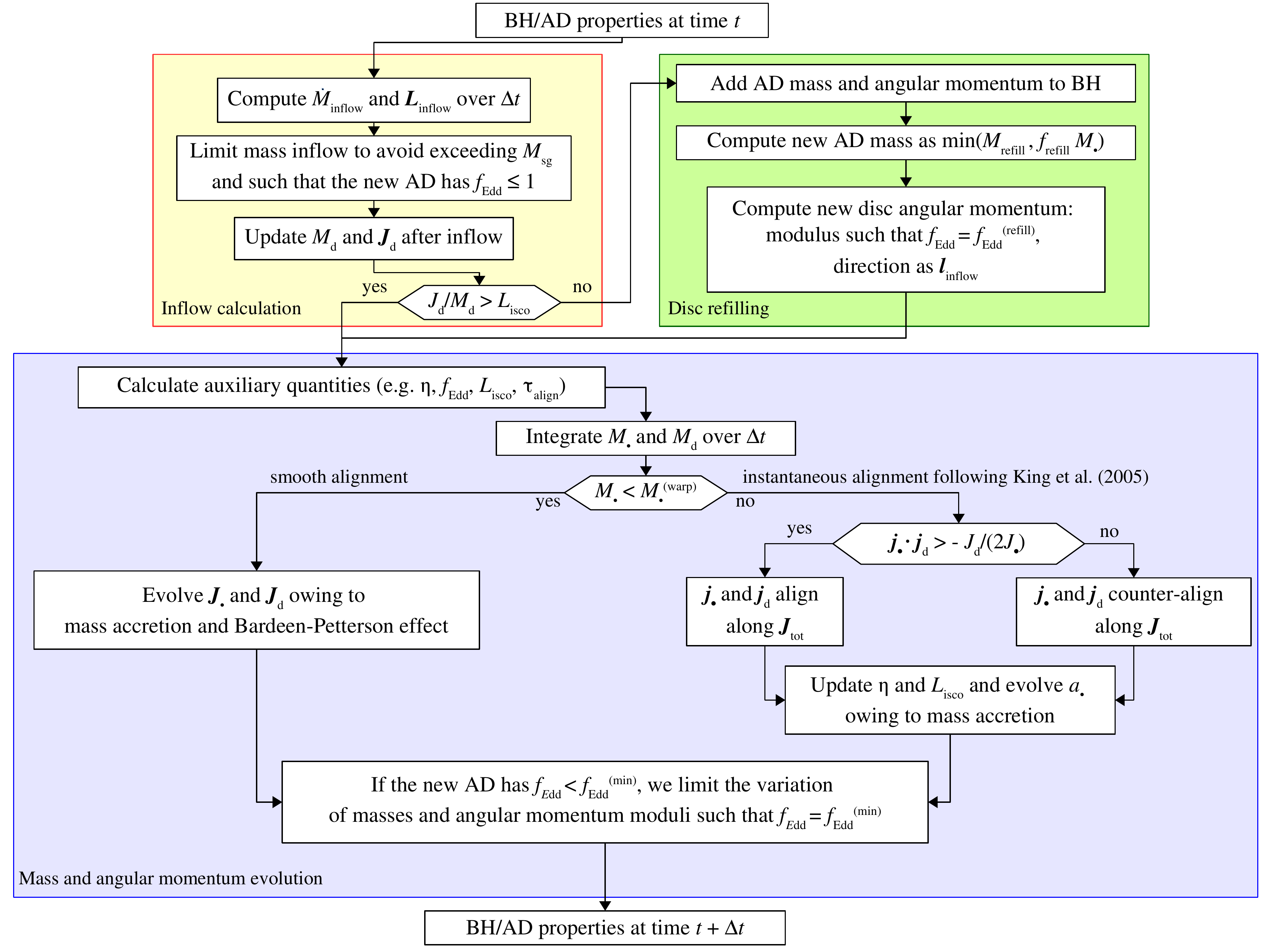}
\caption{This flow chart summarises the main operations of the sub-grid model over a timestep $\Delta t$ during the simulation.
In the figure, AD and BH refer to ``accretion disc'' and ``black hole'', respectively.
There are three main blocks, as discussed in Section~\ref{subsec_implem}: first we calculate the properties of the inflow, i.e. $\dot{M}_{\rm inflow}$ and $\bmath{L}_{\rm inflow}$.
After updating the disc properties to account for inflow, we either (i) reconstruct the disc in case the new $L_{\rm d} < L_{\rm isco}$, or otherwise (ii) we evolve the masses and the angular momenta as described in Sections~\ref{subsec_accretion} and \ref{subsec_angmom}.
We then update the black hole mass and spin considering either the limiting case of a smooth alignment of the disc with the black hole or the instantaneous alignment of the two.
}
\label{fig_model_flow_chart}
\end{center}
\end{figure*}

The effective area $A$ should ideally be related to a meaningful physical scale that therefore does not depend directly on resolution.
The most relevant physical scale is the radius of the accretion disc $R_{\rm d}$; however, this can often be below our spatial resolution and therefore we unavoidably have to rely on a mass accretion rate calculated through our smallest resolution length.
Specifically, we define the effective area as $A = 4 \pi \langle r_{\rm inflow} \rangle^2$, where $\langle r_{\rm inflow} \rangle$ is the kernel-weighted average spherical size of the hydro cells near the black hole, 
\begin{equation} \label{eq_r_inflow_SPH}
\langle r_{\rm inflow} \rangle = \frac{\sum_{j=1}^{N_{\rm neighbour}} (3 V_{j}/(4 \pi))^{1/3} W(d_j)}{\sum_{j=1}^{N_{\rm neighbour}} W(d_j)},
\end{equation}
where $V_{j}$ is the volume of the $j$-th cell.
We have checked that in all the simulations presented below $R_{\rm d} < \langle r_{\rm inflow} \rangle$ at all times.
We note that, despite some similarities, our approach does not follow the typical sink particle implementations based on a characteristic accretion radius as often used e.g. in simulations of star-forming clouds (e.g. \citealt{bate+95,federrath+10}).
We present in Appendix~\ref{appendix_restest} a resolution study to discuss the robustness of our strategy and the dependency on resolution. 

At the same time, we need to estimate the specific angular momentum $\bmath{L}_{\rm inflow}$ carried by $\dot{M}_{\rm inflow}$ to compute the torque $\dot{\bmath{J}}_{\rm inflow} = \dot{M}_{\rm inflow} \bmath{L}_{\rm inflow}$ on the accretion disc angular momentum associated with external inflow.
We evaluate the specific angular momentum of the inflowing gas as
\begin{equation} \label{eq_L_inflow_SPH}
\bmath{L}_{\rm inflow} = \frac{\dot{\bmath{J}}_{\rm inflow}}{\dot{M}_{\rm inflow}} = \frac{\sum_{j=1}^{N_{\rm neighbour}} \rho_{j} u_{r,j}~\mathbf{L}_{j}~W(d_j)}{\sum_{j=1}^{N_{\rm neighbour}} \rho_{j} u_{r,j}~W(d_j)},
\end{equation}
where $\bmath{L}_{j} = (\bmath{x}_{j} - \bmath{x}_{\bullet}) \times (\bmath{u}_{j} - \bmath{u}_{\bullet})$ is the specific angular momentum of the $j$-th mesh-generating point in the reference frame of the black hole.
This represents the kernel-weighted specific angular momentum where each hydro cell contributes to the angular momentum average as they contribute to mass inflow.
We discuss the numerical robustness and convergence of this estimator in Appendix~\ref{appendix_restest}.


\subsection{Implementation of the model} \label{subsec_implem}

In so far we have discussed the physical framework that our accretion sub-grid model tries to capture.
Here, we provide additional details on the numerical implementation of the model in {\sc arepo}.
Figure~\ref{fig_model_flow_chart} shows a flow chart that schematises the algorithm of the model over a timestep $\Delta t$ during the simulation.
First, we calculate the black hole smoothing length $h_{\bullet}$ from the $N_{\rm neighbour}$ closest mesh-generating points to evaluate the properties of the inflow, i.e. $\dot{M}_{\rm inflow}$ and $\bmath{L}_{\rm inflow}$, as discussed in Section~\ref{subsec_inflow}.

We limit the value of $\dot{M}_{\rm inflow}$ to satisfy several conditions.
First of all, we check that the mass flux $\langle \Phi_{M} \rangle$ is negative, i.e. the gas is actually inflowing.
Otherwise, we simply impose $\dot{M}_{\rm inflow} = 0$.
Moreover, we check that $L_{\rm inflow} \leq L_{\rm d} = J_{\rm d} / M_{\rm d}$, i.e. the specific angular momentum of the inflowing material must be lower than that of the accretion disc, otherwise the inflowing gas cannot circularise at a distance $< R_{\rm d}$ from the black hole.
In the latter case similarly as before, we assume that there is no actual inflow and we set $\dot{M}_{\rm inflow} = 0$.
This assumption can be viewed as conservative since it extends the expected behaviour of the gas with $L_{\rm inflow} \gg L_{\rm d}$ to the limit $L_{\rm inflow} \gtrsim L_{\rm d}$, where we neglect physical mechanisms (e.g. self-gravity) that could potentially make the gas lose the slight excess of angular momentum and join the accretion disc.
This choice may somewhat modulate the amount of matter reaching the accretion disc and eventually the central black hole, but we think it represents the simplest way to consistently account for the unresolved gas between $\langle r_{\rm inflow} \rangle$ and $R_{\rm d}$.
We also limit the mass inflow in order to guarantee that (i) $M_{\rm d} \leq M_{\rm sg}$, and (ii) $f_{\rm Edd} \leq 1$.
Specifically, (i) we limit the inflow rate at $\dot{M}_{\rm inflow} = \min(\dot{M}_{\rm inflow}, (M_{\rm sg} - M_{\rm d}) / \Delta t)$, where $M_{\rm sg}$ and $M_{\rm d}$ are the values at the beginning of the timestep, and (ii) we compute the fraction $0 \leq \beta \leq 1$ of $\dot{M}_{\rm inflow}$ that can join the mass and angular momentum (with specific angular momentum $\bmath{L}_{\rm inflow}$) of the accretion disc over $\Delta t$ such that $f_{\rm Edd} \leq 1$.
$\beta$ is inferred by linearly interpolating equation~(\ref{eq_f_edd}) between $(M_{\rm d}, J_{\rm d})$ and  $(M_{\rm d} + \beta \dot{M}_{\rm inflow} \Delta t, |\bmath{J}_{\rm d} + \beta \dot{M}_{\rm inflow} \bmath{L}_{\rm inflow} \Delta t |)$ for given black hole properties $(M_{\bullet}, \bmath{J}_{\bullet})$ and solving for the value of $\beta$ to ensure that $f_{\rm Edd}$ cannot exceed 1.
This procedure enforces not only the constraint on the Eddington limit, but also the self-consistency between $(M_{\bullet}, \bmath{J}_{\bullet}, M_{\rm d}, \bmath{J}_{\rm d})$ and $f_{\rm Edd}$ through the $\alpha$-disc solution.

After we calculate the inflow properties and appropriately limit them, we update the mass and angular momentum of the accretion disc as
\begin{equation}
\begin{aligned}
M_{\rm d} & \mapsto M_{\rm d} + \dot{M}_{\rm inflow}~\Delta t, \\
\bmath{J}_{\rm d} & \mapsto \bmath{J}_{\rm d} + \dot{M}_{\rm inflow}~\bmath{L}_{\rm inflow}~\Delta t.
\end{aligned}
\end{equation}
Since angular momentum is a vectorial quantity, the mass inflow may effectively reduce the modulus of the accretion disc angular momentum.
However, $L_{\rm disc}$ cannot be smaller than the specific angular momentum $L_{\rm isco}$ required by a circular orbit at $R_{\rm isco}$, otherwise the accretion disc would not be able to remain stable on nearly circular orbits and it would just fall on to the central black hole on a dynamical time.
Therefore, after updating $\bmath{J}_{\rm d}$, we check that $L_{\rm d} > L_{\rm isco}$.
When this is not satisfied, we instantaneously add the mass and the angular momentum of the accretion disc to $M_{\bullet}$ and $\bmath{J}_{\bullet}$, respectively.
As a consequence, the black hole may remain without a surrounding accretion disc, and a new accretion disc might form after a new accretion event.
However, the formation of an accretion disc as a result of the accumulation and circularisation of gas around a black hole cannot be easily captured by simple equations of our sub-grid model.
Therefore, we adopt a simplistic approach, deferring a more detailed modelling of this phase to a forthcoming work; we assume that a new accretion disc immediately forms.
The mass of the new accretion disc is set to $M_{\rm d} = \min(M_{\rm refill}, f_{\rm refill} M_{\bullet})$, where $M_{\rm refill}$ and $f_{\rm refill}$ are two phenomenological free parameters, representing a fixed initial mass and a fraction of the black hole mass, respectively.
The angular momentum is initialised by taking the direction of the inflowing material, i.e. $\bmath{j}_{\rm d} = \bmath{l}_{\rm inflow}$, whereas the modulus $J_{\rm d}$ is set to enforce an initial $f_{\rm Edd}^{\rm (refill)}$, which by default is $f_{\rm Edd}^{\rm (refill)} = f_{\rm Edd}^{\rm (min)}$, but can be modified. 
We empirically found that high values of $f_{\rm Edd}^{\rm (refill)}$ can lead to frequent disc draining episodes followed by possibly many, artificial, disc reconstruction events.
This happens only in case of very peculiar conditions, namely when the inflow forms the accretion disc with $\arccos \left( \bmath{j}_{\bullet} \cdot \bmath{j}_{\rm d} \right) \approx 90\degr$ and $J_{\rm d} \ll J_{\bullet}$.
In these cases, (counter-)alignment and accretion require a significant reduction of $J_{\rm d}$ over a timestep, possibly imposing to rebuild the disc over the next timestep.
To avoid this numerical artefact, we limit the value of $f_{\rm Edd}^{\rm (refill)}$ just after the disc is reconstructed.
Specifically, we use the value of $f_{\rm Edd}$ that guaranties a final $J_{\rm d}/M_{\rm d} \geq 3 L_{\rm ISCO}$, where 3 is a safety factor, after that the disc (counter)-aligns from the initial misalignment caused by the disc reconstruction.
We tested that this procedure cures the problem by minimally changing the initial properties of a reconstructed accretion disc.

We then start the actual time evolution of the black hole and the accretion disc.
We check first whether $M_{\bullet} < M_{\bullet}^{\rm (warp)}$; in case the inequality is satisfied, we evolve the masses and the angular momenta according to equations~(\ref{eq_bh_mass_evol}), (\ref{eq_disc_mass_evol}), (\ref{eq_bh_am_evol}) and (\ref{eq_disc_am_evol}).
We use the second-order, predictor-corrector Heun's scheme to integrate the masses and the angular momenta, capping $a_{\bullet}$ to 0.998 both in the predictor and corrector phases.
Otherwise, if $M_{\bullet} > M_{\bullet}^{\rm (warp)}$, we first align the angular momenta, and then we evolve the masses and the angular momenta as described above, but in equations~(\ref{eq_bh_am_evol}) and (\ref{eq_disc_am_evol}) we retain only the accretion term.
This provides us with the black hole and accretion disc properties at $t + \Delta t$.
We conclude the step by further imposing the constraint $f_{\rm Edd} \geq f_{\rm Edd}^{\rm (min)}$.
Specifically, we adopt a strategy similar to limiting the Eddington ratio to 1: we linearly interpolate the quantities $(M_{\bullet}, \bmath{J}_{\bullet}, M_{\rm d}, \bmath{J}_{\rm d})$ and the resulting $f_{\rm Edd}$ between their values at $t$ and $t + \Delta t$ and we calculate the fraction $\beta$ of the variation $\Delta B$, where $B$ is any of $(M_{\bullet}, \bmath{J}_{\bullet}, M_{\rm d}, \bmath{J}_{\rm d})$, that satisfies $f_{\rm Edd} = f_{\rm Edd}^{\rm (min)}$, and we update the properties at $t + \Delta t$ accordingly.

Finally, we note that the our sub-grid model introduces some physical timescales that must be properly resolved during the simulated evolution.
Therefore, we have added an additional constraint on the timestep $\Delta t$ for black hole particles; unless already constrained to a smaller value (e.g. from the hydrodynamics or the gravity), we limit the timestep as follows: 
\begin{equation}
\Delta t = 0.1 \min(\tau_{\rm align}, M_{\rm d} / \dot{M}),
\end{equation}
where $M_{\rm d} / \dot{M}$ represent the draining timescale for the accretion disc, and 0.1 is a safety factor.
In case the system satisfies $M_{\bullet} > M_{\bullet}^{\rm (warp)}$ when we compute the timestep (note $\tau_{\rm align}$ is therefore not properly defined) we just use $\Delta t = 0.1 M_{\rm d} / \dot{M}$.
We note that this requirement is not very stringent and does not impose any appreciable slow down of the simulations; in fact, this constraint typically requires timesteps between a few thousandth's to a tenth of a Myr.
These timesteps are not prohibitive and often already required by the hydrodynamics and the gravity in small-scale, high-resolution simulations as those presented below, as well as in state-of-the-art isolated galactic discs, galaxy mergers, or zoom-in cosmological simulations.


\section{Results} \label{sec_results}


\subsection{Spin evolution in circumnuclear discs} \label{subsec_cnd}

\begin{table*}
\caption{Summary of the cicumnuclear disc runs.
From left to right: run label, initial black hole mass $M_{\bullet, 0}$, initial disc mass $M_{\rm d, 0}$, initial spin parameter $a_{\bullet, 0}$, initial angular momentum ratio $J_{\rm d,0}/J_{\bullet, 0}$, initial angle between black hole and accretion disc angular momenta, initial angle between the black hole angular momenta and the $z$ axis, and initial angle between the accretion disc angular momenta and the $z$ axis.
The last column marks whether the initial conditions satisfy the \citet{king+05} criterion ($\checkmark$, i.e. alignment) or not ($\times$, i.e. counter-alignment).
}
\label{tab_cnd_runs}
\begin{tabular}{lcccc ccccc}
\hline
Label & $M_{\bullet,0}$ & $M_{\rm d,0}$ & $a_{\bullet,0}$ & $J_{\rm d, 0} / J_{\bullet, 0}$ & $\cos^{-1} ( \mathbfit{j}_{\bullet,0} \cdot \mathbfit{j}_{\rm d,0} )$ & $\cos^{-1} ( \mathbfit{j}_{\bullet,0} \cdot \mathbfit{e}_{\rm z} )$ & $\cos^{-1} ( \mathbfit{j}_{\rm d,0} \cdot \mathbfit{e}_{\rm z} )$ & $\mathbfit{j}_{\bullet,0} \cdot \mathbfit{j}_{\rm d,0} > - J_{\rm d, 0} / (2 J_{\bullet, 0})$ \\
 & (M$_{\sun}$) & (M$_{\sun}$) & & & (\degr) & (\degr) & (\degr) &  \\
\hline
cnd1 & $10^6$ & $10^{3}$ & $0.67$ & 0.2 & $37.9$ & $35.9$ & $71.7$ & \checkmark \\
cnd2 & $10^6$ & $10^{4}$ & $0.67$ & 5.0 & $37.9$ & $35.9$ & $71.7$ & \checkmark \\
cnd3 & $10^6$ & $10^{4}$ & $0.32$ & 10.6 & $170.9$ & $161.2$ & $20.4$ & \checkmark \\
cnd4 & $10^7$ & $10^{5}$ & $0.67$ & 1.7 & $37.9$ & $35.9$ & $0.0$ & \checkmark \\
cnd5 & $10^7$ & $10^{5}$ & $0.32$ & 3.5 & $170.9$ & $161.2$ & $20.4$ & \checkmark \\
cnd6 & $10^7$ & $10^{5}$ & $0.8$ & 1.4 & $170.9$ & $161.2$ & $20.4$ & $\times$ \\
\hline
\end{tabular}
\flushleft
\end{table*}

\begin{figure}
\begin{center}
\includegraphics[width=\columnwidth]{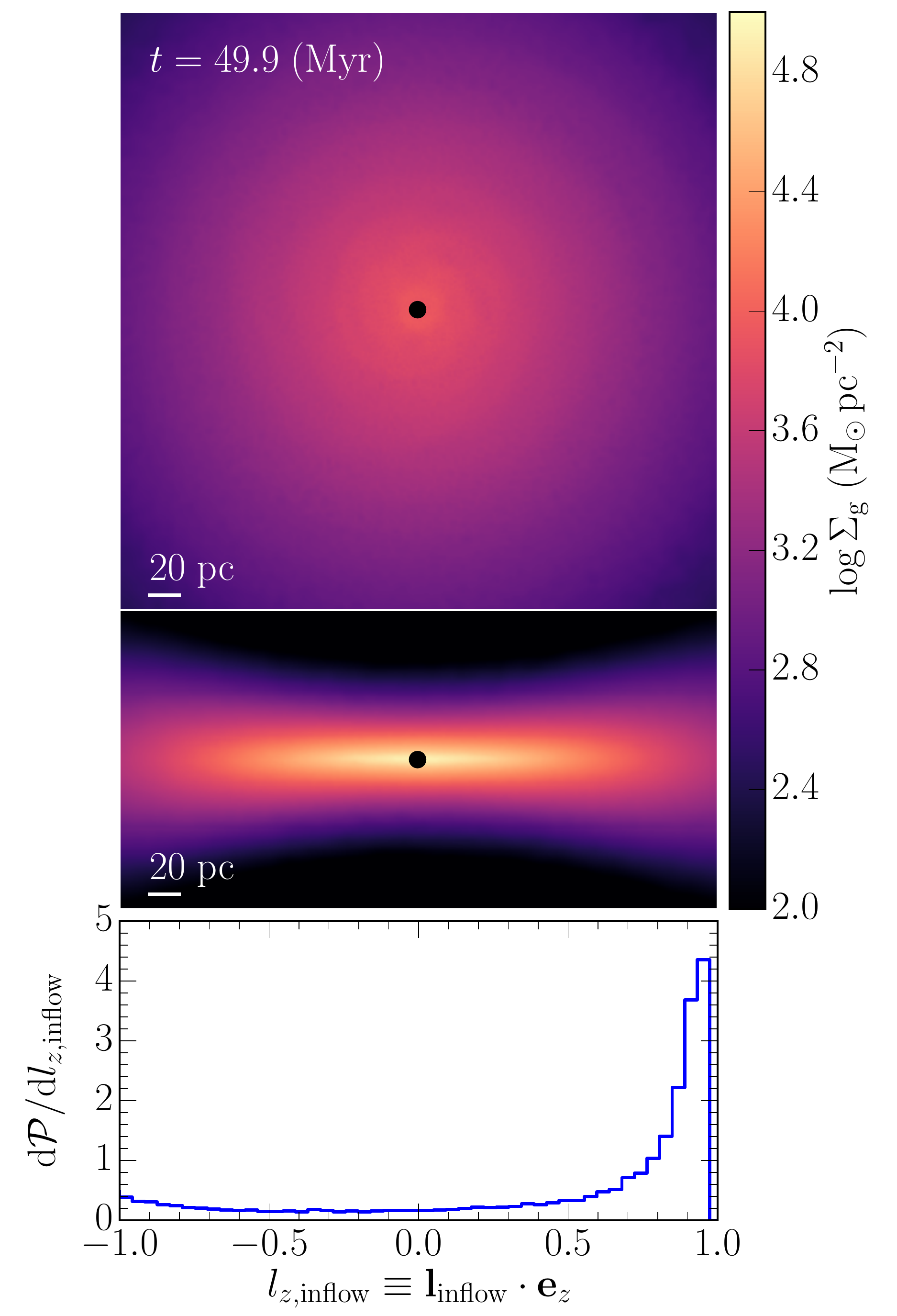}
\caption{Sample snapshot at $t \approx 50$~Myr from the cnd1 run of a supermassive black hole embedded in a circumnuclear disc. The top and middle panels show the face-on and edge-on gas surface density projections, respectively.
The bottom panel show the probability distribution function for the direction of the inflowing specific angular momentum $\bmath{L}_{\rm inflow}$ with respect to the $z$ axis, i.e. the rotation axis of the circumnuclear disc.
}
\label{fig_map_example}
\end{center}
\end{figure}

\subsubsection{Properties of the runs}

Some quasar activity is likely triggered by galaxy mergers \citep{barnes+91,springel+05, hopkins+06, hopkins+08} as well as by secular evolution, for instance through the formation of a bar \citep{laine+02, laurikainen+04, fanali+15}.
Indeed, both mechanisms are able to promote the accumulation of gas in the nucleus of a galaxy, despite the ongoing debate whether one dominates over the other (e.g.  \citealt{lee+12, oh+12, alonso+13, cisternas+13}).
In both cases, the gas may settle in a circumnuclear disc $\sim 100$~pc in size and $\sim 10^{8-9}$~M$_{\sun}$ in mass, as sometimes observed in massive galaxies with clear features of recent mergers \citep{downes+98, medling+14}, as well as in some unperturbed, disc-like Seyfert galaxies \citep{schinnerer+99, chou+07}.

Therefore, we here focus on the evolution of the black hole spin in idealised but physically motivated conditions, namely a supermassive black hole embedded in a circumnuclear gaseous disc within the $x$-$y$ plane at the centre of a stellar spheroid that represents the inner region of a bulge \citep{fiacconi+13,maio+13, lupi+15}.
Specifically, the stellar spheroid follows a \citet{hernquist+90} profile with total mass $M_{\rm b} = 5 \times 10^{8}$~M$_{\sun}$ and scale radius $r_{\rm b} = 100$~pc.
The gaseous circumnuclear disc is rotationally supported, extends for $\approx 200$~pc, has total mass $M_{\rm g} = 10^{8}$~M$_{\sun}$, and follows the density profile \citep{hernquist+93} 
\begin{equation}
\rho_{\rm g}(R, z) = \frac{M_{\rm g}}{4 \pi R_{\rm g}^2 z_{\rm g}(R)} \exp\left(- \frac{R}{R_{\rm g}} \right) \cosh^{-2}\left(- \frac{z}{z_{\rm g}(R)} \right),
\end{equation}
where the scale radius $R_{\rm g} = 50$~pc, while the local scale height $z_{\rm g}(R)$ is calculated by solving the vertical hydrostatic equilibrium under the assumption that the gas is ideal and the temperature is initially uniform at 20,000~K.

\begin{figure*}
\begin{center}
\includegraphics[width=2.1\columnwidth]{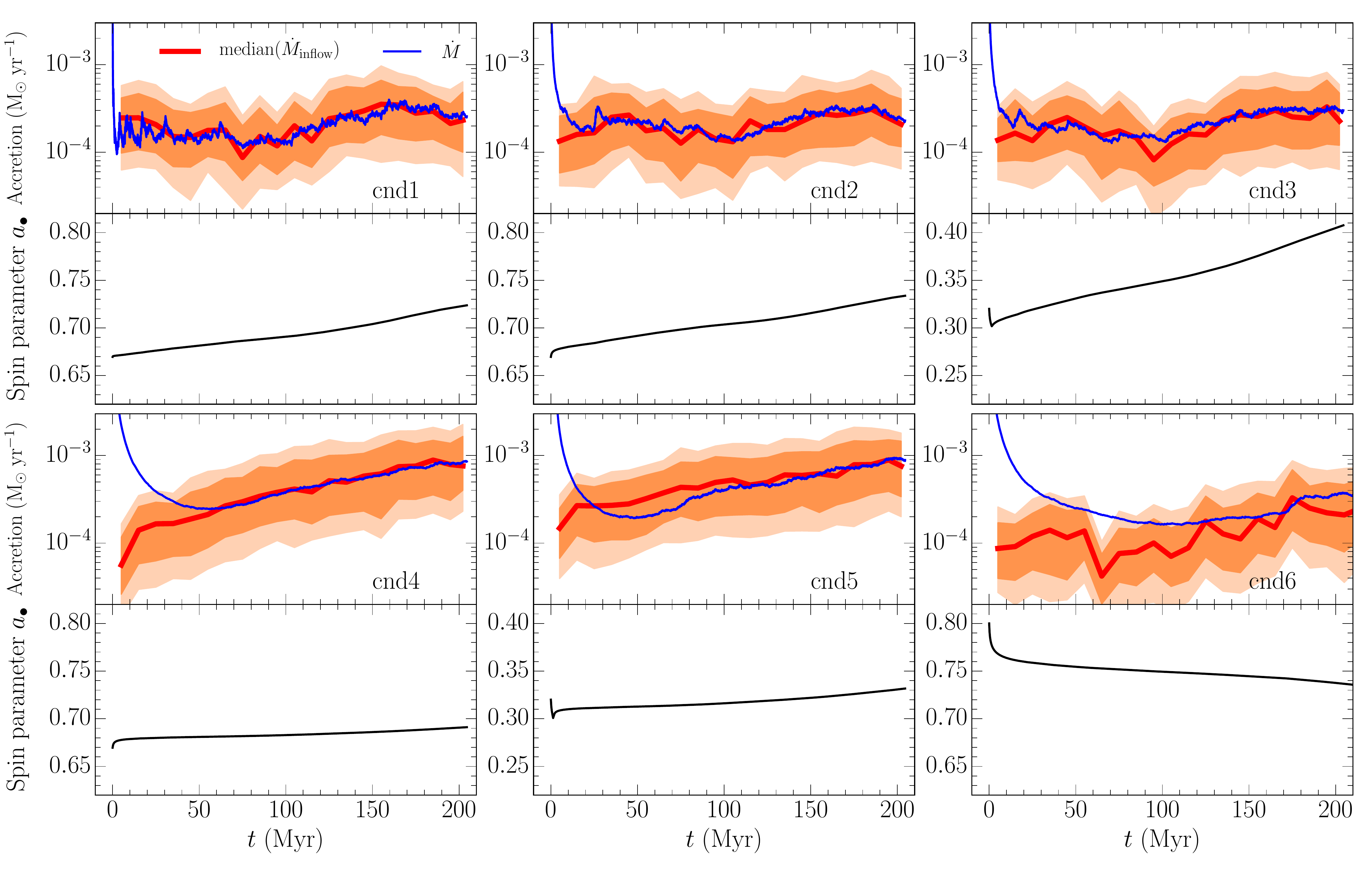}
\caption{Accretion rates and spin parameter evolution as a function of time.
For each plot, the upper panel shows the median of $\dot{M}_{\rm inflow}$ over time bins of 10~Myr (thick red solid curve), and the accretion rate $\dot{M}$ on to the black hole (thin blue solid curve).
The dark and light orange shaded regions mark the 68\% and 90\% regions of $\dot{M}_{\rm inflow}$.
The lower set of panels shows the time evolution of the spin parameter $a_{\bullet}$.
}
\label{fig_mdot_a}
\end{center}
\end{figure*}

We set up the initial conditions by sampling the stellar spheroid with $2,000,000$ collisionless particles with mass $m_{\star} = 250$~M$_{\sun}$ and gravitational softening $\epsilon_{\star} = 0.25$~pc.
The gaseous disc is initially represented by $400,000$ mesh-generating points with target mass $m_{\rm g}^{\rm target} = 250$~M$_{\sun}$.
The gravitational pull from the cells is softened using an adaptive gravitational softening whose minimum value is $\epsilon_{\rm g} = 0.25$~pc.
The gas component is evolved as an ideal gas with equation of state $P = (\gamma - 1) \rho u$, where $\gamma = 5/3$.
For the sake of simplicity, we do not include neither radiative cooling nor star formation and feedback.
We relax the initial conditions for $\approx 20$~Myr, corresponding to about 2 disc rotations at $R_{\rm g}$, to let the disc dissipate some weak transient features such as over-dense rings.
After that, the disc remains stable and smooth with time and it shows only very weak spiral structures.

An example from run cnd1 (see Table~\ref{tab_cnd_runs} for further details and for the properties of all runs) is illustrated in Figure~\ref{fig_map_example} after $\approx 50$~Myr evolution.
The flaring structure of the circumnuclear disc is required by hydrostatic equilibrium, with a typical aspect ratio $z_{\rm g}(R)/R$ that goes from $\approx 0.3$ close to central black hole to $\approx 0.12$ in the outer part of the disc.
The inner circumnuclear disc is rather thick because of the gas temperature, corresponding to a typical sound speed $c_{\rm s}\approx 35$~km~s$^{-1}$ within $R_{\rm g}$.
While a very thin disc would provide a narrow distribution of $\bmath{l}_{\rm inflow} \cdot \bmath{e}_{z}$ peaked around 1, i.e. aligned with the circumnuclear disc angular momentum, the inner thickness causes some broadening in the distribution of $\bmath{l}_{\rm inflow} \cdot \bmath{e}_{z}$ as shown in Figure~\ref{fig_map_example}, with a small tail of uniformly distributed values (i.e. ${\rm d} \mathcal{P} / {\rm d}l_{z, \rm inflow} \approx$constant).
Nonetheless, the system retains a well defined symmetry axis, corresponding to the rotational axis of the circumnuclear disc, and most of the accreted material is aligned with it.

The circumnuclear disc hosts a supermassive black hole at its centre represented by a sink particle implementing the sub-grid accretion model previously described.
The black hole particle has a gravitational softening $\epsilon_{\bullet} = 4$~pc, as given by the scaling $\epsilon_{\bullet} = \epsilon_{\rm g} (M_{\bullet}/m_{\rm g}^{\rm target})^{1/3}$.
We varied the masses and the angular momenta of both the black hole and the accretion disc among several different runs as reported in Table~\ref{tab_cnd_runs}.
The black hole masses varies between $10^6$ and $10^7$~M$_{\sun}$, exploring typical masses inferred for Seyfert galaxies,which are common hosts of circumnuclear discs \citep{wandel+99,cracco+16,rakshit+17}.
The initial accretion disc masses are $10^{-2} M_{\bullet}$ or $10^{-3} M_{\bullet}$ in order to fulfil the $M_{\rm sg}$ constraint from the beginning.
The total mass $M_{\bullet} + M_{\rm d}$ contributes to the gravitational potential of the black hole.
We choose $M_{\rm refill}$ equal to the initial accretion disc mass.
The angular momenta moduli and orientations are initially chosen at random, but in order to intentionally explore different situations: the black hole and accretion disc angular momenta are initially at less than 90\degr misalignment and full alignment is expected (runs cnd1, cnd2 and cnd4); the black hole and accretion disc angular momenta are initially almost counter-aligned but they are expected to align (runs cnd3 and cnd5); the black hole and accretion disc angular momenta are initially almost counter-aligned and they are expected to find an equilibrium counter-aligned configuration (run cnd6).

\subsubsection{Accretion rate and spin parameter evolution}

Figure~\ref{fig_mdot_a} shows the time evolution of the mass accretion rates $\dot{M}$ and $\dot{M}_{\rm inflow}$.
The accretion rate on to the black hole varies between $2 \times 10^{-4}$ and $\sim 10^{-3}$~M$_{\sun}$~yr$^{-1}$ across the different runs; this corresponds to $f_{\rm Edd} \sim 10^{-2}$ and $f_{\rm Edd} \lesssim 10^{-3}$ for runs with $M_{\bullet} = 10^6$ and $M_{\bullet} = 10^7$~M$_{\sun}$, respectively, which is in accord with observed Seyfert galaxies \citep{onken+03,komossa+07,ho+09}.
All the simulations show a transient decrease of $\dot{M}$ at the beginning of the calculations due to the initial arbitrary value of $M_{\rm d}$.
Then, accretion disc properties (i.e. $M_{\rm d}$ and $J_{\rm d}$) tend to readjust to provide an approximate equilibrium with the inflow rate $\dot{M}_{\rm inflow}$ \footnote{Here, we recall that $\dot{M}_{\rm inflow}$ may be set equal to 0 during some timesteps if required (see Section~\ref{subsec_implem}).
In order to take that into account in Figure~\ref{fig_mdot_a}, we have binned  the values of $\dot{M}_{\rm inflow} > 0$ in time bins of 10~Myr, and we have reduced the obtained values by the factor $(1-\delta)$, where $\delta$ is the fraction of each time bin for which $\dot{M}_{\rm inflow} = 0$.}. 

The value of $\dot{M}_{\rm inflow}$ is mostly set by the properties of the circumnuclear disc and it is indeed similar among different simulations; however, the runs with $M_{\bullet} = 10^{7}$~M$_{\sun}$ develop an $m=1$ spiral mode in the inner $~20$ pc after about 75~Myr evolution owing to a small periodic motion of the central black hole. 
While the black hole in runs cnd1-3 is not heavy enough to perturb the gas distribution, in runs cnd4-6 the perturbation excites this spiral structure that transfers angular momentum outward more efficiently and brings more material in, as shown by the increasing $\dot{M}_{\rm inflow}$ in the lower row of Figure~\ref{fig_mdot_a}.
However, also the accretion disc properties can have subtle effects, as it appears in the lower $\dot{M}_{\rm inflow}$ of run cnd6 than that of runs cnd4 and cnd5.
This is because $L_{\rm d}$ is typically lower in the latter runs than in run cnd6, which tends to prefer more inflowing gas to reach the accretion disc (given the conditions discussed in Section \ref{subsec_inflow} and \ref{subsec_implem}) and therefore favours overall larger values of $\dot{M}_{\rm inflow}$.
In later stages of run cnd6, $L_{\rm d}$ decreases and becomes comparable to the final values of runs cnd4 and cnd5, and so does the value of $\dot{M}_{\rm inflow}$.
In turn, the accretion disc mass grows and this results in a larger $\dot{M}$ that tries to follow $\dot{M}_{\rm inflow}$, although limited by the concurrent growth of $J_{\rm d}$ (see Figure~\ref{fig_am_comp} below), which implies a more extended and less dense accretion disc.

\begin{figure*}
\begin{center}
\includegraphics[width=1.75\columnwidth]{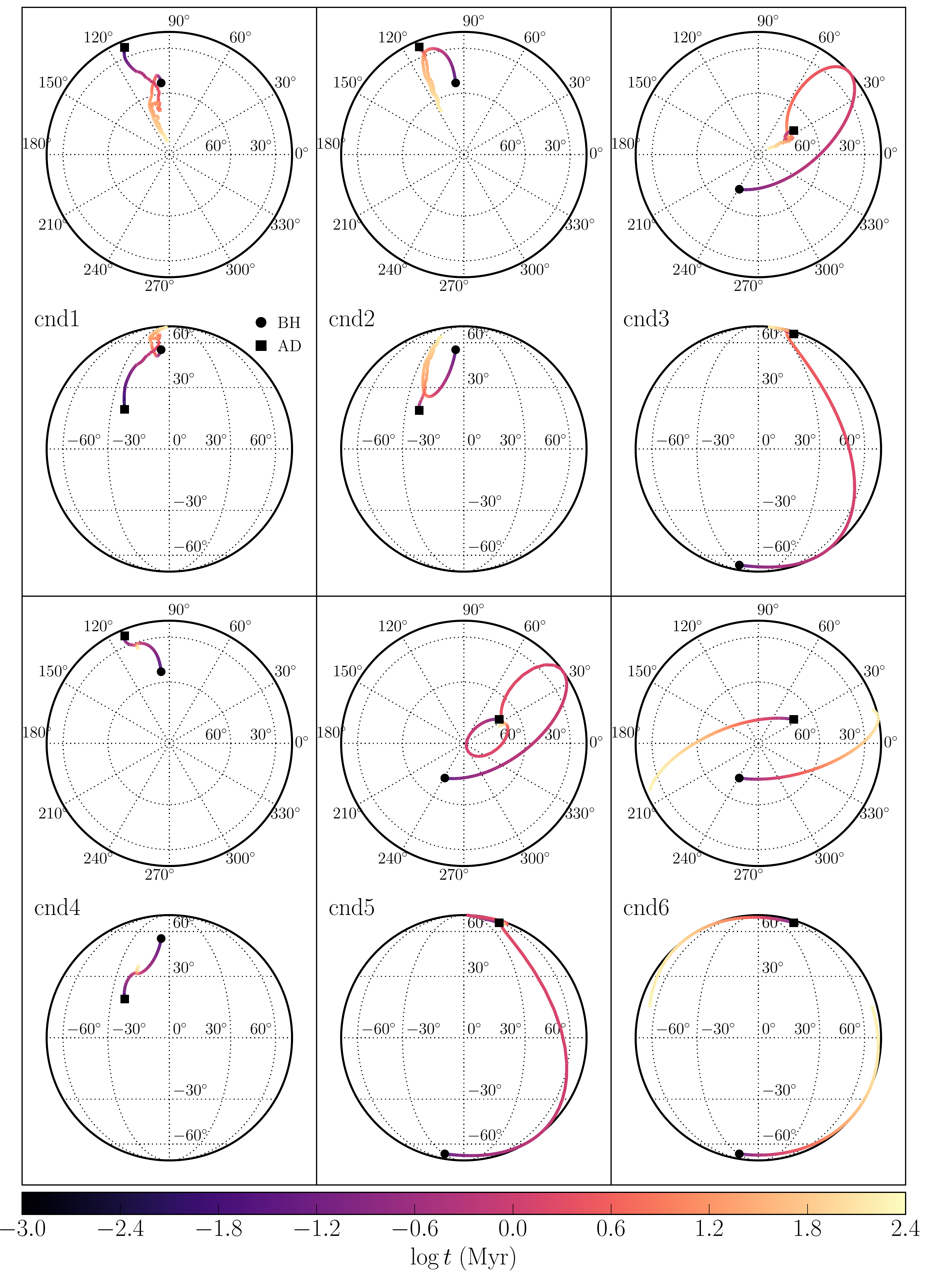}
\caption{Spherical projections of the black hole and the accretion disc angular momentum directions.
The colour of the curves indicates the time coordinate.
In each panel, the black hole is identified by a black filled circle corresponding to the beginning of the evolution, whereas a black square marks the initial orientation of the accretion disc.
The upper row shows the view along the ``N-S'' axis, which corresponds to the $z$ axis in the simulation domain, whereas the lower row shows the ``equatorial'' view of the projection, corresponding to the $y$ axis in the simulation domain.
The Bardeen-Petterson effect efficiently (counter)aligns the black hole and accretion disc angular momenta for low mass black hole; afterwards, the evolution is dictated by the torque from the inflowing material. 
} 
\label{fig_am_proj}
\end{center}
\end{figure*}

\begin{figure*}
\begin{center}
\includegraphics[width=2.1\columnwidth]{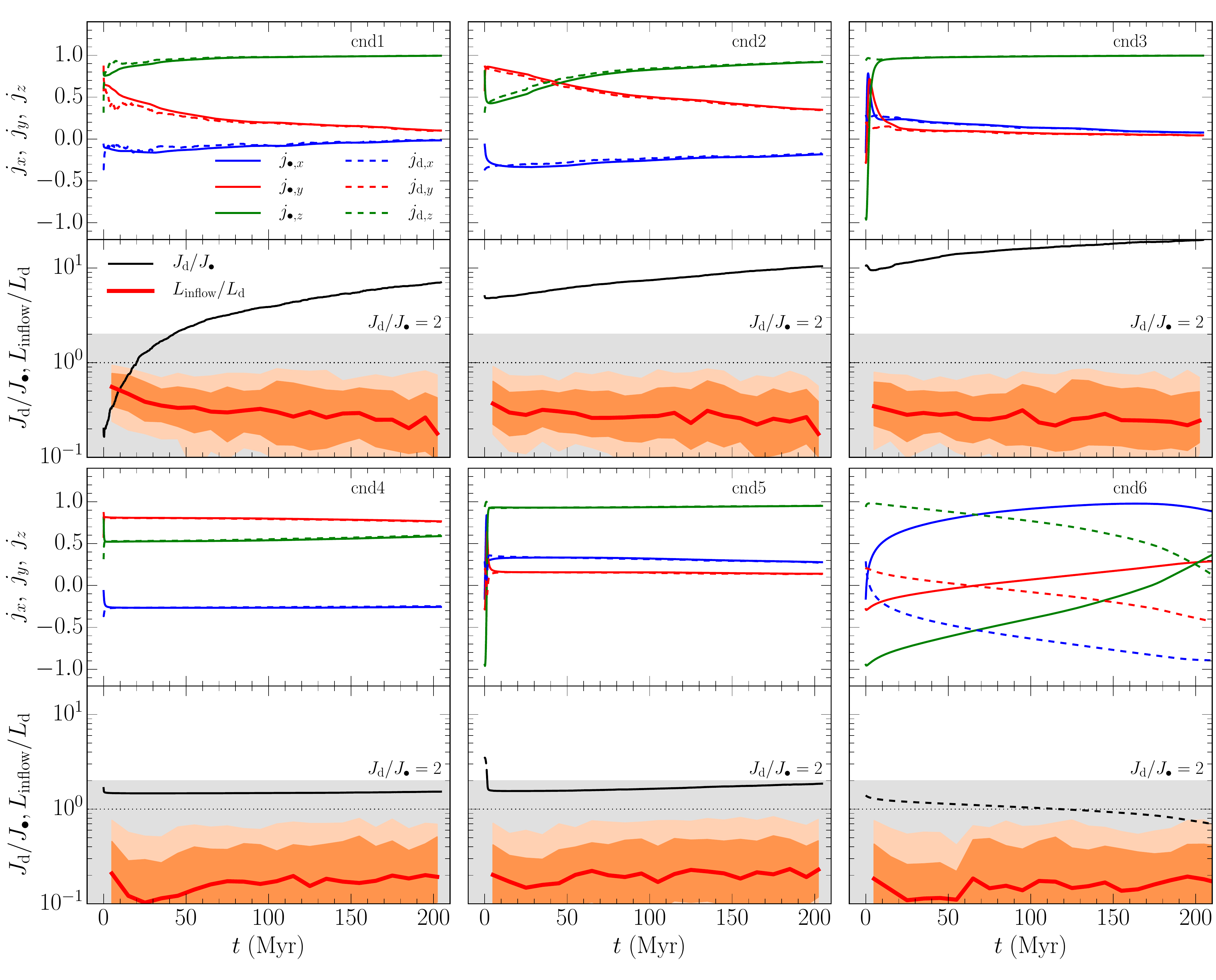}
\caption{Upper panels, from left to right and from top to bottom: time evolution of the Cartesian components ($x$=blue, $y$=red, $z$=green) of $\bmath{j}_{\bullet}$ (solid curves) and $\bmath{j}_{\rm d}$ (dashed curves).
Lower panels, from left to right and from top to bottom: time evolution of the ratio $J_{\rm d}/J_{\bullet}$ (thin black solid curve, whose dashed part indicates when $\bmath{j}_{\bullet} \cdot \bmath{j}_{\rm d} < 0$) and of the ratio $L_{\rm inflow} / L_{\rm d}$ (thick red solid curve), with dark and light orange to indicate the 68\% and 90\% regions of $L_{\rm inflow} / L_{\rm d}$.
The horizontal dotted line marks $y = 1$, while the grey shaded region shows $J_{\rm d} / J_{\bullet} < 2$, i.e. the region where it is possible to have stable counter-aligned configuration.
After alignment, the external inflow drives the growth of $J_{\rm d}/J_{\bullet}$ with typical $L_{\rm inflow} \sim 0.2-0.3 L_{\rm d}$ (except for run cnd6, see Section \ref{subsub_counter}).
}
\label{fig_am_comp}
\end{center}
\end{figure*}

It is instructive to compare the accretion rate calculated by our model with naive predictions based on the Bondi inflow solution, which would be $\sim 1-10$~M$_{\sun}$~yr$^{-1}$ in our circumnuclear disc runs.
The striking difference between the latter estimate and both $\dot{M}_{\rm inflow}$ and $\dot{M}$ illustrates the crucial impact of angular momentum in the mass transport captured by the usage of the mass flux for the inflow rate and by our accretion model (see also e.g. \citealt{hopkins+11,curtis+16b}).

The mass growth of the central black hole is rather slow over the simulated evolution because of the small values of $f_{\rm Edd}$.
Indeed, the black hole masses increase by about 5\% and 1.4\% for initial $M_{\bullet} = 10^6$~M$_{\sun}$ (i.e. run cnd1-3) and $M_{\bullet} = 10^{7}$~M$_{\sun}$, respectively.
In particular, the black holes in run cnd3 and cnd5 grow a bit more than in the other cases.
This is due to the initial very brief counter-rotating phase in cnd3 and cnd5 when the black hole mass quickly increases for about 1~Myr owing to the lower radiative efficiency (we discuss in more detail the angular momentum alignment and the special case of run cnd6 below). 

Similar considerations also apply to the spin parameter $a_{\bullet}$, whose evolution is shown in Figure~\ref{fig_mdot_a}.
Indeed, from equation~(\ref{eq_bh_am_evol}) we can see that $a_{\bullet}$ evolves over a timescale $\tau_{a_{\bullet}} = a_{\bullet} / \dot{a}_{\bullet} \sim \tau_{M_{\bullet}} = M_{\bullet}/\dot{M}$, i.e. of order of the timescale needed to significantly increase the black hole mass, modulo a factor $\mathcal{O}(1)$ that depends on $a_{\bullet}$.
All the simulated black holes eventually spin up except for cnd6, where the system attains a counter-rotating stable configuration.
This behaviour is a consequence of the initial $J_{\rm d} / J_{\bullet}$, which in most cases is dominated by the accretion disc (or the two vectors are close to be aligned from the very beginning), as more likely expected for low mass black holes \citep{dotti+13}. 

\subsubsection{Angular momentum alignment: Bardeen-Petterson effect and external inflow}

The evolution of the directions of $\bmath{J}_{\bullet}$ and $\bmath{J}_{\rm d}$ is shown in Figures~\ref{fig_am_proj} and \ref{fig_am_comp}.
Specifically, Figure \ref{fig_am_proj} shows the time evolution of the polar and equatorial projections of $\mathbf{j}_{\bullet}$ and $\mathbf{j}_{\rm d}$ for all the runs reported in Table~\ref{tab_cnd_runs}.
The polar projection is along the $z$ axis, corresponding to the rotational axis of the large scale circumnuclear disc, while the equatorial is done along the $y$ axis.
With the exception of run cnd6, that will be discussed in further detail below in Section~\ref{subsub_counter}, all the other runs show the expected alignment between the black hole and the accretion disc angular momenta.
Initially, the two angular momenta align rather quickly on a timescale that goes from a fraction of a Myr to a few Myr, as expected from $\tau_{\rm align}$.
For similar accretion rates as shown in Figure~\ref{fig_mdot_a}, the alignment takes longer for black holes with larger masses.
Physically, the torque originates mostly from the material flowing through the warp at $R_{\rm warp}$, but $R_{\rm warp}$ grows highly sub-linearly with $M_{\bullet}$.
As a consequence, the alignment timescale increases for heavier black hole accreting at a similar $\dot{M}$ \citep{lodato+06,martin+07,dotti+13}.

The evolution of the direction proceeds in a combination of precession and alignment with respect of $\bmath{j}_{\rm tot}$, where $\bmath{j}_{\bullet}$ mostly aligns with $\bmath{j}_{\rm d}$ because $J_{\rm d}/J_{\bullet} \gg 1$, as shown in Figure~\ref{fig_am_comp}.
The precession and alignment motions are particularly visible for e.g. run cnd5, because gas inflow modifies $\bmath{J}_{\rm tot}$ rather slowly.
On the other hand, the total angular momentum in runs cnd1-3 with lighter black holes varies more significantly over time.
Indeed, after the initial Bardeen-Petterson alignment, $\bmath{j}_{\bullet}$ and $\bmath{j}_{\rm d}$ move together toward the polar axis in Figure~\ref{fig_am_proj}, i.e. the rotation axis of the circumnuclear disc.
This is a consequence of the coherent angular momentum inflow onto the central region (see Figure~\ref{fig_map_example}) that progressively forces the alignment of the joint black hole and accretion disc system with the larger scale circumnuclear disc angular momentum.
Figure~\ref{fig_am_comp} shows that $L_{\rm inflow}$ has a larger dispersion, but it roughly stays between $\approx 20$ and $\approx 40 \%$ of the disc specific angular momentum.
However, a small degree of misalignment is visible during the migration of $\bmath{J}_{\bullet}$ and $\bmath{J}_{\rm d}$ for runs cnd1-3 both in Figure~\ref{fig_am_proj} and \ref{fig_am_comp}, while it is less evident for runs cnd4 and cnd5.

We estimate the evolution timescale for $\bmath{J}_{\bullet}$ to align with the rotation axis of the circumnuclear disc as follows.
For the sake of simplicity, we assume that (i) the Bardeen-Petterson effect is effective enough in maintaining $\bmath{J}_{\bullet} \parallel \bmath{J}_{\rm d}$ (this is a fair assumption even for the light $M_{\bullet}$ cases), and (ii) the torque caused by inflowing material is perfectly coherent and always aligned with the circumnuclear disc axis.
Therefore, we can simply write the total angular momentum evolution equation as 
\begin{equation}
\frac{{\rm d} \bmath{J}_{\rm tot}}{{\rm d}t} = \frac{{\rm d}}{{\rm d}t} \left[ (J_{\rm d} + J_{\bullet} ) \bmath{j}_{\rm tot} \right] = \dot{M}_{\rm inflow}(t) L_{\rm inflow}(t) \mathbf{e}_{z}.
\end{equation}
If we project the above equation first along $\bmath{j}_{\rm tot}$, then along $\bmath{e}_{z}$, and we finally combine the results, we can write a single evolution equation for $\mu = \cos \theta$, where $\theta$ is the angle between $\bmath{j}_{\rm tot}$ and $\bmath{e}_{z}$, namely
\begin{equation}
\frac{{\rm d} \mu}{{\rm d}t} =  \frac{\dot{M}_{\rm inflow}(t) L_{\rm inflow}(t)}{J_{\rm d}(t) + J_{\bullet}(t)} (1 - \mu^2) \equiv \frac{1 - \mu^2}{T(t)},
\end{equation}
where we have defined the timescale $T(t) = (J_{\rm d} + J_{\bullet})/(\dot{M}_{\rm inflow} L_{\rm inflow})$, which is time-dependent in general.
If we simply consider $T(t)$ as a constant and we neglect the initial value of $\theta$, the time evolution of $\theta$ is $\theta(t) \approx\arccos[\tanh(t/T)]$, which implies that alignment should be nearly completed after $\approx 4 T$.
Therefore, we can calculate directly $T(t)$ from the simulation to have an estimate of the time that $\bmath{j}_{\rm tot}$ requires to align with the circumnuclear disc rotation axis\footnote{Similarly to what we have described above regarding $\dot{M}_{\rm inflow}$ in Figure~\ref{fig_mdot_a}, we boost the estimate of $4 T(t)$ by the factor $1/(1 - \delta)$ to account for timesteps during which $\dot{M}_{\rm inflow} = 0$.}.
We find that $4 T(t)$ typically fluctuates between $\sim 100$~Myr and $\sim 1$~Gyr for runs cnd1, cnd2, and cnd3, while it is longer for runs cnd4 and cnd5, always ranging between 1 and a few Gyr.
We note a posteriori that the assumption that $T(t) = {\rm constant}$ is fairly accurate for runs cnd1-3, while it is less appropriate for runs cnd4 and cnd5, where $T(t)$ decreases by a factor $\sim 3$ over time, likely because of the steady increase of $\dot{M}_{\rm inflow}$.
For consistency, we estimate an evolution timescale by calculating numerical derivatives of the Cartesian components of $\bmath{j}_{\rm tot}$ and we find similar values.

The timescale $T(t)$ is significantly larger than the typical values of $\tau_{\rm align}$.
This explains why the Bardeen-Petterson effect is effective in maintaining alignment between the two angular momenta, whose evolution is ultimately due to the gas inflow coming from larger scales, at least for the rather low mass black holes that we explored thus far in this set of simulations (see also Section~\ref{subsec_turb_cloud}).
Finally, we note that we did not computed the value of $4 T(t)$ for run cnd6.
Indeed, the evolution of the components of $\bmath{j}_{\bullet}$ and $\bmath{j}_{\rm d}$ in Figure~\ref{fig_am_comp} show a completely different dynamics that does not follow the consideration above and that we explore more specifically in the following section.

\subsubsection{Counter-rotating accretion disc}\label{subsub_counter}

\begin{figure}
\begin{center}
\includegraphics[width=0.99\columnwidth]{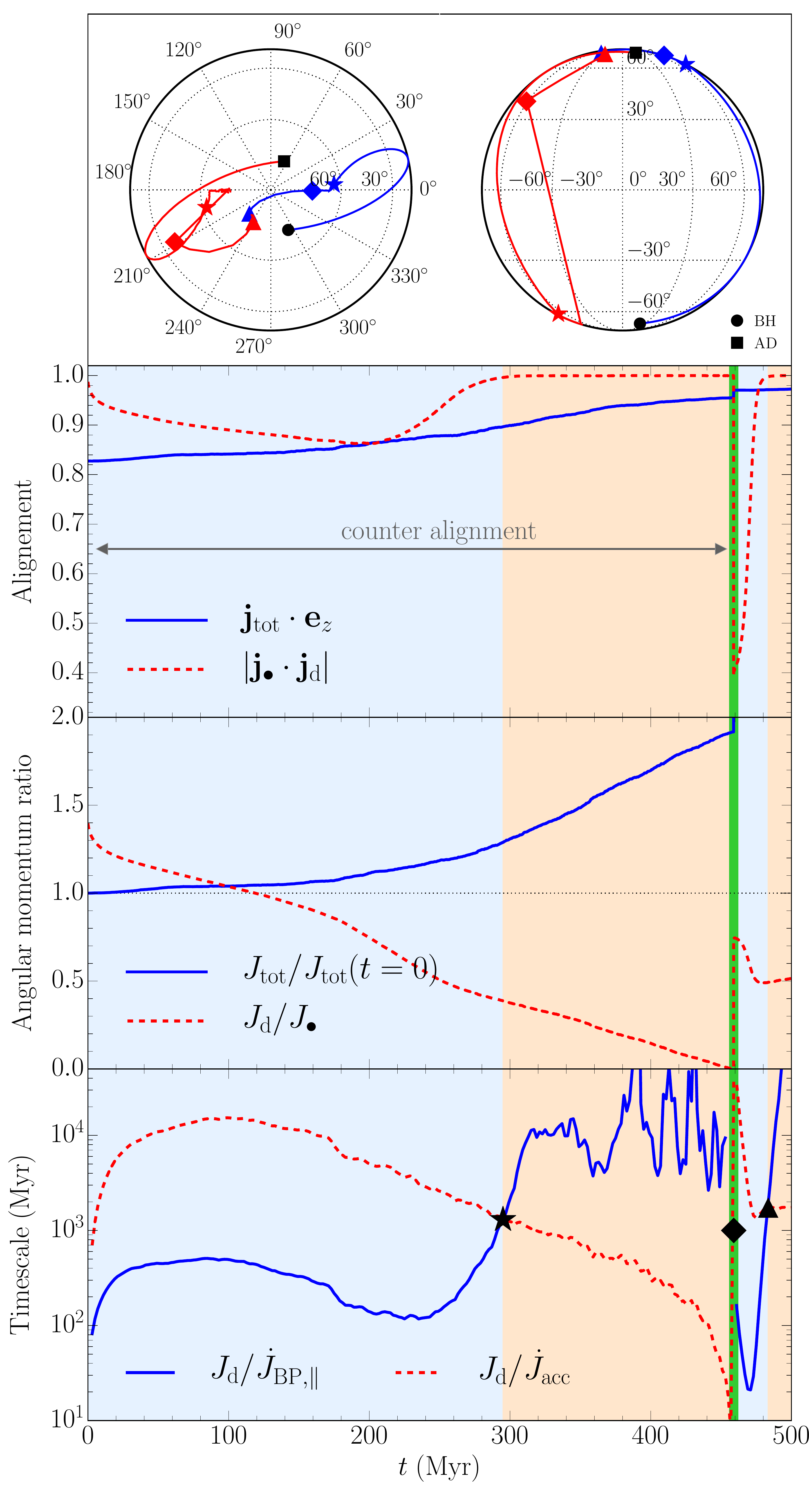}
\caption{Full time evolution of run cnd6 up to 500~Myr.
First row: polar and equatorial projection of $\bmath{j}_{\bullet}$ (blue curve, from the black circle) and $\bmath{j}_{\rm d}$ (red curve, from the black square).
Second row: $z$-component of $\bmath{j}_{\rm tot}$ (blue solid curve) and $|\bmath{j}_{\bullet} \cdot \bmath{j}_{\rm d}|$ (red dashed curve).
Third row: $J_{\rm tot} / J_{\rm tot}(t=0)$ (blue solid curve) and $J_{\rm d}/J_{\bullet}$ (red dashed curve).
Fourth row: timescale for the Bardeen-Petterson effect ($J_{\rm d}/\dot{J}_{\rm BP, \parallel}$, blue solid curve) and accretion ($J_{\rm d}/\dot{J}_{\rm acc}$, red dashed curve) to modify $J_{\rm d}$.
The light blue and light orange shaded regions highlight when $J_{\rm d}/\dot{J}_{\rm BP, \parallel} < J_{\rm d}/\dot{J}_{\rm acc}$ and $J_{\rm d}/\dot{J}_{\rm BP, \parallel} > J_{\rm d}/\dot{J}_{\rm acc}$, respectively, while the green vertical stripe indicates when the accretion disc is rebuilt.
The corresponding times are indicated by the star, diamond, and triangle in the first and fourth row.
For $t \leq 450$~Myr, $\bmath{j}_{\bullet} \cdot \bmath{j}_{\rm d} < 0$ as indicated in the second row.
}
\label{fig_cnd6}
\end{center}
\end{figure}

The evolution of the spin parameter and of the black hole and accretion disc $\bmath{j}$'s of run cnd6 shows a qualitatively different dynamics from the other cases as summarised in Figure~\ref{fig_cnd6}.
As noted earlier, the initial configuration of run cnd6 is expected to reach a stable equilibrium with the disc angular momentum counter-aligned with respect to the black hole angular momentum.
This is indeed shown by the upper panel of Figure~\ref{fig_cnd6}: the projected $\bmath{j}_{\bullet}$ and $\bmath{j}_{\rm d}$ are nearly counter-aligned from the beginning and they remain so for about $450$~Myr. 
However, they change orientation at the same time by roughly 180\degr, almost swapping in direction, with the black hole and accretion disc angular momentum eventually pointing to the ``north'' and the ``south'' poles, respectively. 

Such dynamics can be understood by looking simultaneously at the evolution of the total angular momentum.
The second row of Figure~\ref{fig_cnd6} shows that $\bmath{j}_{\rm tot}$ is almost aligned with the $z$ axis, i.e. the rotation axis of the large scale circumnuclear disc.
Over time, the direction of the total angular momentum does not change much, at least for the first $\approx 300$~Myr.
Therefore, we can consider $\bmath{j}_{\rm tot}$ as nearly constant.
The third row of Figure~\ref{fig_cnd6} shows instead that $J_{\rm tot}$ increases with time by $\approx 20-30\%$ after $\approx 300$~Myr from the beginning of the run.
Such a variation is ultimately related to the mass and angular momentum inflow. 
The latter is rather coherent with respect to $\bmath{J}_{\rm tot}$, namely $\bmath{j}_{\rm tot} \cdot \bmath{l}_{\rm inflow} \approx 1$.
Note, however, that the variation of $\bmath{J}_{\rm tot}$ is rather small and for the sake of simplicity, we can assume that $\bmath{J}_{\rm tot}$ is roughly conserved as if the external inflow were negligible and the system evolved in isolation.
Then, it is easy to realise that the swapping between the directions of the two angular momenta is related to the rapid decline of the ratio $J_{\rm d}/J_{\bullet}$ shown in Figure~\ref{fig_cnd6}.
If the total angular momentum has to be conserved while $\bmath{j}_{\bullet}$ and $\bmath{j}_{\rm d}$ remain nearly counter-aligned, they must move such that the largest lie roughly along $\bmath{j}_{\rm tot}$.
Initially, this is $J_{\rm d}$, but as $J_{\rm d}/J_{\bullet}$ drops significantly below 1, the two vectors must swap in direction.

The ratio $J_{\rm d}/J_{\bullet}$ evolves because of the effect of mass accretion and the Bardeen-Petterson effect. 
Mass accretion modifies the modulus of both the angular momentum of the black hole and of the accretion disc.
On the other hand, while the Bardeen-Petterson effect does not modify $J_{\bullet}$, it affects $J_{\rm d}$ owing to its dissipative nature \citep{king+05}.
Indeed, the torque $\dot{\bmath{J}}_{\bullet, \rm BP} = - \dot{\bmath{J}}_{\rm d, BP}$ produced by the Bardeen-Petterson effect is perpendicular to $\bmath{J}_{\bullet}$, but it must have a component $\dot{J}_{\rm BP, \parallel}$ along $\bmath{J}_{\rm d}$ if (counter-)alignment is not exact.
This is the case during the initial $\approx 300$~Myr of the simulation (see second row of Figure~\ref{fig_cnd6}).
Therefore, we can estimate the timescale for the Bardeen-Petterson effect to reduce $J_{\rm d}/J_{\bullet}$ as $J_{\rm d} / \dot{J}_{\rm BP, \parallel}$. 
Similarly, we can estimate the equivalent timescale for accretion as $J_{\rm d} / \dot{J}_{\rm acc}$, where $\dot{J}_{\rm acc}$ is the torque due to transfer of matter from the accretion disc to the black hole. 

The bottom row of Figure~\ref{fig_cnd6} compares the two timescales.
Over the first $\approx 300$~Myr, $J_{\rm d} / \dot{J}_{\rm BP, \parallel}$ is much shorter than $J_{\rm d} / \dot{J}_{\rm acc}$ and of the same order, i.e. $\sim 300$~Myr, of the observed timescale for the vectors' swapping described above.
This suggests that the dissipative component of the Bardeen-Petterson effect is the main driver of the spin dynamics in run cnd6.
This can be qualitatively understood also by considering that the Bardeen-Petterson torque originates around $R_{\rm warp}$, while the accretion torque is related to $L_{\rm isco}$.
For a Keplerian disc, their ratio must be $\sim \left( R_{\rm warp} / R_{\rm isco} \right)^{1/2} \sim 25~(M_{\bullet}/10^6~{\rm M}_{\sun})^{2/35}~f_{\rm Edd}^{-3/35}$ \citep{lodato+06, martin+07, perego+09}.
Therefore, the misalignment $\theta$ between $\bmath{j}_{\bullet}$ and $\bmath{j}_{\rm d}$ must satisfy $\sin(\pi - \theta) \lesssim \left( R_{\rm isco} / R_{\rm warp} \right)^{1/2} \sim 0.01-0.05$ for the accretion torque to become comparable to or to dominate over $\dot{J}_{\rm BP,\parallel}$.

This is indeed shown in Figure~\ref{fig_cnd6} at $t \approx 300$~Myr, when the timescale associated to $\dot{J}_{\rm BP, \parallel}$ becomes significantly longer than $J_{\rm d} / \dot{J}_{\rm acc}$.
Thereafter, the two vectors are almost exactly counter-aligned, the accretion torques dominates, and the ratio $J_{\rm d}/J_{\bullet}$ quickly goes down with time. The decrease of $J_{\rm d}/J_{\bullet}$ is also aided by angular momentum inflow that after the swap is mostly counter-aligned with $\bmath{j}_{\rm d}$, i.e. $\bmath{j}_{\rm d} \cdot \bmath{l}_{\rm inflow} \approx -1$.
After about 150~Myr, the angular momentum of the disc decreases so much that it hits the threshold $J_{\rm d}/M_{\rm d} = L_{\rm ISCO}$ and the disc is drained by the black hole.
We then reconstruct the accretion disc with initial mass $10^4$~M$_{\sun}$, initial angular momentum such that $J_{\rm d}/J_{\bullet} \approx 0.74$, and initial misalignment of $\theta \approx 66\degr$.
According to the \citet{king+05} criterion, $\cos\theta = 0.4 > -0.37 = -J_{\rm d}/(2 J_{\bullet})$, the black hole and the accretion disc angular momenta should realign as indeed happens in about 10~Myr.
During this time, we have again that $\dot{J}_{\rm BP,\parallel} > \dot{J}_{\rm acc}$ and $J_{\rm d}/J_{\bullet}$ decreases.
After alignment is complete, accretion starts to dominate and $J_{\rm d}/J_{\bullet}$ begins to raise slowly because $\bmath{J}_{\rm tot}$ points within 90\degr from the circumnuclear disc rotation axis, and therefore the angular momentum inflow adds up rather coherently to $\bmath{J}_{\rm d}$.


\subsection{Spin evolution in turbulent environments} \label{subsec_turb_cloud}

\subsubsection{A toy model of a bulge: run set-up}

\begin{table*}
\caption{Summary of the turbulent cloud runs.
From left to right: run label, mass of the background potential component, radius of the background potential component, initial radius of the gas cloud, initial black hole mass, initial accretion disc mass, fraction of solenoidal modes, rotation-over-turbulence ratio, Eddington ratio used to reconstruct the accretion disc after a draining event, turbulence decay timescale, dynamical time of the gas cloud [calculated as $t_{\rm dyn} = r_{\rm g} / V$, where $V$ is the total velocity scale of the gas+background isothermal sphere, see equation (\ref{eq_sis_velscale})], final simulation time in units of $t_{\rm dyn}$.
We use $M_{\rm g}/M_{\star} = 0.01$ and $r_{\rm c} = 50$~pc in all simulations; $f_{\rm Edd}^{\rm (min)} = 10^{-4}$.
}
\label{tab_tc_runs}
\begin{tabular}{lccccccccccc}
\hline
Label & $M_{\star}$ & $r_{\star}$ & $r_{\rm g}$ & $M_{\bullet,0}$ & $M_{\rm d,0}$ & $f_{\rm sol}$ & $V_{\phi}/\sigma$ & $f_{\rm Edd}^{\rm (refill)}$ & $t_{\rm decay}$ & $t_{\rm dyn}$ & $t_{\rm fin} / t_{\rm dyn}$ \\
 & (M$_{\sun}$) & (kpc) & (kpc) & (M$_{\sun}$) & (M$_{\sun}$) &  & & & (Myr) & (Myr) & \\
\hline
tc1     & $2 \times 10^{10}$ & 2.5 & 1 & $10^{6}$ & $10^4$ & 0.75 & 0 & $f_{\rm Edd}^{\rm (min)}$ & 7.6  & 7.4 & 9.1 \\
tc1\_LF$^{a}$ & $2 \times 10^{10}$ & 2.5 & 1 & $10^{6}$ & $10^4$ & 0.75 & 0 & $f_{\rm Edd}^{\rm (min)}$ & 7.6  & 7.4 & 13.3 \\
tc2     & $2 \times 10^{10}$ & 2.5 & 1 & $10^{6}$ & $10^4$ & 0.75 & 5 & $f_{\rm Edd}^{\rm (min)}$ & 37.2 & 7.4 & 10.1 \\
tc3     & $2 \times 10^{10}$ & 2.5 & 1 & $10^{6}$ & $10^4$ & 0.25 & 0 & $f_{\rm Edd}^{\rm (min)}$ & 7.6  & 7.4 & 4.0 \\
tc4     & $2 \times 10^{10}$ & 2.5 & 1 & $10^{6}$ & $10^4$ & 0.25 & 5 & $f_{\rm Edd}^{\rm (min)}$ & 37.2 & 7.4 & 10.6 \\
\hline
tc5     & $2 \times 10^{11}$ & 12 & 1.5 & $2 \times 10^{8}$ & $10^5$ & 0.75 & 0 & $f_{\rm Edd}^{\rm (min)}$ & 8.0 & 7.5 & 8.5 \\
tc5\_LF$^{b}$ & $2 \times 10^{11}$ & 12 & 1.5 & $2 \times 10^{8}$ & $10^5$ & 0.75 & 0 & $f_{\rm Edd}^{\rm (min)}$ & 8.0 & 7.5 & 4.2 \\
tc5\_HE & $2 \times 10^{11}$ & 12 & 1.5 & $2 \times 10^{8}$ & $10^5$ & 0.75 & 0 & $10^{-1}$ & 8.0 & 7.5 & 9.9 \\
tc6\_HE & $2 \times 10^{11}$ & 12 & 1.5 & $2 \times 10^{8}$ & $10^5$ & 0.75 & 5 & $10^{-1}$ & 39.8 & 7.5 & 7.6 \\
tc7\_HE & $2 \times 10^{11}$ & 12 & 1.5 & $2 \times 10^{8}$ & $10^5$ & 0.25 & 0 & $10^{-1}$ & 8.0 & 7.5 & 4.3 \\
\hline
\end{tabular}
\flushleft
Notes: $^{a}$ the forcing field amplitude of run tc1\_WF is $(1/3) \times$ that of run tc1; 
$^{b}$ the forcing field amplitude of run tc5\_LF is $(1/5) \times$ that of run tc5.
\end{table*}

The simulations described in Section~\ref{subsec_cnd} are useful tools to understand the connection between the large scale inflow and the black hole spin in simplified conditions.
However, additionally to our set-up being likely too idealised, it is worth to consider more general conditions, where we simultaneously wish to retain the character of a controlled numerical experiment to minimise the numerical impact of e.g. additional sub-grid modelling.
Therefore, we devised initial conditions to model a galactic bulge or a spherical early-type galaxy with a variety of gas kinematics. 
Such models are made of three components: (i) a stellar background spheroid, (ii) a gaseous medium, and (iii) a central supermassive black hole.
Since we are not interested in the dynamics of the stellar bulge itself, we model the stellar spheroid as a fixed background potential of an isothermal sphere with a central core, corresponding to the density profile
\begin{equation}\label{eq_lowered_sis}
\rho_{\star}(r) = \frac{M_{\star}}{4 \pi r_{\star}^3}~\frac{3 (r_{\rm c}/r_{\star})^2 + (r/r_{\star})^2}{\left[(r_{\rm c}/r_{\star})^2 + (r/r_{\star})^2 \right]^2},
\end{equation}
where $M_{\star}$ is the total stellar mass which is approximately contained inside $r_{\star}$, while $r_{\rm c} \ll r_{\star}$ is the radius of the central constant-density core.
For $r \gg r_{\rm c}$, the density profile follows the usual $\propto r^{-2}$ scaling.
The gaseous component initially follows the same profile of equation~(\ref{eq_lowered_sis}) with the same $r_{\rm c}$ but different total mass $M_{\rm g}$ and radial extent $r_{\rm g}$, and then is let evolve according to the gravitational pull exerted by the background stellar potential and its own self-gravity.

We explore diverse physical conditions by changing the initial dynamical state of the gas
Specifically, we initialise several models\footnote{The code to initialise the models described in this section is freely available at \url{https://bitbucket.org/fiacconi/turbulent_cloud}.} in approximate virial equilibrium by distributing different amounts of energy between thermal and kinetic, and subdividing further the kinetic energy between rotation and turbulence.
Such an approach motivates why we chose a density profile that mostly follows an isothermal sphere.
Indeed, the latter is characterised by a unique velocity scale that is constant across $r$, namely $V^2 \approx GM_{\star}/r_{\star} + G M_{\rm g}/r_{\rm g}$.
Therefore, we can set the redistribution of energy components simply by specifying two dimensionless number, (i) the Mach number $\mathcal{M} = \sigma / c_{\rm s}$ that relates the turbulent velocity dispersion $\sigma$ and the isothermal gas sound speed $c_{\rm s}$, and (ii) the ratio $V_{\phi} / \sigma$ between the rotational velocity $V_{\phi}$ and $\sigma$.
All components are then specified by the following relation
\begin{equation}\label{eq_sis_velscale}
V^2 = c_{\rm s}^2 \left\{ 1 + \mathcal{M}^2 \left[ 1 + \left( \frac{V_{\phi}}{\sigma} \right)^2 \right] \right\}.
\end{equation}

We assume that the gas is isothermal and the Mach number identifies a unique value for the velocity dispersion $\sigma$.
However, it is very likely that the interplay of several physical processes such as gravity and star formation and feedback may induce a multi-scale turbulent velocity field, as almost ubiquitously observed in the insterstellar medium \citep{hennebelle+12,falceta+14}.
It is beyond the purpose of this work to self-consistently include all these processes or to investigate their mutual relevance in shaping the interstellar medium.
Instead, we assume that their overall effect on large scales can be captured by an initially imposed turbulent velocity field that is maintained over time by an effective forcing field, as customary done in local simulations of turbulence \citep{schmidt+06, bauer+12,konstandin+12, federrath+13}.  

Following previous work \citep{hobbs+11, mapelli+12}, we initialise the velocity field in Fourier components on a $512^3$ Cartesian grid $2 r_{\rm g}$ per side that surrounds the gas cloud.
The Fourier components $\bmath{u}_{\bmath{k}}$ follow a power spectrum $P_{\bmath{u}}(\bmath{k}) \propto |\bmath{k}|^{-4}$ that extends between $k_{\rm min} = \pi/r_{\rm g}$ to $k_{\rm max} = 256 \pi /r_{\rm g}$, i.e. the Nyquist frequency associated to the grid sampling.
This choice for the power spectrum is appropriate to describe supersonic turbulence \citep{federrath+13}.
In order to control the relative amount of compressive (i.e. $\bmath{\nabla} \times \bmath{u} = 0$) and solenoidal (i.e. $\bmath{\nabla} \cdot \bmath{u} = 0$) modes, we do not initialise the velocity field directly.
Instead, we initialise a scalar Gaussian field $\phi$ and a vectorial Gaussian field $\bmath{A}$, both following a power spectrum $P(\bmath{k}) \propto |\bmath{k}|^{-6}$.
We calculate the solenoidal and the compressive modes of the velocity field that in real space correspond to $\bmath{\nabla} \times \bmath{A}$ and $\bmath{\nabla}\phi$, respectively.
Then, we compute the velocity field by summing up the solenoidal and compressive modes by means of the parameter $0 \leq f_{\rm sol} \leq 1$ that describes the fraction of solenoidal modes with respect to the total.
We Fourier transform $\bmath{u}_{\bmath{k}}$ back to real space, and we finally normalise the resulting velocity field according to the initial value of the Mach number $\mathcal{M}_{0} \approx 5$.
Some simulations also include a net rotation specified by the value of $V_{\phi}/\sigma$; the rotational velocity field is initially uniform around the $z$ axis. 

\begin{figure*}
\begin{center}
\includegraphics[width=2.1\columnwidth]{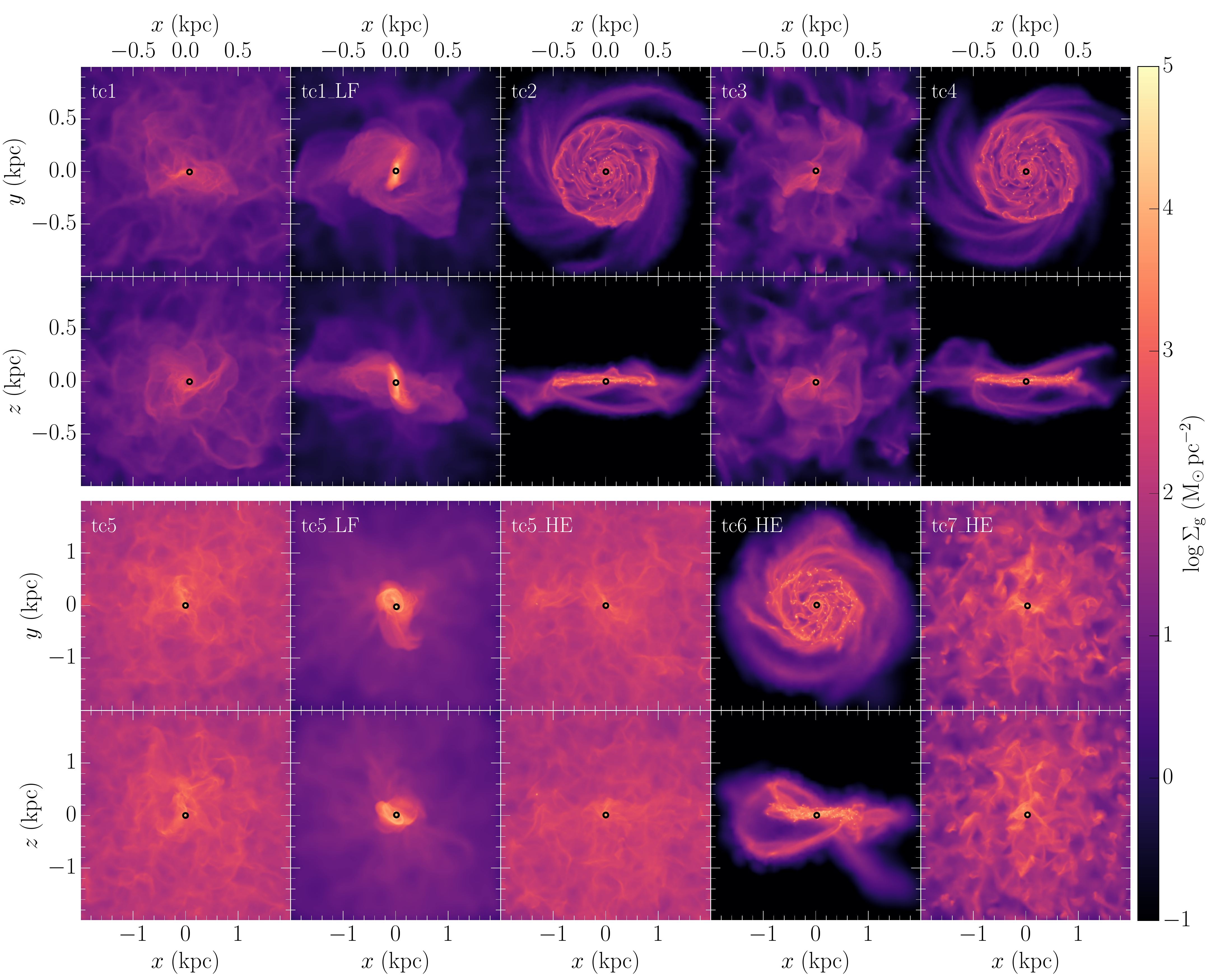}
\caption{Gas surface density maps of the ``bulge'' runs (top two rows) and of the ``elliptical'' runs (bottom two rows) at $t = 3 t_{\rm dyn}$. 
For each set, the top row is face-on ($xy$ plane), while the bottom row is edge-on ($xz$ plane).
In each map, the black circle marks the position of the central black hole.
Solenoidal forcing in spherical systems (tc1, tc5, tc5\_HE) produces curly and tenuous structures, while compressive forcing (tc3, tc7\_HE) produces thicker and denser plumes of gas.
Systems dominated by rotational energy (tc2, tc4,tc6\_HE) quickly settle into a clumpy disc, while systems with weaker turbulent forcing (tc1\_LF, tc5\_LF) develop denser central structures.
}
\label{fig_tc_comp_maps}
\end{center}
\end{figure*}

We evolve the simulations assuming that the gas is isothermal and including a stochastic acceleration field because otherwise turbulence would decay approximately on a turbulent crossing time $\sim r_{\rm g} / \sigma$.
To accomplish this we mostly follow \citet{bauer+12}.
The acceleration field is sampled in Fourier components $\bmath{a}_{\bmath{k}}$ between $k_{\rm min} = \pi/r_{\rm g}$ and $k_{\rm max} = 2 \pi / (3 r_{\rm g})$.
The amplitude and phase of each component follows a Ornstein-Uhlenbeck stochastic process which in differential form reads \citep{schmidt+06} 
\begin{equation}
{\rm d}\bmath{a}_{\bmath{k}}(t) = - \bmath{a}_{\bmath{k}}(t) \frac{{\rm d}t}{t_{\rm decay}} + \zeta \left( \frac{2 \tilde{a}^2_{k}}{t_{\rm decay}} \right)^{1/2} {\rm d}\bmath{\mathcal{W}}_{t}, 
\end{equation}
where $t_{\rm decay} = r_{\rm g} / \sigma$, ${\rm d}\bmath{\mathcal{W}}_t$ is a stochastic three-components Wiener process, $\tilde{a}^2_{k}$ is the asymptotic variance of the Ornstein-Uhlenbeck process, i.e. ${\rm Var}(\bmath{a}_{\bmath{k}}) = \tilde{a}^2_{k} \approx \sigma^2/t_{\rm decay}^2$, and $\zeta \sim 0.1$ is a dimensionless free parameter to tune the final acceleration field in real space such that the average long-term Mach number $\sim \mathcal{M}_{0}$.
$\tilde{a}_{k}$ also sets the relative amplitude of each mode to scale with $k = |\bmath{k}|$ as $a_{k} \propto -(k-\bar{k})^2$ (with $\bar{k} = (k_{\rm min} + k_{\rm max})/2$).
The resulting components of the acceleration field have zero mean and time correlation $\langle \bmath{a}_{\bmath{k}}(t), \bmath{a}_{\bmath{k}}(t+\Delta t) \rangle = \tilde{a}^2_{k} \exp(-\Delta t/t_{\rm decay})$ between two arbitrary instants separated by $\Delta t$.
This allows stochastic fluctuations but with a ``smoothly'' varying turbulent driving field over timescales $\sim t_{\rm decay}$.
Finally, before we sum up the Fourier components and transform the acceleration field back to real space, we project each Fourier component in solenoidal and compressive modes with the same fraction of solenoidal modes $f_{\rm sol}$ as in the initial velocity field.

We have run two classes of models meant to mimic (i) the bulge of Seyfert spiral galaxies hosting rather light supermassive black holes, and (ii) the inner regions of a massive elliptical galaxies hosting heavier supermassive black holes.
The two sets of simulations respectively explore the evolution of black holes typically with $M_{\bullet} < M_{\bullet}^{\rm (warp)}$ and $M_{\bullet} > M_{\bullet}^{\rm (warp)}$.
The properties of all runs are summarised in Table~\ref{tab_tc_runs}: simulations tc1 to tc4 belong to the first type of models, while runs from tc5 to tc7 belong to the second one.
The Seyfert bulges, are characterised by $M_{\star} = 2 \times 10^{10}$~M$_{\sun}$ enclosed in $r_{\star} = 2.5$~kpc.
This corresponds to a velocity dispersion $\approx 132$~km~s$^{-1}$, consistent with local scaling relations \citep{catinella+12}.
The gas cloud has a mass $M_{\rm g} = 2 \times 10^8$~M$_{\sun}$ within $r_{\rm g} = 1$~kpc, embedding at its centre a supermassive black hole $M_{\bullet} = 10^6$~M$_{\sun}$.
The gas cloud is initially sampled with a million gas cells with $m_{\rm g}^{\rm target} = 200$~M$_{\sun}$ and $\epsilon_{\rm g} = 1.5$~pc, while the black hole softening is 5~pc.
The early-type ellipticals instead consist of a background stellar component with $M_{\star} = 2 \times 10^{11}$~M$_{\sun}$ and $r_{\star} = 12$~kpc, consistently with mass-size relations for local early-type massive galaxies \citep{vanderwel+14}.
The gas content is again $0.01 M_{\star}$ and it initially extends to $r_{\rm g} = 1.5$~kpc, and the black hole mass is $M_{\bullet} = 2 \times 10^8$~M$_{\sun}$. 
Gas cell target mass is $m_{\rm g}^{\rm target} = 2000$~M$_{\sun}$ and the softening is $\epsilon_{\rm g} = 3.25$~pc; the black hole softening is 29~pc. 
In all simulations, the initial $a_{\bullet} = 0.5$ and the black hole and accretion disc angular momenta point to random directions separated by $\approx 124\degr$. We use $M_{\rm refill} = 10^5$~M$_{\sun}$ and $f_{\rm refill} = 10^{-3}$.
In the following, we will sometimes generally refer to the first group of runs as ``bulge'' simulations, to the second as ``elliptical'' simulations, and to all of them as ``turbulent clouds''.
We use outflowing gas boundary conditions and a box size $L=4 r_{\rm g}$, filled with low density and pressure gas.

\begin{figure*}
\begin{center}
\includegraphics[width=2.1\columnwidth]{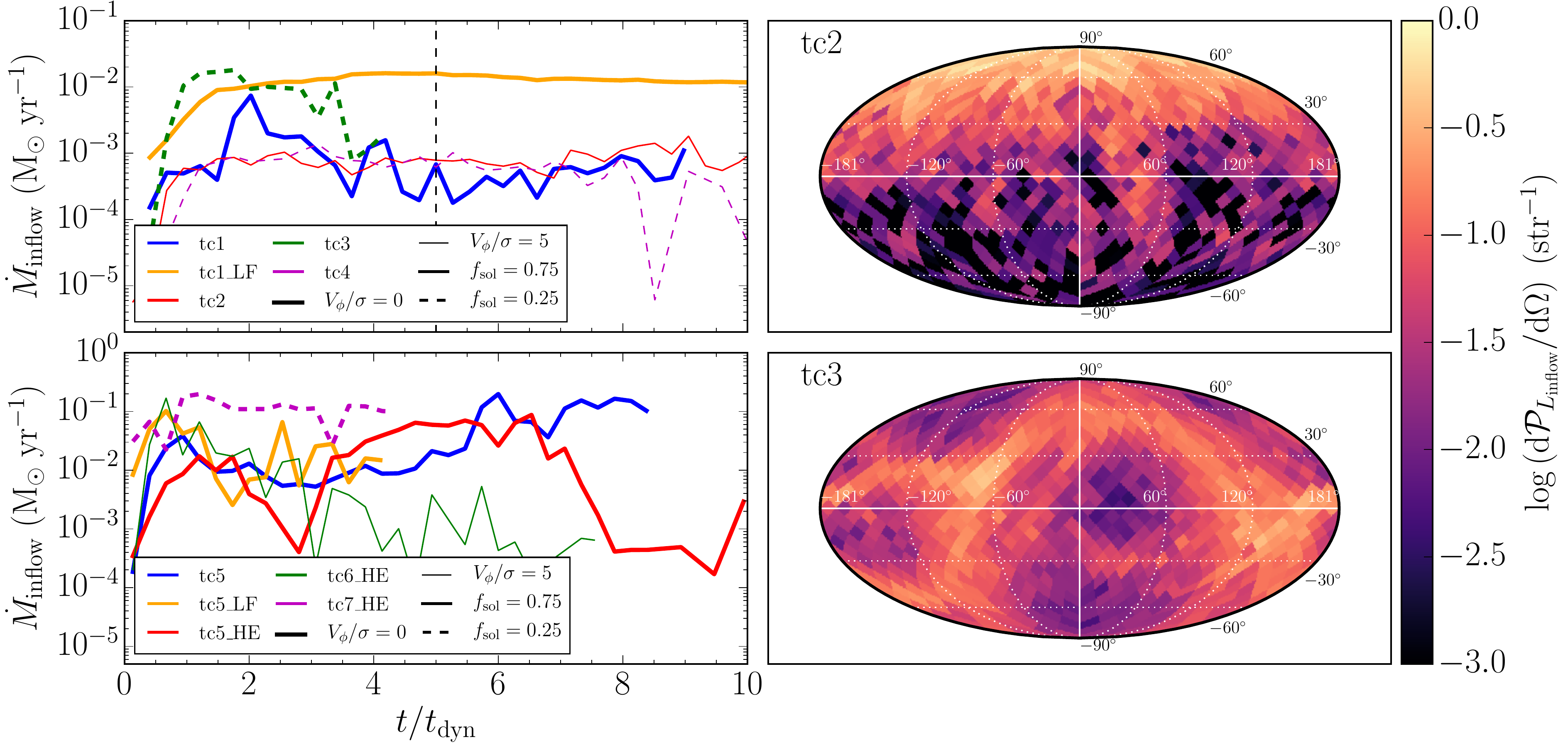}
\caption{Left column: time evolution of $\dot{M}_{\rm inflow}$ for ``bulge'' runs (upper panel) and ``ellipical'' runs (lower panel).
The colours indicate each run as stated in the legend; thick and thin curves are associated to $V_{\phi} / \sigma = 0$ and $V_{\phi} / \sigma = 5$, respectively; 
solid and dashed curves refer to $f_{\rm sol} = 0.75$ and $f_{\rm sol} = 0.25$, respectively.
The vertical dashed line in the top panel indicates when the black hole in run tc1 gets ejected.
Right column: $L_{\rm inflow}$-weighted angular probability over solid angle of the $\bmath{l}_{\rm inflow}$ direction for run tc2 (upper panel) and run tc3 (lower panel).
The equator corresponds with the $x$-$y$ plane in the simulations.
}
\label{fig_tc_inflow}
\end{center}
\end{figure*}

We run the simulations for several dynamical times (see Table~\ref{tab_tc_runs}) until they slow down due to very short timesteps $\lesssim 20$~yr and we cannot evolve them further.
This happens because of our simplified approach of neglecting the actual small-scale feedback processes from the black hole or from stars, while capturing only their large-scale effects through the turbulent driving.
As the turbulent force field stirs the gas, it creates overdense regions that may become self-gravitating and eventually collapse.
However, the collapse is not counteracted by star formation and the local energy injection associated with stellar feedback.
Therefore, the simulations are computationally limited by the short free-fall timescale in high-density regions, which form at different times depending on the properties of the forcing field and of the initial redistribution of kinetic energy between rotation and turbulence.
We defer more realistic setups that include gas cooling, star formation and feedback to a future work.

\subsubsection{Turbulence and gas flow}

The evolution of the gas changes depending on the parameters $f_{\rm sol}$ and $V_{\phi}/\sigma$ as illustrated in Figure~\ref{fig_tc_comp_maps}, which shows the gas distribution of all runs at $t = 3 t_{\rm dyn}$.
Regardless of the exact potential well (i.e. whether we consider the ``bulge'' simulations, runs tc1-4, or the ``elliptical'' simulations, runs tc5-7), the ``turbulent clouds'' whose kinetic energy is dominated by rotational motions (i.e. $V_{\phi}/\sigma = 5$) tend to settle down to a rotationally-supported thick disc in about one dynamical time.
The disc is gravitationally unstable and it fragments in small and dense clumps.
The turbulent forcing mainly stirs the disc in the vertical direction, while the planar motions are always dominated by rotation.
The disc formed in run tc6\_HE is slightly thicker and more turbulent in the vertical direction than in runs tc2 and tc4 because the sound speed and velocity dispersion assume larger values despite the similar $t_{\rm dyn}$ of the ``bulge'' and ``elliptical'' systems.
On the other hand, the runs with no net rotation tend to remain rather spherical with small scales substructures. 

\begin{figure*}
\begin{center}
\includegraphics[width=2.1\columnwidth]{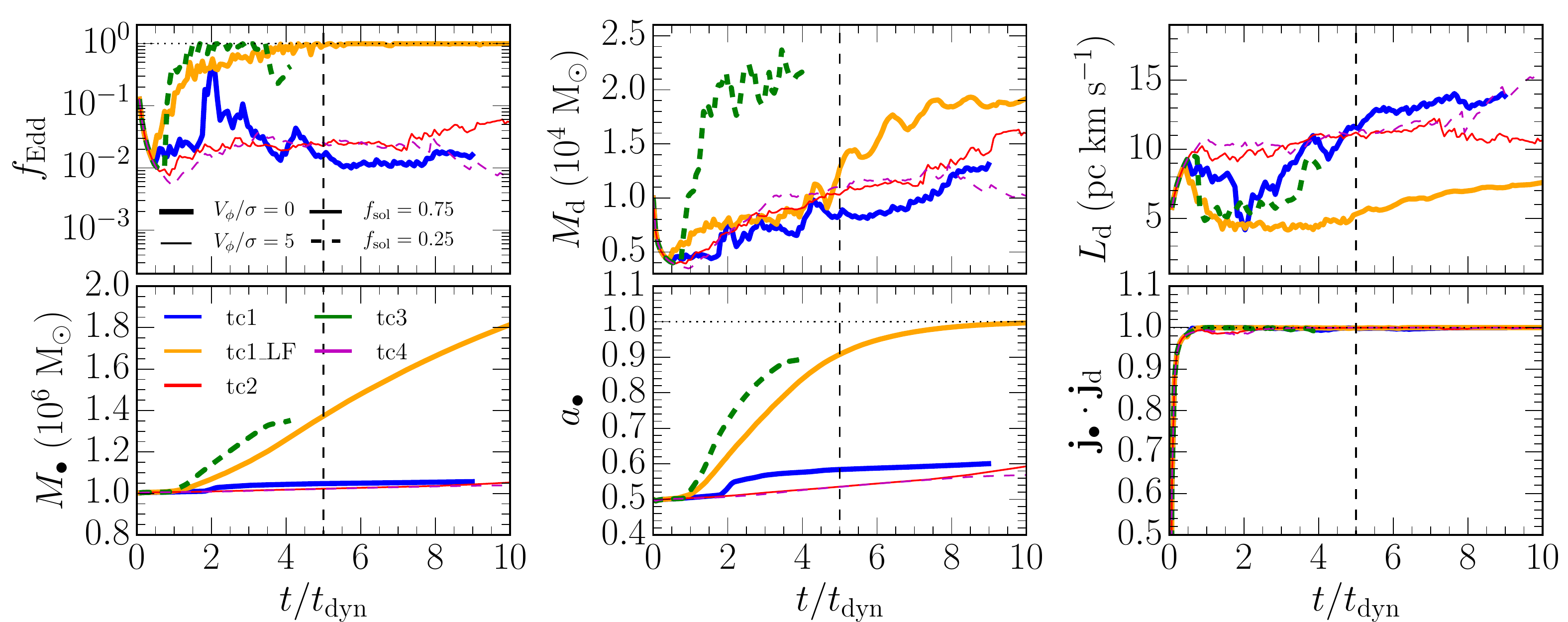}
\caption{From left to right and from top to bottom: time evolution of $f_{\rm Edd}$, $M_{\bullet}$, $L_{\rm d}$, $M_{\rm d}$, $a_{\bullet}$, and $\bmath{j}_{\bullet} \cdot \bmath{j}_{\rm d}$ in the ``bulge'' runs.
In each panel, thin (thick) curves correspond to $V_{\phi}/\sigma=5$ ($V_{\phi}/\sigma=0$), while solid (dashed) curves correspond to $f_{\rm sol}=0.75$ ($f_{\rm sol}=0.25$).
Blue, orange, red, green and magenta corresponds to run tc1, tc1\_LF, tc2, tc3, tc4, respectively.
The vertical dashed lines indicate when the black hole in run tc1 gets ejected. 
For low mass black holes where $J_{\rm d}/J_{\bullet} \gg 1$, the system reach alignment and does spin-up regardless of the details of the mass inflow.
}
\label{fig_tc_bulge_summary}
\end{center}
\end{figure*}

After $\approx1.5-2 t_{\rm dyn}$, the turbulence reaches an approximate steady state, as indicated by the mass-weighted Mach number $\langle \mathcal{M} \rangle$ that becomes rather constant and $\langle \mathcal{M} \rangle \approx \mathcal{M}_{0}$.
The amount of solenoidal vs. compressive modes in simulations without net rotation causes some qualitative differences in the gas flow as shown in Figure~\ref{fig_tc_comp_maps}.
The runs dominated by solenoidal modes develop curly, filamentary structures with lower density constrast than runs dominated by compressional modes; the latter, on the other hand, show thick and dense plumes of gas.
For simulations with $V_{\phi}/\sigma = 0$, the angular momentum imprinted in the gas ultimately leads to the alternate formation and disruption of a nuclear disc in the central core of the background potential on the scale of $\sim 20-30$~pc.
During the formation of these nuclear discs, they often fragments into massive clumps, in particular in the ``elliptical'' simulations.
Run tc1 is special in this respect, because a dense, disc-like clump forms around the black hole.
After $\approx 40$~Myr$\approx 5 t_{\rm  dyn}$, this clump gets ejected from the centre because of the dynamical interaction with the surrounding gas clumps and wonders at a few hundreds of pc from the centre, taking away the black hole that remains bound to it and keeps accreting from it.
Similarly, in run tc4, the black hole gets ejected out to $\approx 100$~pc from the centre after a 2-body encounter at $t\approx 54$~Myr with a massive clump and then it slowly sinks back owing to dynamical friction \citep[see e.g.][]{fiacconi+13,roskar+15}.

The strength of the turbulence field is set roughly to maintain the virial equilibrium, except for runs tc1\_LF and tc5\_LF where the stochastic acceleration is weaker.
As a consequence, in these runs the gas contracts and flows in during the initial $\approx 2 t_{\rm dyn}$.
It forms a dense circumnuclear disc with developed spiral arms around the black hole on the scale of $\approx200-300$~pc.
This circumnuclear disc is self-gravitating and it fragments into dense clumps, especially in run tc5\_LF.
The circumnuclear disc is surrounded by a spherical cloud of low density gas that is stirred by the turbulent field.
The circumnuclear disc also changes orientation in response to infalling streams of gas and torquing from the turbulent field.

Figure~\ref{fig_tc_inflow} shows the time evolution of $\dot{M}_{\rm inflow}$ (calculated as in Figure~\ref{fig_mdot_a}) for all ``turbulent cloud'' runs.
Focussing on the ``bulge'' simulations first, we note that the inflow that reaches the accretion disc is rather constant ($\approx 10^{-3}$~M$_{\sun}$~yr$^{-1}$) in runs with $V_{\phi}/\sigma = 5$, because mass inflow is slowly sustained by the coherent transport of angular momentum
in the disc due to spiral arms and shearing gas clumps.
We find that $\dot{M}_{\rm inflow}$ fluctuates more in runs with $V_{\phi}/\sigma = 0$.
In runs tc1 and tc3, the inflow is initially almost an order of magnitude larger than in the $V_{\phi}/\sigma = 5$ runs because of streams of gas infalling from different directions that lead to angular momentum cancellation and larger inflows\footnote{Note that the inflow decreases in run tc1 after the black hole is ejected from the centre.}.
Instead, run tc1\_LF sustains a rather constant but high $\dot{M}_{\rm inflow} \approx 0.01$~M$_{\sun}$~yr$^{-1}$ as a result of mass transport in the surrounding circumnuclear disc.
The mass inflow in the ``elliptical'' simulations does not show a clear trend among different runs.
In fact, $\dot{M}_{\rm inflow}$ fluctuates significantly between $\sim 10^{-3}$ and $\sim 10^{-1}$~M$_{\sun}$~yr$^{-1}$ in response to the evolution of the inner region that is locally dominated by massive clumps and spiral arms in the nuclear discs.
However, when we look at the mass inflow at $\gtrsim 100$~pc, we recover in each run the overall average time evolution and we find similar trends as for the ``bulge'' simulations, e.g. run tc6 has lower average mass inflow than run tc5 or tc7.

\begin{figure*}
\begin{center}
\includegraphics[width=2.1\columnwidth]{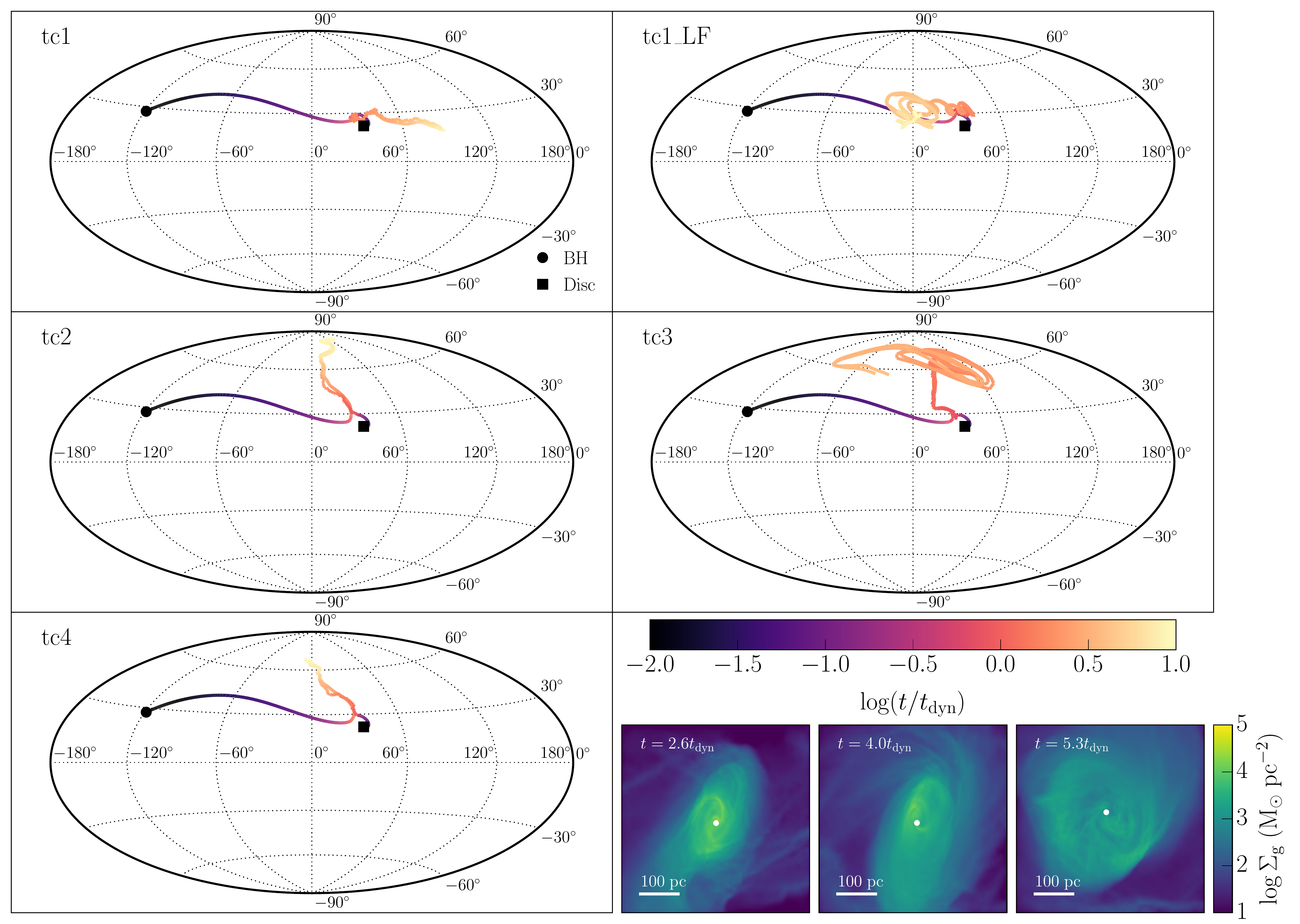}
\caption{Hammer ``full-sky'' projections of the black hole and the accretion disc angular momentum directions for the ``bulge'' simulations.
The colour of the curves indicates the time coordinate.
In each panel, the black hole is identified by a black circle corresponding to the beginning of the evolution, whereas a black square marks the initial orientation of the accretion disc.
The sequence of images shows the evolution of the inclination of the circumnuclear disc that forms in run cnd1\_LF as a representative example.
Each images shows the gas surface density on a scale of 500~pc; the white circle indicates the position of the central black hole.
After effective alignment between $\bmath{J}_{\bullet}$ and $\bmath{J}_{\rm d}$, the two vectors tends to align with the large scale angular momentum vector for rotation-dominated systems (tc2, tc4), while they erratically change direction in turbulence-dominated systems (tc1, tc1\_Lf, tc3).
}
\label{fig_tc_bulge_amproj}
\end{center}
\end{figure*}

Figure~\ref{fig_tc_inflow} also shows the $L_{\rm inflow}$-weighted angular distribution\footnote{We use the tessellation of the sphere provided by {\sc healpy}, available at \url{https://github.com/healpy/healpy}.} of $\bmath{l}_{\rm inflow}$ for two representative example runs.
Most of the specific angular momentum of the inflowing gas that reaches the accretion disc in run tc2 (representative of the $V_{\phi}/\sigma = 5$ cases) is aligned with the rotation axis of the circumnuclear disc within $\approx 60\degr$.
This distribution reflects the rather ordered motion of the gas in the disc formed after the collapse of the rotating spherical cloud, while the spread around the rotation axis accounts for the thickness and the turbulence in the disc.
On the other hand, simulations with no net rotation are characterised by a more isotropic distribution of $\bmath{L}_{\rm inflow}$, as shown, for example, by run tc3.
There are however regions that show an excess of probability where more specific angular momentum is coming from.
Those are associated with streams of gas or likely with the directions of the rotation axis of the central nuclear discs that often forms around the central black hole.

\subsubsection{Evolution of light supermassive black holes in galactic bulges}

The evolution of the black hole and accretion disc properties in the ``bulge'' simulations is summarised in Figure~\ref{fig_tc_bulge_summary}.
The black hole and accretion disc systems in all ``bulge'' simulations evolve smoothly because there are no events of disc draining and rebuilding.
All quantities show an initial transient of $\approx 0.5 t_{\rm dyn}$ due to the initial set up and the contemporary development of turbulence.
Thereafter, we observe clear differences among the runs in terms of the accretion rate $\dot{M}$ in units of the Eddington rate.
$f_{\rm Edd}$ is on average lower and rather constant, $f_{\rm Edd} \approx 2-3 \times 10^{-2}$, in runs tc2 and tc4 (i.e. $V_{\phi} / \sigma = 5$).
This behaviour reflects (i) the evolution emphasised above regarding $\dot{M}_{\rm inflow}$, and (ii) the tendency for $\dot{M}$ and $\dot{M}_{\rm inflow}$ to follow each other, as already noticed for the circumnucler disc simulations in Section~\ref{subsec_cnd}.
Indeed, the values of $\dot{M}$ in physical units are similar to the time-averaged values of $\dot{M}_{\rm inflow}$ in time bins of 1 Myr, although brief fluctuations in $\dot{M}_{\rm inflow}$ and timesteps with $\dot{M}_{\rm inflow} = 0$ may lead to differences between $\dot{M}_{\rm inflow}$ and $\dot{M}$ on a single timestep basis.
Similar considerations also apply to the runs without net rotation (as well as to the ``elliptical'' simulations), where $\dot{M}$ fluctuates more and it is capped to the Eddington rate for prolonged periods of time in response to the external inflow.

The accretion rate on to the black hole is ultimately set by the accretion disc mass and angular momentum, which evolve according to $\dot{M}_{\rm inflow}$ and $\dot{\bmath{J}}_{\rm inflow}$.
Figure~\ref{fig_tc_bulge_summary} shows the time evolution of $M_{\rm d}$ and $L_{\rm d}$.
The accretion disc mass tends to grow in all runs with some fluctuations, while $L_{\rm d}$ evolves differently in each simulation.
The rather coherent direction of $\bmath{L}_{\rm inflow}$ in run tc2 and tc4 forces a steady increase in $L_{\rm d}$, which implies a more extended accretion disc.
The evolution of $L_{\rm d}$ counterbalances the growth of $M_{\rm d}$ such that the accretion rate on to the black hole remains rather constant.
On the other hand, $L_{\rm d}$ remains initially lower and fluctuates more in the runs with no net rotation.
This behaviour, together with the increase of $M_{\rm d}$, favours larger values of $\dot{M}$.
In run tc3, the accretion disc mass doubles its value within $2 t_{\rm dyn}$, quickly boosting $\dot{M}$ untill it hits the Eddington limit.
Then, fluctuations in both $M_{\rm d}$ and $L_{\rm d}$ modulate the evolution of $f_{\rm Edd}$ that remains close to unity.
Run tc1 and tc1\_LF show how the evolution of the black hole and accretion disc system may respond to different boundary conditions.
During the first $\approx 4 t_{\rm dyn}$, the accretion disc mass grows rather similarly in the two simulations, but slightly faster in tc1\_LF at early times.
However, the angular momentum in tc1\_LF remains initially lower and less fluctuating than in tc1, which makes $\dot{M}$ grow faster and more steadily in tc1\_LF than in tc1.
The dense circumnuclear disc keeps dumping mass on the accretion disc in run tc1\_LF, whereas $L_{\rm d}$ grows more slowly relative to $M_{\rm d}$.
This explains why the accretion rate onto the black hole remains close to the Eddington limit.
Instead, the mass of the accretion disc in run tc1 increases slowly after that the black hole gets ejected from the inner region, while the specific angular momentum of the disc becomes comparable to the tc2 and tc4 cases.
Therefore, the accretion disc readjusts to a more extended and less dense configuration that can only sustain a lower accretion rate, explaining the low values of $f_{\rm Edd}$ for run tc1 after $\approx 5 t_{\rm dyn}$.

The black hole mass and spin parameter evolve directly under the effect of mass accretion.
As expected, the growth of $M_{\bullet}$ simply reflects the capability of the disc to transfer mass onto the black hole.
The black holes in run tc1\_LF and tc3 grow quickly almost constantly at the Eddington rate, while $M_{\bullet}$ grows only by about 5\% in the other runs; run tc1\_LF almost doubles $M_{\bullet}$ in $10 t_{\rm dyn}$, corresponding to $\approx 75$~Myr, but the growth reduces slightly after $\approx 6 t_{\rm dyn}$ because of the simultaneous increase of the radiative efficiency.
The radiative efficiency evolves as a consequence of change in the black hole spin.

Figure~\ref{fig_tc_bulge_summary} shows that $\bmath{J}_{\bullet}$ and $\bmath{J}_{\rm d}$ quickly align to the direction of the total angular momentum in about half $t_{\rm dyn}$, i.e. $\approx 3.5$~Myr.
After that, the Bardeen-Petterson effect maintains the two vectors aligned, i.e. $\bmath{j}_{\bullet} \cdot \bmath{j}_{\rm d} \approx 1$, as already seen in the circumnuclear disc simulations in Section~\ref{subsec_cnd}.
Indeed, $J_{\rm d} / J_{\bullet}$ always remains $\gtrsim 3$ in all ``bulge'' simulations and it grows to about 15 in runs tc1, tc2, and tc4, i.e. only alignment is possible.
As a consequence, the accretion disc remains always co-rotating with the black hole and matter accretion drives the growth of the spin parameter $a_{\bullet}$ similarly to the growth of $M_{\bullet}$.
In run tc1\_LF and tc3, $a_{\bullet}$ reaches $\approx 0.9$ and the limiting value 0.998 from the initial value 0.5, respectively, while in the other simulations the spin parameter only grows to $\approx 0.6$.

\begin{figure*}
\begin{center}
\includegraphics[width=2.1\columnwidth]{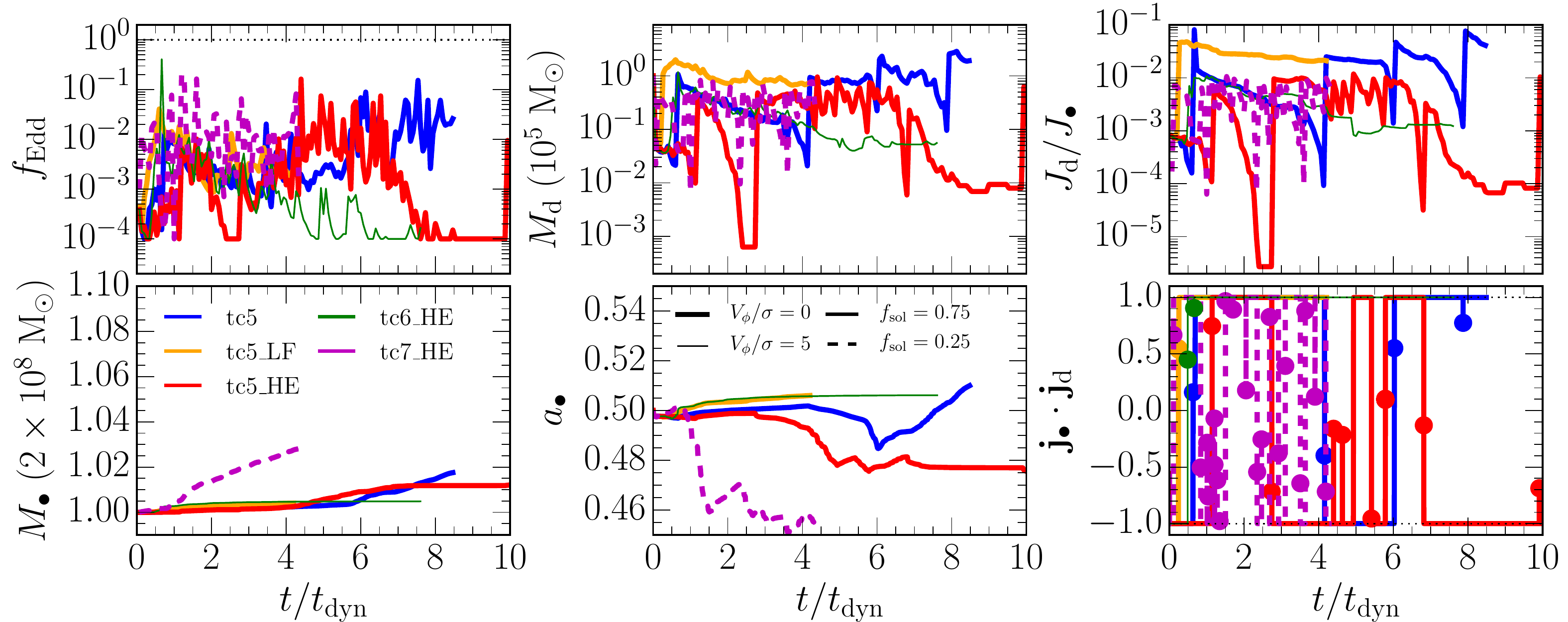}
\caption{From left to right and from top to bottom: time evolution of $f_{\rm Edd}$, $M_{\bullet}$, $J_{\rm d}/J_{\bullet}$, $M_{\rm d}$, $a_{\bullet}$, and $\bmath{j}_{\bullet} \cdot \bmath{j}_{\rm d}$ in the ``elliptical'' runs.
The big circles in the last panel indicate the moments of accretion disc draining and reconstruction.
In each panel, thin (thick) curves correspond to $V_{\phi}/\sigma=5$ ($V_{\phi}/\sigma=0$), while solid (dashed) curves correspond to $f_{\rm sol}=0.75$ ($f_{\rm sol}=0.25$).
Blue, orange, red, green, and magenta corresponds to run tc5, tc5\_LF, tc5\_HE, tc6\_HE, tc7\_HE, respectively.
For massive black holes with $M_{\rm d}/M_{\bullet} \ll 1$ and $J_{\rm d}/J_{\bullet} \ll 1$, the accretion history is more episodic, with frequent events of draining and reconstruction of the accretion disc.
Every newly formed accretion disc may align or counter align with the black hole spin, leading either to spin-up or spin-down depending on the coherency of the external inflow.
}
\label{fig_tc_elliptical_summary}
\end{center}
\end{figure*}

While the Bardeen-Petterson effect maintains an effective coupling between the black hole and the accretion disc angular momenta, the overall evolution of their directions is dictated by the external inflow.
This is shown in Figure~\ref{fig_tc_bulge_amproj}, where we plot the time evolution of $\bmath{j}_{\bullet}$ and $\bmath{j}_{\rm d}$ projected over the full-sky sphere with Hammer projections.
The equatorial plane corresponds with the $x$-$y$ plane, i.e. the disc plane in runs tc2 and tc4.
The initial part of the evolution is very similar across all simulations.
As already indicated by $\bmath{j}_{\bullet} \cdot \bmath{j}_{\rm d}$, the two vectors quickly align in a few Myr from the beginning of the runs.
Then, they both follow the direction of $\bmath{J}_{\rm tot}$ as it changes after torquing from matter inflow.
Once the disc forms from the collapse of the initial cloud, the behaviour of runs tc2 and tc4 is similar to the set of circumnuclear disc simulations.
The coherent adding of angular momentum to $\bmath{J}_{\rm d}$ rather aligned with the disc rotation axis forces $\bmath{J}_{\bullet}$ and $\bmath{J}_{\rm d}$ to migrate together to align with the large scale disc angular momentum.
The alignment is faster than in the circumnuclear disc simulations because the typical $\dot{M}_{\rm inflow}$ is higher by a factor $\approx 5$ (i.e. compare Figure~\ref{fig_mdot_a} and \ref{fig_tc_inflow}).
Indeed, alignment is close to completion in $\approx 80$~Myr of evolution for run tc2, while it slows down after $\approx 7 t_{\rm dyn}$ in tc4, i.e. when the black hole is scattered away from the disc centre and it eventually sinks back slowly.
In both cases, fluctuations of about 10-20\degr account for the thickness and vertical turbulence in the disc.
Furthermore, we observe wide motions in all the runs with no net rotation, as the direction of both $\bmath{J}_{\bullet}$ and $\bmath{J}_{\rm d}$ varies by more than 60\degr over the simulated timescales.
In run tc1, the direction initially changes because of the formation of a small nuclear disc, until the central dense knot of gas is ejected with the black hole bound to it; then, the reorientation of $\bmath{j}_{\rm tot}$ slows down.
Instead, both run tc1\_LF and tc3 describe curly curves in Figure~\ref{fig_tc_bulge_amproj} as a consequence of the evolution of the gas structures in the inner regions, showed by the sequence of images in the same figure for run cnd1\_LF as an example.

\subsubsection{Evolution of heavy supermassive black holes in early-type ellipticals}

The suite of ``elliptical'' simulations explores the evolution of supermassive black hole for which $R_{\rm warp}$ may be larger than $R_{\rm d}$ (see Section~\ref{subsec_angmom}).
In this regime, the evolution may significantly differ from what we have seen so far in the ``bulge`` runs.
The results of our computations are shown in Figure~\ref{fig_tc_elliptical_summary}.
The major difference with respect to the ``bulge'' runs is that the accretion disc contain less mass and angular momentum relative to the black hole than for $M_{\bullet} \sim 10^6$~M$_{\sun}$.
This can be seen by comparing the evolution of $M_{\rm d}$ and $M_{\bullet}$: the ratio $M_{\rm d}/M_{\bullet}$ is always smaller
than $\approx 10^{-3}$ in order to maintain the accretion disc mass below the gravitational instability threshold $M_{\rm sg}$. 
Similarly, the angular momentum content of the system is dominated by the black hole.
Indeed, the ratio $J_{\rm d}/J_{\bullet}$ is always lower than unity, which allows for counter-alignment of $\bmath{J}_{\bullet}$ and $\bmath{J}_{\rm d}$ \citep{king+05}. 
The combined evolution of the disc mass and agular momentum makes the accretion disc able to sustain very fluctuating accretion rates between a significant fraction of the Eddington limit, i.e. $\sim 0.2 \dot{M}_{\rm Edd}$, and our imposed threshold $f_{\rm Edd}^{\rm (min)} = 10^{-4}$.
Nonetheless, the average time behaviour of $f_{\rm Edd}$ approximatively follows that of $\dot{M}_{\rm inflow}$, confirming once more that the external inflow ultimately drives the long-term evolution of the system.

The lower amount of mass and angular momentum in the accretion disc relative to the black hole makes the draining timescale for the accretion disc shorter since it scales with $M_{\rm d}/M_{\bullet}$.
Therefore, the black hole and accretion disc system undergoes several episodes of disc draining and reconstruction.
They can be spotted in the corresponding spikes of $M_{\rm d}$ and $J_{\rm d}/ J_{\bullet}$ which follow longer time periods of reduction in these quantities.
Regardless of the simple assumptions that we made to treat these events, this makes the black hole growth intrinsically more episodic.
After the formation of a new accretion disc, it may align or counter-align, as indicated by the frequents jumps of $\bmath{j}_{\bullet} \cdot \bmath{j}_{\rm d}$ from -1 to +1 in Figure~\ref{fig_tc_elliptical_summary}.
Alignment and counter-alignment become almost equally probable because $J_{\rm d} / J_{\bullet} \ll 1$.
In fact, the accretion disc and the black hole switch between alignment and counter-alignment most of the times when a new accretion disc forms after draining. Since the new orientation follows the angular momentum of the inflowing gas at that time, the evolution of the alignment is ultimately set by the details and the episodic character of the inflow from larger scales.
This is quite different from what we have discussed regarding the ``bulge'' simulations, where $J_{\rm d} / J_{\bullet} \gg 1$ forces alignment through the Bardeen-Petterson effect in all cases and promotes spin-up.
Instead, the different behaviour leaves an imprint in the evolution of the spin parameter.
Indeed, $a_{\bullet}$ alternates more or less frequent phases of spin-up and spin-down according to the properties of the simulated system.
Time intervals of spin-down are more recurrent in runs tc5, tc5\_HE, and tc7\_HE where the angular momentum of the gas is rather isotropically distributed.
Instead, run tc5\_LF and tc6\_HE show a more coherent spin-up.
This is because the rotation axes of the disc and the circumnuclear disc (in run tc6\_HE and tc5\_LF, respectively) end up being roughly aligned within 90\degr with $\bmath{j}_{\rm tot}$.
Therefore, most of the gas transported from the large scale disc or the circumnuclear disc eventually adds up coherently with the accretion disc gas and increases $J_{\rm d}$.
However, if the opposite had happened (i.e. misalignment larger than 90\degr), we would have observed a more coherent spin-down until the torque caused by inflow would have reduced the misalignment between $\bmath{J}_{\bullet}$ and the large scale structure to less than 90\degr; then, a phase of spin up would have followed.
Moreover, we also note that the conservation of angular momentum requires that the accretion disc angular momentum mainly changes its direction, while $\bmath{J}_{\bullet}$ only wobbles around its original direction within $\approx 10\degr$ over the simulated time in all runs.

After alignment or counter-alignment, the accretion disc dumps mass and angular momentum to the black hole.
The contribution of each of these events is modest to the mass growth of the black hole.
Indeed, $M_{\bullet}$ grows by about 1-3\% in all runs, but faster in run tc7\_HE that sustains on average a larger accretion rate over the simulated time.

Finally, we caution that the black hole evolution might be affected by the frequent disc draining and reconstruction because this requires the choice of
a few free parameters, namely $M_{\rm refill}$, $f_{\rm refill}$, and $f_{\rm Edd}^{\rm (refill)}$.
We expect that the impact of both $M_{\rm refill}$ and $f_{\rm refill}$ is similar to that of $f_{\rm Edd}^{\rm (refill)}$, because they all change the draining timescale of the accretion disc and therefore the frequency of the draining and reconstruction episodes given the same properties of the external inflow.
Therefore, we assume the same $M_{\rm refill}$ and $f_{\rm refill}$ in all runs, but we probe the impact of $f_{\rm Edd}^{\rm (refill)}$ in runs tc5 and tc5\_HE.
Despite the three orders of magnitude difference in $f_{\rm Edd}^{\rm (refill)}$, the evolution of all the black hole and accretion disc quantities is qualitatively similar among the two simulations, with some quantitative differences.
As expected, run tc5\_HE undergoes more draining and reconstruction episodes than run tc5 over the simulated timescale.
However, we note that this is more sensitive to the inflow properties than to the parameter choice.
Indeed, while the chosen parameters may imply a longer or shorter draining timescale, the properties of the accretion disc may change quickly enough to lose memory of the initialisation after a draining event.
Therefore, we conclude that the qualitatively similar behaviour of runs tc5 and tc5\_HE suggests that the impact of the phenomenological parameters used in the simplified disc draining and reconstruction approach do not have a major impact in the black hole evolution.
In all cases, the latter is ultimately dictated by the properties and kinematics of the large scale gas reservoir.


\section{Discussion} \label{sec_discussion}

In this study we present a new black hole accretion model implemented in the moving-mesh code {\sc arepo} that (i) takes into account mass accretion from a sub-grid thin $\alpha$-disc, and (ii) self-consistently accounts for the evolution of the black hole spin.
While this certainly represents a step ahead in trying to merge the knowledge from small-scale theoretical investigations into more physically-motivated sub-grid recipes for galaxy formation simulations, our modelling necessarily relies on some assumptions.

The most crucial assumption we made is that the sub-grid accretion disc of mass $M_{\rm d}$ and angular momentum $\bmath{J}_{\rm d}$ is instantaneously in steady state and it follows the solution by \citet{shakura+73}.
This is a working assumption and it is mainly motivated by the extensive use of this solution in the literature owing to its simplicity and success. 
Indeed, the thin disc model successfully describes the broad-band features -- such as the ``big blue bump'' -- in the optical/UV spectra of luminous broad-line AGN or flat-spectrum radio quasars with accretion rates $\gtrsim 10^{-2} \dot{M}_{\rm Edd}$ (e.g. \citealt{ghisellini+10,davis+11,capellupo+15,sbarrato+16}).
The thin disc model can be used to obtain an estimate of the central black hole mass by fitting the big blue bump and this method returns masses in reasonable agreement with single-epoch virial estimates \citep{zheng+95, calderone+13, castignani+13}.
Discrepancies however appear when the same method is applied to narrower line AGN such as radio-loud narrow-line Seyfert 1 galaxies.
Indeed, the estimated masses are larger than from single-epoch virial methods, but they are apparently in better agreement with broader line AGN of the same kind \citep{calderone+13}.

Detailed observations however suggest that the thin disc model may be oversimplified.
Indeed, careful comparisons between thin disc model spectra (including relativistic corrections and departures from local thermodynamic equilibrium) and quasar spectra show that the theoretical optical part is often bluer than observed, while the UV luminosity is sometimes under predicted (e.g. \citealt{blaes+01, davis+07}).
Moreover, microlensing observations have constrained the accretion disc sizes, indicating that they are $\sim 4$ times larger than expected from the thin disc model at optical/UV wavelengths \citep{pooley+07, dai+10, morgan+10}.
Finally, quasar spectra are aperiodically variable by $\approx 10$-20\% in the optical/UV over a wide range of timescales, from days to years, with short time lags ($\approx 1$-2 days) between different wavelengths and the tendency to be bluer when brighter (e.g. \citealt{vandenberk+04, sesar+07, meusinger+11, ruan+14}).
All this peculiarities are difficult to reconcile with the idea of a smooth and steady-state thin accretion disc, whereas they can be better accounted for by a non-steady, inhomogeneous disc model with localised temperature fluctuations \citep{kelly+09, dexter+11, ruan+14, cai+16}.

Despite the limitations of the thin disc model to describe the vast phenomenology observed in real AGN, we argue that it can still be regarded as an effective model to account for the fundamental role of gas angular momentum in accretion and in the evolution of black hole spin, usually neglected in galaxy formation models.
We note that our computations effectively extend the usage of the thin disc model down to very low accretion rates, as we impose a numerical lower limit $f_{\rm Edd}^{\rm (min)} = 10^{-4}$.
At low accretion rates ($f_{\rm Edd} \lesssim 10^{-3}$), the accretion disc density decreases and radiative cooling becomes inefficient.
Then, the disc puffs up and a significant fraction of the viscous heating is advected inward by the gas flow \citep{narayn+94, narayan+95, blandford+99}.
These hot, optically-thin, quasi-spherical, advection-dominated accretion disc are radiatively inefficient and have been often advocated in the literature to interpret low luminosity sources (e.g. \citealt{lasota+96}).
If this is the common nature of low $f_{\rm Edd}$ accretion disc (we note though that the thin disc solution holds at low $f_{\rm Edd}$ as well; \citealt{chen+95}), the alignment and evolution of the black hole spin may actually change. 
When the disc becomes geometrically thick and the aspect ratio $H/R \gtrsim \alpha$, the perturbation induced by the Lense-Thirring precession does not propagate diffusively.
Instead, it propagates as bending waves \citep{papaloizou+83, papaloizou+95}.
In this regime, the inner part of the disc does not completely align with the black hole spin and the tilt oscillates and precesses around the angular momentum of the hole \citep{lubow+02, fragile+07, nealon+15}.
This different behaviour possibly makes our treatment invalid at small $f_{\rm Edd}$.
However, a qualitative but general result of our simulations is that the evolution of a black hole and accretion disc system is ultimately dictated by the boundary conditions provided by inflows from large scales.
Unless the host galaxy is extremely gas poor and the central black hole remains quiescent for long time, even a brief inflow of gas may bring enough mass to force higher accretion rates in the disc thus possibly ``restoring'' efficient radiative cooling.
Therefore, we conclude that accounting for this different behaviour may lead to some corrections on the black hole angular momentum evolution, but these corrections likely have a minor impact over the long term evolution of a rather luminous quasar.
However, we will attempt to model more accurately the behaviour at low $f_{\rm Edd}$ in a future work.

Our model takes into account the fundamental role of angular momentum in feeding supermassive black holes.
Despite the limitations of the precise assumptions, we have shown that our model captures the basic expected behaviour of an accretion disc.
When matter falls in, gas can join the accretion disc when its specific angular momentum is lower than that of the disc.
This condition becomes more restrictive for compact accretion discs and it can affect the evolution of the accretion disc itself since more radial inflow is required to actually join the accretion disc, whose mass is otherwise consumed over time.
This approach is rather conservative, but it can be relaxed in future work (e.g. to account unresolved physical processes) by allowing the infalling gas to join the accretion disc even with $L_{\rm inflow} \gtrsim L_{\rm d}$.

Mass accretion through the accretion disc is modulated by the disc mass and angular momentum evolution. 
When $M_{\rm d}$ increases or $J_{\rm d}$ decreases owing to external inflow leading to angular momentum cancellation, the accretion rates grows.
Instead, $\dot{M}$ consistently decreases when the disc mass decreases or the angular momentum grows.
Recall that at the same time, we couple the spin evolution to the disc evolution through the Bardeen-Petterson effect.
We have based our model on the well established body of analytical work that has explored the consequences of the Bardeen-Petterson effect in the limit of the thin disc model (e.g. \citealt{king+05, lodato+06, martin+07, perego+09}).
However, this approach does not allow us to capture entirely the phenomenology that numerical simulations have recently unveiled. \citet{nixon+12a} used 1D, time-dependent numerical models of warped accretion disc and found that the disc can break into discrete rings that independently follow Lense-Thirring precession.
The break keeps propagating and it enhances mass transfer locally in the disc, building up spikes in the surface density distribution.
This feature lasts until the disc is fully aligned or counter-aligned with the black hole spin.
This has been confirmed with hydrodynamical simulations by \citet{nixon+12b}, showing that disc tearing might be common and leads to bursty enhancements of the accretion rate.

Our results are based on idealised numerical experiments meant (i) to test the capabilities of the model, and (ii) to explore simplified evolutionary scenarios for the spin of supermassive black holes.
While we plan to apply our model to more realistic systems in a future work, some interesting conclusions can be already derived. 
The Bardeen-Petterson effect represents a mechanism to link the evolution of the black hole spin to the accretion disc. 
The outcome of this coupling is mainly set by the ratio $J_{\rm d} / J_{\bullet}$.
Indeed, this ratio is key to decide whether $\bmath{j}_{\bullet}$ and $\bmath{j}_{\rm disc}$ will end up being aligned or counter-aligned, as first shown by \citet{king+05}.
In agreement with previous semi-analytic calculations by \citet{dotti+13}, we find that supermassive black holes with $10^{6} \lesssim M_{\bullet}/{\rm M}_{\sun} \lesssim {\rm a~few} \times 10^{7}$, preferentially have $J_{\rm d} \gg J_{\bullet}$.
In such circumstances, it has been shown that the Bardeen-Petterson effect quickly leads to alignment of the black hole angular momentum to the disc angular momentum \citep{lodato+06, martin+07, perego+09}.
The short alignment timescale effectively transforms any accretion disc that originally formed with some degree of misalignment into a co-rotating configuration, as shown, for example, by Figure~\ref{fig_tc_bulge_summary}.
On the other hand, heavier supermassive black hole, i.e. $M_{\bullet} \gtrsim 10^{8}$~M$_{\sun}$, preferentially have $J_{\rm d} \ll J_{\bullet}$, which makes counter-alignment as likely as alignment, and the final configuration is sensitive to the initial direction of the accretion disc angular momentum as set in each accretion event (see e.g. Figure~\ref{fig_tc_elliptical_summary}).

Our results are qualitatively in agreement with previous work that has also pointed out this dichotomy in spin evolution between low and high mass systems by means of different techniques, ranging from semi-analytic models \citep{sesana+14} to hydrodynamical simulations \citep{maio+13,dubois+14a, dubois+14b}.
For what concerns numerical simulation, \citet{maio+13} studied the spin evolution of black holes with masses $\sim 10^{6}$~M$_{\sun}$ in detailed idealised simulations of circumnuclear discs, finding the preference for the black hole to spin up and to eventually align with the circumnuclear disc.
The spin evolution was calculated in post-processing through non-steady, 1D warped accretion disc models whose boundary conditions were set from the simulation outputs.
On the other hand, \citet{dubois+14a, dubois+14b} developed a simpler spin evolution model that has been applied on the fly as well as in post-processing to both isolated and cosmological simulations.
They also highlighted (i) the tendency for heavy supermassive black holes to have lower spins (although the contribution of black hole mergers becomes relevant at high masses), and (ii) the tendency for lower mass black holes to align with the angular momentum of gas-rich hosts.
However, the latter model relies on the \citet{bondi+52} inflow solution, which does not account for angular momentum to estimate mass accretion.
The same accretion rate is then assigned to a sub-grid thin disc whose angular momentum orientation is extrapolated from that of the gas at a few parsec separation.
Comparing with the previous work, our new model represents an attempt to make a step ahead with respect to these different approaches by incorporating accretion disc physics in a live model for black hole spin evolution.

This dual behaviour of the Bardeen-Petterson effect has important consequences on the way spin evolution is connected to the gas reservoir provided by the host galaxy.
Two bracketing cases have been envisaged to describe how the host galaxy may fuel the central supermassive black hole \citep{king+06,king_pringle+07}.
The first is the ``chaotic accretion'' scenario: gas parcels fall onto the black hole with an isotropic distribution of angular momenta such that $\bmath{l}_{\rm inflow}$ changes on a timescales comparable to the disc draining time, $t_{\rm drain} \sim M_{\rm d}/\dot{M}\sim 4.5 \times 10^{7}~f_{\rm Edd}^{-1} (M_{\rm d}/M_{\bullet})~{\rm yr}$.
In this case, the accretion process is more episodic and each accretion episode is represented by an accretion disc whose angular momentum may point to a direction completely unrelated to the previous one \citep{king+06,king_pringle+07}.
The second possibility is the ``coherent accretion'' scenario: the black hole accretes gas with angular momentum along a well-defined, almost constant direction.
The ``chaotic'' case potentially leads to low values for $a_{\bullet}$ because a counter-aligned configuration corresponds to larger values of $|L_{\rm isco}|$ (see Figure~\ref{fig_spin_factors}).
Instead, ``coherent accretion'' keeps adding angular momentum to the black hole along the same direction and should result in high spins.
Both scenarios are qualitatively represented in both the ``bulge'' and ``elliptical'' simulation suites by the $V_{\phi} / \sigma = 0$ and $V_{\phi} / \sigma = 5$ cases, respectively.
The two expected behaviours outlined above are indeed recovered in the ``elliptical'' runs.
Simulations with no net rotation alternates intermittent phases of spin up and spin down, with the tendency for the spin parameter to decrease over time, while simulations with a preferential direction for $\bmath{L}_{\rm inflow}$, despite not perfectly coherent, tend to favour spin up.
Preferential amounts of ``coherent'' or ``chaotic'' accretion may be possibly associated with the morphology of the host galaxy, as suggested by \citet{sesana+14}, or with the mass scale and environment of the host, such as, for example, in the case of chaotic cold accretion enhanced by thermal instability in the centre of galaxy clusters \citep{pizzolato+10,mccourt+12,sharma+12,gaspari+13}.

Our results suggest that the most massive black holes probably carry the clearer imprint of their accretion history in their spin distribution.
This could be tested in principle with the available ($\approx 20$) observational estimates of $a_{\bullet}$ from modelling of the broad K$\alpha$ iron line at 6.4 keV \citep{brenneman+13, reynolds+14}.
Although the uncertainties in the observational measurements are still large and prevent such a test from being conclusive, almost all estimates point to $a_{\bullet} \gtrsim 0.5$, with larger spread in the value of the spin parameter for $M_{\bullet} \gtrsim 10^7$~M$_{\sun}$ and values closer to 1 for lighter black holes.
This is at least qualitatively consistent with our results (see also \citealt{sesana+14}).

Orthogonal constraints on these findings may come from the fact that larger spins require higher radiative efficiency, although significant deviations from the canonically assumed $\eta \approx 0.1$ start only beyond $a_{\bullet} \approx 0.9$.
\citet{shankar+13} used semi-empirical model to constrain the time-dependent distribution of Eddington ratios to recover several observational constrains on the fraction of active galaxies and the AGN luminosity function.
They found that a radiative efficiency that is increasing with $M_{\bullet}$ is preferable to describe the data, possibly suggesting higher spins at higher black hole masses.
However, the typical $a_{\bullet}$ implied for $\sim 10^{9}$-$10^{10}$~M$_{\sun}$ black holes would be $\sim 0.7$-$0.9$, therefore compatible with the broad spin distribution potentially expected at high masses as well as with the current observations. 
Interestingly, \citet{trakhtenbrot+14} inferred high radiative  efficiencies for a sample of luminous AGN at $1.5 \lesssim z \lesssim 3.5$ with typical masses $> 10^{9}$~M$_{\sun}$, suggesting that those objects harbour highly spinning black holes and therefore disfavouring a purely chaotic accretion scenario for the early assembly of massive quasars.

The different coupling provided by the Bardeen-Petterson effect between the black hole spin and the accretion disc angular momentum for light and heavy supermassive black holes might also have implications for the degree of misalignment with the host galaxy.
The accretion disc surrounding $\sim 10^{6-7}$~M$_{\bullet}$ black holes dominates the angular momentum budget in a relative sense. 
However, $J_{\rm tot}$ is low enough to allow accretion to significantly torque the total angular momentum through the material dumped in the disc.
If the large scale gas flow is mostly rotationally supported, such as in our circumnuclear disc simulations and in the $V_{\phi}/\sigma = 5$ ``bulge'' runs, $\bmath{j}_{\rm tot}$ tends to align with the large scale angular momentum on timescale that can range between $\sim 100$~Myr to a few Gyr, depending on the mass inflow rate and the black hole mass.
Otherwise, if the gas kinematics is less ordered, like in the $V_{\phi}/\sigma = 0$ ``bulge'' simulations, we find that the direction of the total angular momentum (and effectively both the black hole and accretion disc angular momentum) can change erratically by up to $\sim 50$-$70$\degr on rather short timescales.
Note however that $\bmath{j}_{\rm tot}$ (and therefore $\bmath{j}_{\bullet}$) may change much more slowly for heavy supermassive black holes $\gtrsim 10^8$~M$_{\sun}$.
In fact, while the disc may keep flipping direction compared to the black hole spin, the latter only wobbles by several degrees over the simulated timescales.
This is because $M_{\rm d} \ll M_{\bullet}$ and $J_{\rm d} \ll J_{\bullet}$, and therefore a larger amount of mass and angular momentum has to be ultimately supplied from larger scales before $\bmath{J}_{\rm tot}$ can significantly change.

These findings could possibly explain the direction of relativistic jets in radio-loud AGN.
Radio jets are likely launched by extracting the black hole rotational energy stored in the spin \citep{blandford+77,tchekhovskoy+11}.
The statistical analysis of an observed sample of Seyfert 1 and 2 galaxies (i.e. spiral galaxies typically with $M_{\bullet} \lesssim 10^{8}$~M$_{\sun}$; \citealt{onken+03, komossa+07}) by \citet{kinney+00} suggests that radio jets are consistent with being randomly oriented with respect to the plane of the large-scale galactic disc.
\citet{schmitt+03} corroborated these findings by showing that the [\ion{O}{iii}] emission in the nuclear regions of 60 Seyfert galaxies (both type 1 and 2) is well aligned with the radio emission and therefore misaligned with respect to the galactic disc.
They interpret this by hypothesising that the orientation of the gas in the torus determines the orientation of the accretion disc, and ultimately the jet direction.
This is qualitatively in agreement with our findings from the circumnuclear disc and ``bulge'' runs: the Bardeen-Petterson effect represents only a tight connection between the black hole spin and the accretion disc angular momentum, but the gas inflow on a few pc scale (i.e. the gas that feeds the torus) ultimately imposes the orientation of both $J_{\bullet}$ and $J_{\rm d}$ according to the gas kinematics.
However, our results suggest that $\bmath{j}_{\bullet}$ and $\bmath{j}_{\rm d}$ would eventually align with the large-scale galactic disc over a timescale between $\sim 100$~Myr and a few Gyr if the gas is rotationally supported down to $\lesssim 100$~pc.
This may appear in contrast with the observational evidences, unless the longer alignment timescales are preferred or the gas kinematic in the inner part of the bulge is sufficiently complicated and turbulent to behave qualitatively similarly to the $V_{\phi}/\sigma = 0$ ``bulge'' simulations, e.g. through non-circular motion triggered by local stellar and AGN feedback episodes \citep{riffel+08,lena+15}. 

For larger supermassive black hole masses, jets seem to be slightly more aligned with the galaxy semi-minor axis \citep{schmitt+02}.
\citet{verdoeskleijn+05} found that dust ellipses in jetted radio galaxies are typically aligned with the major axis of the galaxy and randomly oriented with respect to the jet, while dust lane closer to the black hole are coaxial with the jet within $\sim 20$-30\degr.
Unfortunately, we could not evolve our ``elliptical'' runs for long enough to estimate any typical timescale for alignment and settling of the black hole and accretion disc angular momenta.
We only observe wobbling of the black hole spin up to $\sim 10$\degr, posing an upper limit to the timescale for any significant black hole spin reorientation of $\sim 100$~Myr.
However, we recall that we assumed a geometrically thin disc; if the disc were thick, the jet direction would likely be modulated by the disc funnel in case of incomplete alignment, possibly causing precession on shorter timescales \citep{liska+17}.
Moreover, note that our model has a maximum cap on the accretion disc mass, i.e. $M_{\rm sg}$, and this may affect the timescale of black hole fuelling.
For example, in the case of rapid growth of high-$z$ quasars, accretion may proceed on very short timescales owing to substantial gas inflows which would lead to different evolution for both the mass and spin with respect to the predictions of our model.

The simulations presented in this work do not include any feedback from supermassive black holes because we intentionally focussed on probing just the coupling between the black hole mass and spin with the surrounding accretion disc.
Nonetheless, black hole feedback certainly represents an important ingredient to model black hole evolution in realistic simulations.
Indeed, it can impact both the thermal properties and the bulk motion of the gas in the proximity of the black hole, possibly modulating the availability of fresh material for accretion and affecting gas orbits.
Therefore, we plan to explore the role of feedback by self-consistently coupling our new accretion model with accretion disc winds in the form of bipolar outflows and with the production of jets in a forthcoming work \citep{curtis+16a,bourne+17}.

Finally, while our work focussed on the role of gas accretion in the supermassive black hole spin evolution, we recall that also black hole mergers are expected to contribute.
For example, \citet{berti+08} used semi-analytic methods to discuss the interplay of different modes of accretion and merger configurations in dictating spin evolution, finding that many mergers can lead to large spin parameters $a_{\bullet} \gtrsim 0.9$ only if the progenitors already had high spin and they mostly underwent minor mergers.
\citet{barausse+12} also used a semi-analytic model that includes prescriptions for galaxy evolution, finding indeed that the dichotomy already discussed for accretion is fostered by black hole mergers happening preferentially in gas-poor environments for high mass black holes at low redshift, therefore contributing to the spin-down of massive black holes.
Similar findings have been confirmed by the post-processing analysis of a large cosmological volume performed by \citet{dubois+14a}.


\section{Summary and conclusions} \label{sec_conclusions}

We have implemented a new black hole accretion model in the moving-mesh code {\sc arepo} that self-consistently couples mass and angular momentum flow from large scales as dictated by the hydro solver to an analytic thin $\alpha$-disc which ultimately delivers mass to the central supermassive black hole.
In addition to the black hole mass the model also evolves black hole spin and accounts for the coupling between the black hole spin and the accretion disc angular momentum provided by the Bardeen-Petterson effect.
We have tested our model in a series of idealised yet physically-motivated simulations that bracket several possible conditions of fueling the nuclear regions of a galaxy and ultimately the central supermassive black hole.
We summarise our findings as follows. 
\begin{enumerate}
\item The Bardeen-Petterson effect leads to black hole spin and accretion disc alignment when the black hole mass is not too large ($M_{\bullet} \lesssim 10^7$~M$_{\sun}$) on timescales of order of few Myr because the accretion disc typically dominates the total angular momentum budget (i.e. $J_{\rm d} \gg J_{\bullet}$).
Therefore, longer periods of spin up are likely expected for supermassive black holes in this mass range. 
\item At larger masses ($M_{\bullet} \gtrsim 10^8$~M$_{\sun}$), the angular momentum of the accretion disc is typically smaller than that of the black hole (i.e. $J_{\rm d} \ll J_{\bullet}$) and the accretion disc is proportionally lighter than for lower mass black holes.
Consequently, the accretion history is intrinsically more episodic and the accretion disc may either align or counter-align with respect to the black hole spin, potentially leading to either spin up or spin down.
Therefore, we expect a wider distribution of spin parameters at high masses. 
\item As a consequence heavier black holes are likely more sensitive to the properties of the gas kinematic on large scales.
A systematic campaign to measure $a_{\bullet}$ in the most massive black holes may shed light on the average accretion history of the popolation and enable us to distinguish whether supermassive black holes are preferentially fed through a series of isotropic accretion events leading to spin down, or through sustained accretion along the same direction that may eventually favour spin up. 
\item While the Bardeen-Petterson effect represents the connection between $\bmath{J}_{\bullet}$ and $\bmath{J}_{\rm d}$, the direction of $\bmath{J}_{\rm tot}$ evolves according to the inflow coming from larger scales.
For $M_{\bullet} \lesssim 10^7$~M$_{\sun}$ black holes (i.e. more typical of Seyfert galaxies), this may change erratically by large angles ($\sim 60$-70\degr) over short timescales ($\lesssim 100$~Myr) if the gas kinematic is dominated by turbulence that enhances bursty inflows towards the galactic nucleus.
If however the gas kinematics is rotation-dominated, the black hole spin and the accretion disc rotation axis may align with the angular momentum at kpc scales on a timescale of $\sim 100$~Myr to a few Gyr.
Heavier supermassive black holes require much more mass to feed the accretion disc before the direction of $\bmath{J}_{\bullet}$ changes significantly.
Thus, on short timescales ($\lesssim 100$~Myr), the direction of the spin only wobbles by $\sim 10\degr$, but it may change more substantially over longer timescales, depending on the coherence of the infalling gas. 
\end{enumerate}

Finally, note that we have presented a series of idealised, small-scale, high-resolution simulations to be able to study in detail the spin evolution in the presence of accretion only. Our present setup however neglects more complex gas thermodynamics, star formation and associated feedback as well as AGN feedback in a realistic cosmological environment.
The methodology that we have developed can be directly applied to such more comprehensive simulations where it will be possible to study, for example, the interplay between gas consumption by star formation, gas accretion onto the black hole and AGN-driven wind launched along the direction of the black hole spin.
Study of these effects in cosmological simulations presents largely uncharted territory and will allow us to gain much deeper insight into how galaxies and black holes evolve in tandem.


\section*{Acknowledgements}

We thank the anonymous referee for useful comments that helped us improve the quality of the paper.
We thank Massimo Dotti and Giuseppe Lodato for useful and inspiring discussions.
D.F. and D.S. acknowledge support by European Research Council Starting Grant 638707 ``Black holes and their host galaxies: coevolution across cosmic time''.
This  work  was  performed  on  the  following:  DiRAC  Darwin  Supercomputer  hosted  by  the  University  of  Cambridge  High  Performance  Computing  Service  (http://www.hpc.cam.ac.uk/),  provided  by  Dell  Inc. using  Strategic  Research  Infrastructure  Funding  from  the Higher Education Funding Council for England and funding from the Science and Technology Facilities Council; DiRAC Complexity  system,  operated  by  the  University  of  Leicester IT Services.
This equipment is funded by BIS National E-Infrastructure  capital  grant  ST/K000373/1  and  STFC DiRAC  Operations  grant  ST/K0003259/1;  COSMA  Data Centric system at Durham University, operated by the Institute for Computational Cosmology on behalf of the STFC DiRAC  HPC  Facility.
This  equipment  was  funded  by  a BIS National E-infrastructure capital grant ST/K00042X/1, STFC  capital  grant  ST/K00087X/1,  DiRAC  Operations grant  ST/K003267/1  and  Durham  University.
DiRAC  is part of the National E-Infrastructure.


\bibliographystyle{mnras}
\bibliography{bh_spin}


\appendix

\section{Derivation of the expressions based on the accretion disc structure} \label{appendix_formulae}

We assume that the accretion disc is thin, in steady state, and described by the $\alpha$-disc solution where the gas pressure dominates over the radiation pressure and the main opacity source is free-free absorption \citep{shakura+73}.
This is not generally true across the entire disc, since the latter conditions may break down close to the centre.
However, the quantities derived below depend on the assumptions above through the radial viscosity $\nu_{1}$, whose scalings do not change much across different parts of the solution.
Therefore, for the sake of simplicity, we consider $\nu_{1} = C R^{3/4}$ everywhere, with \citep{frank+02,perego+09}
\begin{equation}\label{eq_app_nu}
C = 9 \times 10^{6} \left( \frac{\alpha}{0.1} \right)^{4/5} \left( \frac{M_{\bullet}}{10^6~{\rm M}_{\sun}} \right)^{1/20} \left( \frac{f_{\rm Edd}}{\eta_{0.1}} \right)^{3/10}~{\rm cm^{5/4}~s^{-1}},
\end{equation}
where the radiative efficiency $\eta = \eta_{0.1} 0.1$.
The relation $\dot{M} = 3 \pi \nu_{1} \Sigma / [1 - (R/R_{\rm isco})^{-1/2}] \approx 3 \pi \nu_{1} \Sigma$ between the mass accretion rate $\dot{M}$, the radial viscosity $\nu_{1}$, and the gas surface density $\Sigma$, implies that $\Sigma \propto R^{-3/4}$.
By means of equation (\ref{eq_app_nu}), we can calculate the disc mass enclosed in a cylindrical radius $R$ as
\begin{equation} \label{eq_app_md}
\begin{aligned}
M_{\rm d}(R) & = 2 \pi \int_{R_{\rm isco}}^{R} \Sigma(R) R {\rm d}R = \frac{8 \dot{M}}{15 C} \left( R^{5/4} - R_{\rm isco}^{5/4} \right) \\
& \approx 10^{-2} \left( \frac{\alpha}{0.1} \right)^{-4/5} \left( \frac{M_{\bullet}}{10^6~{\rm M}_{\sun}} \right)^{11/5} \left( \frac{f_{\rm Edd}}{\eta_{0.1}} \right)^{7/10} \left( \frac{R}{R_{\rm S}} \right)^{5/4} {\rm M_{\sun}},
\end{aligned}
\end{equation}
where in the last passage we assumed $R \gg R_{\rm isco}$ and we normalised $R$ with the Schwarzschild radius $R_{\rm S} = 2 G M_{\bullet} / c^2$.

Similarly, by recalling that the specific angular momentum of the thin $\alpha$-disc solution is $L(R) = \sqrt{G M_{\bullet} R}$, we can calculate the total angular momentum of the disc within a cylindrical radius $R$ as
\begin{equation}
J_{\rm d}(R) = 2 \pi \int_{R_{\rm isco}}^{R} \Sigma(R) L(R) R {\rm d}R = \frac{8 \dot{M} \sqrt{G M_{\bullet}}}{21 C} \left( R^{7/4} - R_{\rm isco}^{7/4} \right).
\end{equation}
If we plug in equation (\ref{eq_app_nu}) and we normalise by $J_{\bullet} = G M_{\bullet}^2 a_\bullet / c$, we obtain ($R \gg R_{\rm isco}$)
\begin{equation} \label{eq_app_jratio}
\begin{aligned}
\frac{J_{\rm d}(R)}{J_{\bullet}} & = \frac{8 \dot{M} c}{21 C (G M_{\bullet}^3)^{1/2} a_{\bullet}} \left( R^{7/4} - R_{\rm isco}^{7/4} \right) \\
& \approx 10^{-8} \left( \frac{\alpha}{0.1} \right)^{-4/5} \left( \frac{M_{\bullet}}{10^6~{\rm M}_{\sun}} \right)^{6/5} \left( \frac{f_{\rm Edd}}{\eta_{0.1}} \right)^{7/10} \left( \frac{R}{R_{\rm S}} \right)^{7/4} a_{\bullet}^{-1}.
\end{aligned}
\end{equation}
Finally, we can combine equation (\ref{eq_app_md}) and (\ref{eq_app_jratio}) first by isolating $R/R_{\rm S}$ as function of $M_{\rm d}$ from equation (\ref{eq_app_md}), and then by substituting the result in equation (\ref{eq_app_jratio}).
We thus obtain an expression of the ratio $J_{\rm d}/J_{\bullet}$ as function of $\alpha$, $M_{\bullet}$, $M_{\rm d}$, $a_{\bullet}$, and $f_{\rm Edd}/\eta_{0.1}$, namely
\begin{equation}
\frac{J_{\rm d}}{J_{\bullet}} \approx 2.8 \left( \frac{\alpha}{0.1} \right)^{8/25} \left( \frac{M_{\rm d}}{10^4~{\rm M}_{\sun}} \right)^{7/5} \left( \frac{M_{\bullet}}{10^6~{\rm M}_{\sun}} \right)^{-47/25} \left( \frac{f_{\rm Edd}}{\eta_{0.1}} \right)^{-7/25} a_{\bullet}^{-1},
\end{equation}
and the latter equation can be inverted to obtain equation (\ref{eq_f_edd}).

The expression for the self-gravitating mass $M_{\rm sg}$, equation (\ref{eq_M_sg}), can also be derived starting from equation (\ref{eq_app_md}).
If we imagine an arbitrarily extended accretion disc with $\Sigma \propto R^{-3/4}$, its mass could grow enough, according to equation (\ref{eq_app_md}), to violate the assumption that the central black hole dominates the local gravitational potential and then the disc self-gravity should be taken into account.
Self-gravity induces local instabilities where the Toomre parameter $Q(R) = \Omega c_{\rm s} / (\pi G \Sigma)$ goes below the critical value $\approx 1$.
Here $c_{\rm s}$ is the local sound speed and $\Omega = \sqrt{G M_{\bullet} / R^3}$ is the Keplerian angular velocity under the assumption that the accretion disc self-gravity is negligible.
According to the $\alpha$-disc solution, i.e. $\Sigma \propto R^{-3/4}$ and $c_{\rm s} \propto R^{-3/8}$ \citep{shakura+73}, the Toomre parameter $Q \propto R^{-9/8}$ is a monotonically decreasing function of $R$.
This implies that there is a unique radius $R_{\rm sg}$ where $Q(R_{\rm sg}) = 1$ and therefore a threshold mass beyond which gravitational instabilities can arise.
\citet{perego+09} have calculated $R_{\rm sg}$ under the same assumptions,
\begin{equation}
\frac{R_{\rm sg}}{R_{\rm S}} \approx 1.2 \times 10^5 \left( \frac{\alpha}{0.1} \right)^{28/45} \left( \frac{M_{\bullet}}{10^6~{\rm M}_{\sun}} \right)^{-52/45} \left( \frac{f_{\rm Edd}}{\eta_{0.1}} \right)^{-22/45}.
\end{equation}
Therefore, equation (\ref{eq_M_sg}) can be obtained by plugging $R_{\rm sg}/R_{\rm S}$ into equation (\ref{eq_app_md}).

The alignment timescale $\tau_{\rm align}$ has been derived by \citet{martin+07} for arbitrary viscosity laws $\nu_{1} \propto R^{\beta}$, i.e. their Eq. (51) or equally their Eq. (54) multiplied by $\cos\{\pi/[4(1+\beta)]\}$, under the assumption that the vertical viscosity $\nu_{2}$ is proportional to $\nu_{1}$.
Their notation corresponds to ours with $\beta=3/4$ and $\Sigma_0 = \dot{M}/(3 \pi C) R_{\rm warp}^{-3/4}$, therefore we obtain
\begin{equation} \label{eq_app_talign}
\tau_{\rm align} = \frac{3}{4} \left( \frac{4}{7} \right)^{3/7} \frac{\Gamma(2/7)}{\Gamma(5/7)} \frac{C c^2}{\dot{M} (G^3 M_{\bullet})^{1/2}}~R_{\rm warp}^{5/4},
\end{equation}
where we use the $\Gamma$ function.
$R_{\rm warp}$ is the radius where the Lense-Thirring precession period equals the vertical warp diffusion timescale, i.e. $\omega_{\rm LT}^{-1}(R_{\rm warp}) = R_{\rm warp}^2/v_{2}(R_{\rm warp})$,
\begin{equation} \label{eq_app_rwarp}
\begin{aligned}
R_{\rm warp} & = \left( \frac{4 G^2 M_{\bullet}^2 a_{\bullet} \alpha^{2}}{\xi C c^3} \right)^{4/7} \\
& \approx 476 \xi^{-4/7} \left( \frac{\alpha}{0.1} \right)^{24/35}  \left( \frac{M_{\bullet}}{10^6~{\rm M}_{\sun}} \right)^{4/35} \left( \frac{f_{\rm Edd}}{\eta_{0.1}} \right)^{-6/35} a_{\bullet}^{4/7}
R_{\rm S},
\end{aligned}
\end{equation}
where we expressed the proportionality between $\nu_1$ and $\nu_2$ as $\nu_2 / \nu_1 = \xi/(2 \alpha^2)$ (see Section \ref{subsec_angmom}; \citealt{papaloizou+83,lodato+07}).
Therefore, the timescale $\tau_{\rm align}$ in equation (\ref{eq_t_align}) can be obtained by plugging the expression for $R_{\rm warp}$ above into equation (\ref{eq_app_talign}).
We note from equation (\ref{eq_app_rwarp}) that $R_{\rm warp}$ grows with the black hole mass $M_{\bullet}$.
Given an accretion disc of mass $M_{\rm d}$, the warp radius may become even larger than the accretion disc radius $R_{\rm out}$ for sufficiently heavy black holes.
The critical condition $R_{\rm out} = R_{\rm warp}$ sets therefore a threshold black hole mass given $M_{\rm d}$.
For a total disc mass $M_{\rm d}$, $R_{\rm out}$ can be obtained by inverting equation (\ref{eq_app_md}); we can finally derive the threshold mass $M_{\bullet}^{\rm (warp)}$ in equation (\ref{eq_M_warp}) by matching the resulting $R_{\rm out}$ with equation (\ref{eq_app_rwarp}) for $R_{\rm warp}$ \citep{dotti+13}.


\section{Parameters at the innermost stable circular orbit}\label{appendix_afunc}

\begin{figure*}
\begin{center}
\includegraphics[width=2.1\columnwidth]{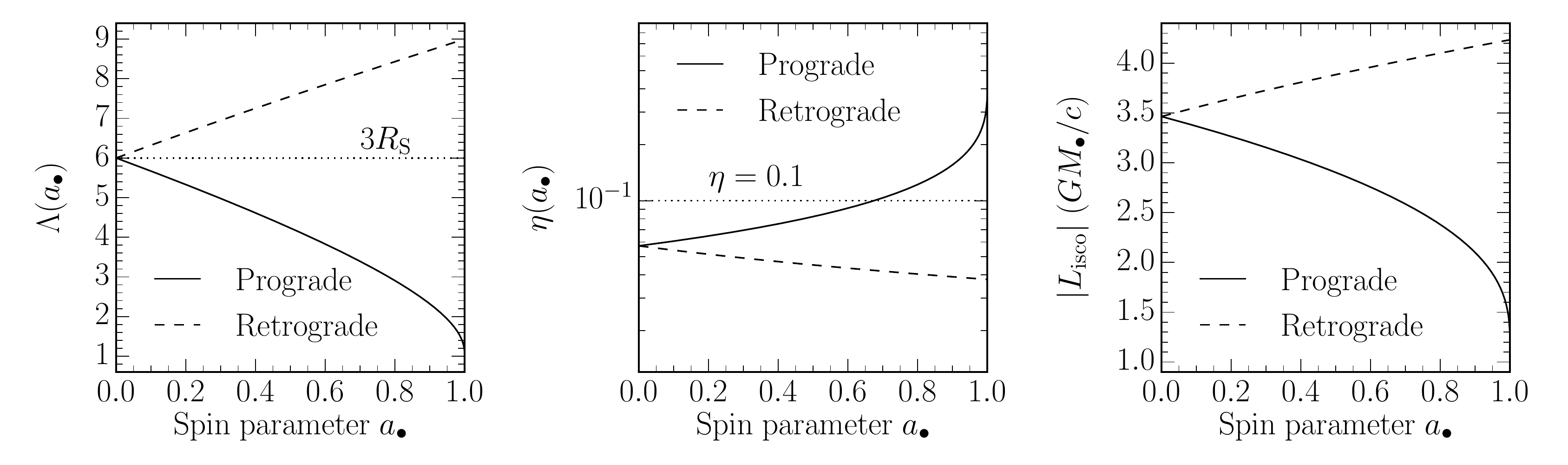}
\caption{Left panel: dimensionless extent of the innermost stable circular orbit $\Lambda(a_{\bullet})$ as a function of the spin parameter $a_{\bullet}$.
The horizontal dotted line indicates three Schwarzschild radii, $3 R_{\rm S} = 6 G M_{\bullet} / c^2$.
Middle panel: radiative efficiency $\eta(a_{\bullet})$ as a function of the spin parameter $a_{\bullet}$.
The horizontal dotted line indicates the reference radiative efficiency $\eta = 0.1$.
Right panel: specific angular momentum at the innermost stable circular orbit $\Lambda(a_{\bullet})$ as a function of the spin parameter $a_{\bullet}$.
In all panels, solid and dashed curves refer to prograde and retrograde cases, respectively.
}
\label{fig_spin_factors}
\end{center}
\end{figure*}

Here, we briefly collect for reference the expressions that we have used to describe $a_{\bullet}$-dependent quantities, namely the size of the innermost stable circular orbit $R_{\rm isco}$, the radiative efficiency $\eta$, and the gas specific angular momentum $L_{\rm isco}$ at $R_{\rm isco}$ \citep{bardeen+72}.
The size of $R_{\rm isco}$ can be written as
\begin{equation}
R_{\rm isco}(a_{\bullet}) = \Lambda(a_{\bullet}) \frac{G M_{\bullet}}{c^2},
\end{equation}
where the factor $\Lambda$ is:
\begin{equation}
\Lambda(a_{\bullet}) = 3 + Z_{2} \mp \sqrt{(3 - Z_{1}) (3 + Z_{1} + 2 Z_{2})},
\end{equation}
where the upper and lower signs refer to prograde and retrograde orbits with respect to the black hole spin, respectively; $Z_{1}$ and $Z_{2}$ are two functions of $a_{\bullet}$,
\begin{equation}
Z_{1}(a_{\bullet}) = 1 + (1 - a_{\bullet}^2)^{1/3} \left[ (1 + a_{\bullet})^{1/3} +  (1 - a_{\bullet})^{1/3} \right],
\end{equation}
and
\begin{equation}
Z_{2}(a_{\bullet}) = \sqrt{3 a_{\bullet}^2 + Z_{1}^2(a_\bullet)}.
\end{equation}

The radiative efficiency $\eta$ varies with the spin parameter $a_{\bullet}$ and it is related to $R_{\rm isco}$ via:
\begin{equation}
\eta(a_{\bullet}) = 1 - \sqrt{1 - \frac{2}{3 \Lambda(a_{\bullet})}}.
\end{equation}

Finally, the specific angular momentum at $R_{\rm isco}$ is a function of $a_{\bullet}$ and $\Lambda$, namely
\begin{equation}
L_{\rm isco}(a_{\bullet}) = \pm \frac{G M_{\bullet}}{c \Lambda} \frac{\Lambda^2 \mp 2 a_{\bullet} \sqrt{\Lambda} + a_{\bullet}^2}{(\Lambda - 3 \pm 2 a_{\bullet} / \sqrt{\Lambda})^{1/2}},
\end{equation}
where the upper and lower signs refer to prograde and retrograde orbits, respectively.
Figure \ref{fig_spin_factors} shows how $\Lambda(a_{\bullet})$, $\eta(a_{\bullet})$, and $L_{\rm isco}(a_{\bullet})$ vary with the spin parameter $a_{\bullet}$.


\section{Resolution tests}\label{appendix_restest}

\begin{table}
\caption{Summary of the simulations used to test resolution effects.}
\label{tab_res_runs}
\begin{tabular}{lccc}
\hline
Label & $N_{\rm neighbour}$ & $m_{\rm g}^{\rm target}$ & $\epsilon_{\rm g}$ \\
 & & (M$_{\sun}$) & (pc) \\
\hline
fr1 & 32 & 500 & 0.32 \\
fr2 & 64 & 500 & 0.32  \\
fr3 & 128 & 500 & 0.32  \\
fr4 & 256 & 500 & 0.32  \\
\hline
fm1 & 32 & 1000 & 0.4 \\
fm2 & 64 & 500 & 0.32 \\
fm3 & 128 & 250 & 0.25 \\
\hline
\end{tabular}
\flushleft
\end{table}

The behaviour of the accretion model that we have presented may be influenced by numerical resolution through the inflow quantities $\dot{M}_{\rm inflow}$ and $\bmath{L}_{\rm inflow}$.
In order to test that, we have run a set of additional simulations based on the circumnuclear disc initial conditions of run cnd2.
Specifically, we test the resolution dependence of the estimators in equations (\ref{eq_mass_flux_sph}), (\ref{eq_r_inflow_SPH}), and (\ref{eq_L_inflow_SPH}).
The simulations are summarised in Table \ref{tab_res_runs} and they are divided in two subsets.
In the first subset, dubbed ``frX'', we explore the effect of the number of neighbours $N_{\rm neighbour}$ used to calculate $\langle \Phi_{M} \rangle$, $\langle \Phi_{\bmath{J}} \rangle$, and $\langle r_{\rm inflow} \rangle$.
We use force and mass resolution of $\epsilon_{\rm g} = 0.32$~pc and $m_{\rm g}^{\rm targer} = 500$~M$_{\sun}$, respectively, and we change $N_{\rm neighbour}$ between 32 and 256 in steps of 2-factors on the same initial conditions.
We note that this is not the same resolution used for the production runs discussed in Section \ref{subsec_cnd}; the production runs have higher mass and force resolution, namely $m_{\rm g}^{\rm target} = 250$~M$_{\sun}$ and $\epsilon_{\rm g} = 0.25$~pc, while in these text simulations we used a slightly coarser reference resolution by consistently rescaling the original mass and force resolution by a factor $2$ and $2^{1/3}$, respectively.
However, this is of minor importance for the purpose of the discussion here, as we are interested in understanding the relative effects of changing resolution in twin runs.
On the other hand, the second subset, dubbed ``fmX'', explores the combined effect of mass and force resolution by changing the value of $m_{\rm g}^{\rm target}$ and $N_{\rm neighbour}$ by the same factor 2 in order to have the same mass within the black hole smoothing volume across different runs (with individual initial conditions).
Simultaneously, we decrease the force resolution by consistently increasing the gravitational softening by factor $2^{1/3}$.

\begin{figure*}
\begin{center}
\includegraphics[width=2.1\columnwidth]{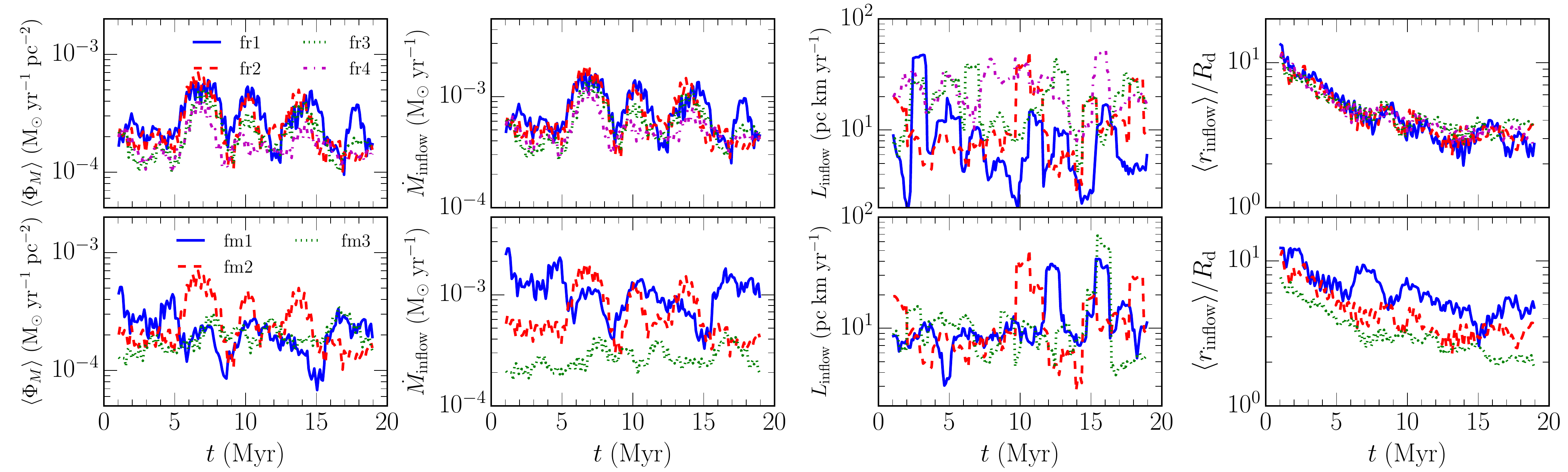}
\caption{Top row: results from runs at fixed resolution fr1 (blue solid curve), fr2 (red dashed curve), fr3 (green dotted curve), and fr4 (magenta dash-dotted curve), exploring the effect of $N_{\rm neighbour}$.
Bottom row: results from runs at fixed mass enclosed by the black hole smoothing radius fm1 (blue solid curve), fm2 (red dashed curve), and fm3 (green dotted curve), exploring the effect of different spatial and mass resolution.
From left to right: mass flux $\langle \Phi_{M} \rangle$, inflow rate $\dot{M}_{\rm inflow}$, modulus of the inflowing specific angular momentum $L_{\rm inflow}$, and $\langle r_{\rm inflow} \rangle / R_{\rm d}$ ratio.
Correspondence between runs and resolution is indicated in Table \ref{tab_res_runs}.}
\label{fig_res_time}
\end{center}
\end{figure*}

Figure \ref{fig_res_time} shows the time evolution of several quantities for both the run set fr and fm.
Specifically, we plot the mass flux $\langle \Phi_{M} \rangle$, the inflow rate $\dot{M}_{\rm inflow}$, the modulus of the specific angular momentum $L_{\rm inflow}$ of the inflowing gas, and the ratio $\langle r_{\rm inflow} \rangle / R_{\rm d}$ between the inflow radius and the accretion disc radius.
For the sake of graphical clarity of the plots, we smoothed all the curves exactly in the same way with a 10~Myr-wide top-hat filter.
The fr runs show that the estimator of $\langle \Phi_{M} \rangle$ depends rather weakly on $N_{\rm neighbour}$, and reasonable convergence seems to be achieved for $N_{\rm neighbour} \leq 64$.
The time evolution of $\dot{M}_{\rm inflow}$ shows very similar features.
This is due to the same mass and spacial resolution across the fr simulations.
Indeed, the ratio $\langle r_{\rm inflow} \rangle / R_{\rm d}$ is very similar across all the simulations because $\langle r_{\rm inflow} \rangle$ is a measure of the typical size of the mesh cells in proximity of the black hole, which is similar across the fr simulations.
We note however that there might be a weak trend with $N_{\rm neighbour}$: for increasing $N_{\rm neighbour}$, $\langle \Phi_{M} \rangle$ and $\langle r_{\rm inflow} \rangle$ respectively tend to decreases and increases slightly.
This is likely due to the larger smoothing volume that includes cells farther from the black hole, biasing therefore the estimate of $\langle r_{\rm inflow} \rangle$ and $\langle \Phi_{M} \rangle$.
On the other hand, the time evolution of $L_{\rm inflow}$ appears to converge much less with $N_{\rm neighbour}$.
Given the properties of the system, it could be reasonable to expect a rather constant value of $L_{\rm inflow}$ with time, with fluctuations due to both numerical noise and thermal motions of the gas.
When we calculate the mean and the variance of the values of $L_{\rm inflow}$ over time, we find that (i) the variance is comparable in all simulations, namely ${\rm Var}(L_{\rm inflow}) \approx 90$~pc$^2$~km$^2$~s$^{-2}$, and (ii) the mean value increases steadily but slightly with $N_{\rm neighbour}$.
This suggests that (i) $L_{\rm inflow}$ might converge only in average sense, and (ii) we are witnessing the effect of the increasing smoothing volume, as $L_{\rm inflow}$ increases including cells farther from the black hole that likely have larger specific angular momentum.

\begin{figure*}
\begin{center}
\includegraphics[width=2.1\columnwidth]{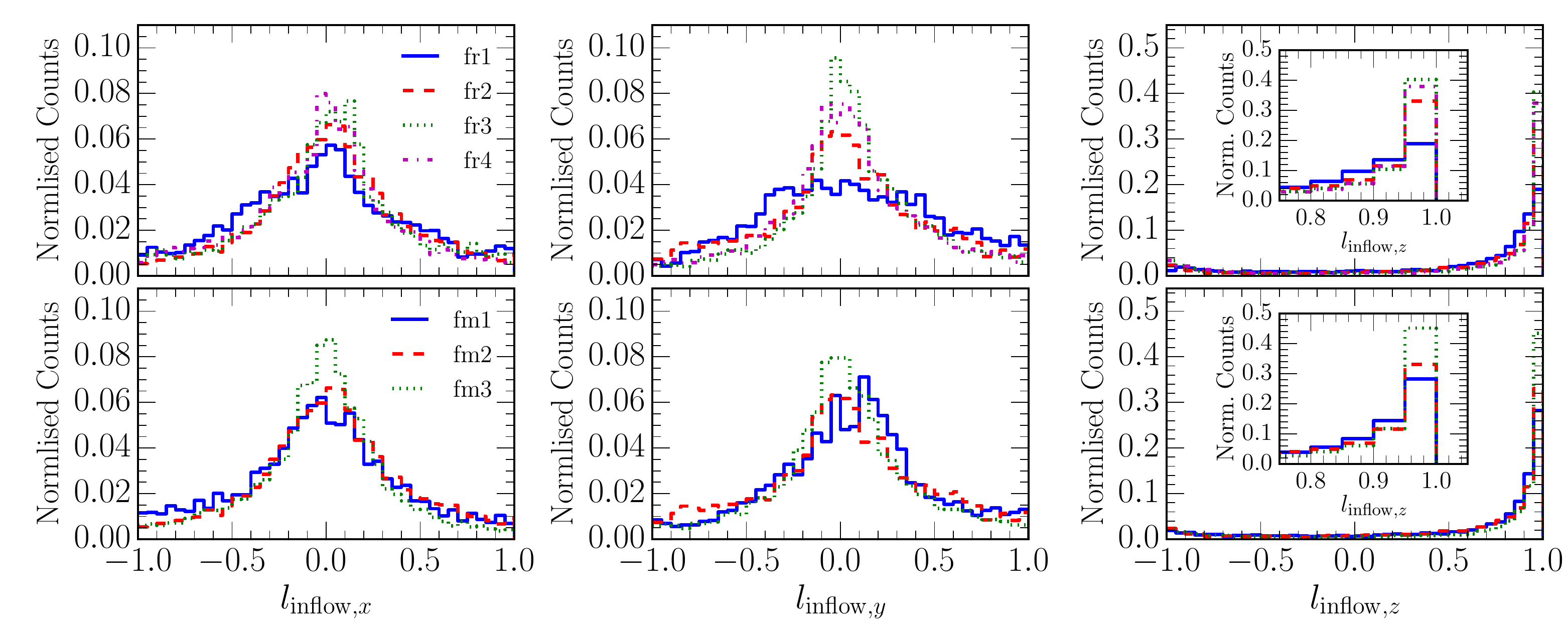}
\caption{Top row: results from runs at fixed resolution fr1 (blue solid curve), fr2 (red dashed curve), fr3 (green dotted curve), and fr4 (magenta dash-dotted curve), exploring the effect of $N_{\rm neighbour}$.
Bottom row: results from runs at fixed mass enclosed by the black hole smoothing radius fm1 (blue solid curve), fm2 (red dashed curve), and fm3 (green dotted curve), exploring the effect of different spatial and mass resolution.
From left to right: $x$, $y$, and $z$ component of the direction $\bmath{l}_{\rm inflow}$ of the inflowing specific angular momentum.
The insets in the rightmost column simply show a zoom around the peak at $l_{{\rm inflow}, z} = 1$.
Correspondence between runs and resolution is indicated in Table \ref{tab_res_runs}.}
\label{fig_res_dist}
\end{center}
\end{figure*}

\begin{figure*}
\begin{center}
\includegraphics[width=2.1\columnwidth]{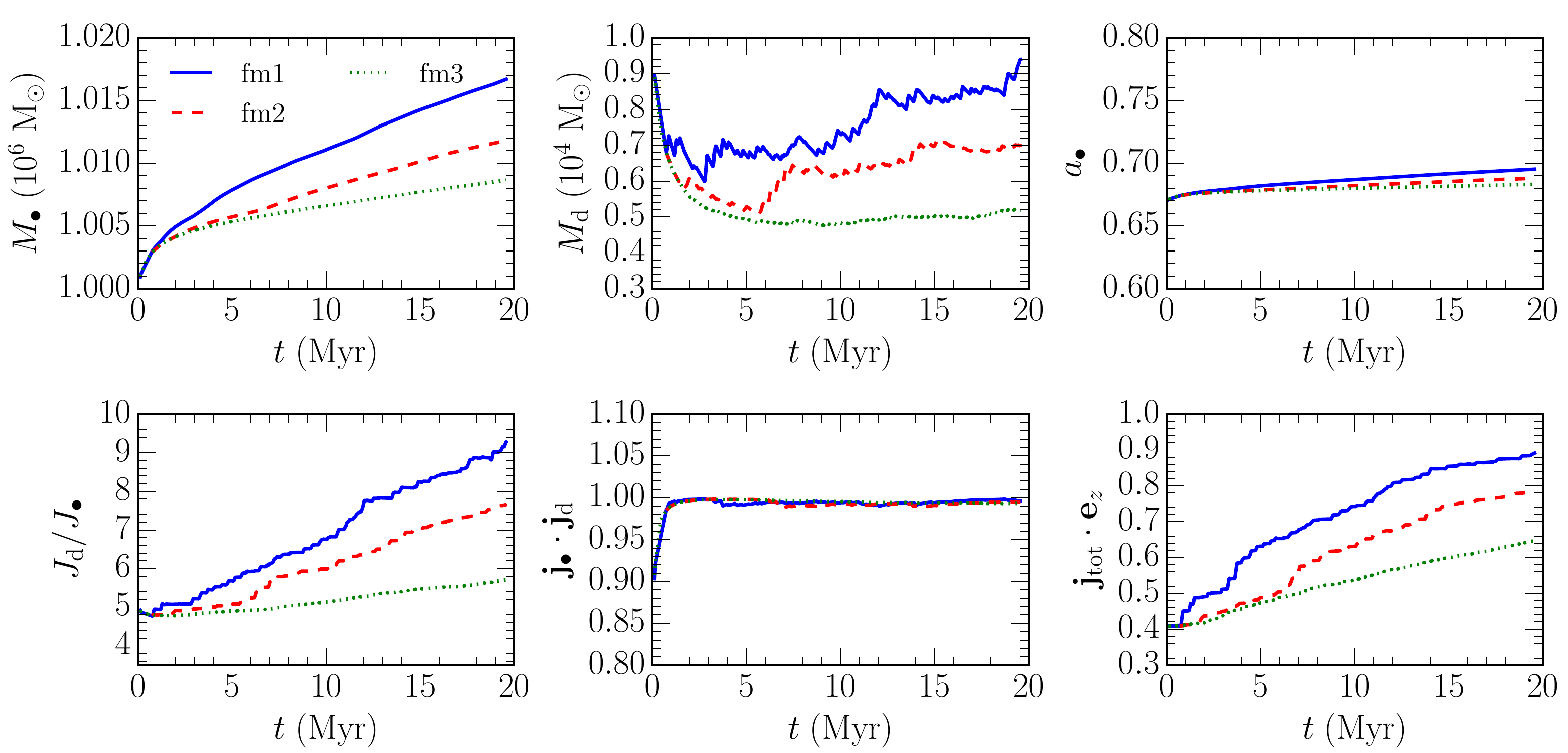}
\caption{Time evolution of several quantities of the accretion model as a function of mass and force resolution.
From left to right and from top to bottom: black hole mass $M_{\bullet}$, accretion disc mass $M_{\rm d}$, spin parameter $a_{\bullet}$, angular momenta ratio $J_{\rm d} / J_{\bullet}$, alignment $\bmath{j}_{\bullet} \cdot \bmath{j}_{\rm d}$, and alignment between the total angular momentum and the $z$ axis $\bmath{j}_{\rm tot} \cdot \bmath{e}_{z}$.
Blue solid curves, red dashed curves, and green dotted curves correspond to run fm1, fm2 and fm3, respectively.
Correspondence between runs and resolution is indicated in Table \ref{tab_res_runs}.}
\label{fig_res_bhspin}
\end{center}
\end{figure*}

The subset of simulations at different resolution with fixed mass enclosed by the black hole smoothing radius shows slightly different results.
Taking into account that the initial conditions of the simulations are different because the clearly include a different initial number of mesh-generating points, both $\langle \Phi_{M} \rangle$ and $L_{\rm inflow}$ converge rather well with resolution.
This is reassuring since these two quantities are intrinsic to the flow and they do not manifestly depend on resolution.
On the other hand, we can see that $\langle r_{\rm inflow} \rangle / R_{\rm d}$ changes with resolution, i.e. coarser runs have larger $\langle r_{\rm inflow} \rangle$.
This is not unexpected, as the inflow radius is explicitly and unavoidably resolution dependent.
Such a dependence enters in both the mass and angular momentum evolution equations [equations (\ref{eq_disc_mass_evol}) and (\ref{eq_disc_am_evol})] through $\dot{M}_{\rm inflow}$ and $\dot{\bmath{J}}_{\rm inflow}$, respectively.
In fact, we note that since $\dot{\bmath{J}}_{\rm inflow} \propto \dot{M}_{\rm inflow}$, the resolution dependence enters equally in the mass and angular momentum evolution, effectively as a rescaling of the amount of mass and angular momentum transferred to the accretion disc.

Figure \ref{fig_res_dist} shows the distributions of the Cartesian components of $\bmath{l}_{\rm inflow}$ for all our simulations.
The distributions qualitatively look as expected for the gaseous circumnuclear disc, namely a broadened peak around 0 for $l_{{\rm inflow}, x}$ and $l_{{\rm inflow}, y}$, and sharp spike around 1 with a tail at lower values for $l_{{\rm inflow}, z}$.
We find reasonable convergence in the distribution both for the suite of frX and fmX runs.
At fixed resolution (runs rfX), convergence is achieved as soon as $N_{\rm neighbour} \geq 64$, otherwise the tails of all distributions are more extended, likely because of numerical noise.
On the other hand, the direction of $\bmath{L}_{\rm inflow}$ seems to be only weakly affected by resolution (runs fmX).
We note however that comparing the distributions indicates convergence in a statistical sense, while the precise time evolution of each component of $\bmath{l}_{\rm inflow}$ may differ, as e.g. for the time evolution of $\dot{M}_{\rm inflow}$.

Different resolutions may indirectly impact the sub-grid accretion model through the differences in the boundary conditions imposed by $\dot{M}_{\rm inflow}$ and $\bmath{L}_{\rm inflow}$.
We explore such effects in Figure \ref{fig_res_bhspin}, where we show the time evolution of several quantities, namely $M_{\bullet}$, $M_{\rm d}$, $a_{\bullet}$, $J_{\rm d}/J_{\bullet}$, $\bmath{j}_{\bullet} \cdot \bmath{j}_{\rm d}$, and $\bmath{j}_{\rm tot} \cdot \bmath{e}_{z}$, for run fm1-3.
All the runs show a qualitatively similar behaviour during the short evolution time; however, they also show quantitative differences that can be clearly understood in terms of the trend seen above for $\dot{M}_{\rm inflow}$.
As the resolution increases, $\dot{M}_{\rm inflow}$ decreases, mostly because of the intrinsic resolution dependence of $\langle r_{\rm inflow} \rangle$.
As a consequence, the accretion disc mass tends to be larger at lower resolution.
As more mass gets dumped on to the accretion disc, also the ratio $J_{\rm d} / J_{\bullet}$ increases faster at lower resolution owing to the coherent direction of $\bmath{L}_{\rm inflow}$.
The net effect is to provide more intense accretion on to the central black hole, whose mass and spin parameter grow faster as the resolution decreases.
On the other hand, the alignment process is only weakly affected by the resolution as shown in the time evolution of $\bmath{j}_{\bullet} \cdot \bmath{j}_{\rm d}$.
However, the evolution of $\bmath{j}_{\rm tot}$ is more sensitive to resolution as it depends directly on $\dot{M}_{\rm inflow}$.
Similarly as above, the total angular momentum of the black hole+accretion disc system tends to align with the large scale circumnuclear disc rotation axis faster at lower resolution.
As a whole, we can summarise that resolution has a sort of ``rescaling'' effect.
As resolution changes the amount of inflowing material (but not too much its angular momentum), it makes the whole evolution of the black hole/accretion disc mass and angular momentum similar but faster or slower, depending on the actual value of the mass inflow rate.
Moreover, we note also that, at lower resolution, the effect of numerical noise is visible in the less smooth evolution of the accretion disc properties; however, this seems to be a secondary effect that does not alter significantly the time evolution predicted by the model.


\bsp
\label{lastpage}

\end{document}